\documentclass{aa}  
\usepackage{hyperref}
\usepackage{placeins}

\usepackage{xcolor}
\hypersetup{    
    colorlinks=true,
    citecolor = blue,
    linkcolor=blue,
    filecolor=magenta,      
    urlcolor=cyan,
}
\usepackage{multirow}

\usepackage{graphicx}
\usepackage{txfonts}
 \def\mso{\,\mathrm{M}_\odot}

 \def\msoy{\, \mso~{\rm yr}^{-1}}
 \def\kms{\, \rm km\,s^{-1}}
 
 \def\Msun{\,\mathrm{M}_\odot}
 \def\Rsun{\,{\rm R}_\odot}
 \def\Lsun{\,{\rm L}_\odot}

\newcommand{\case}[1]{Case\,#1}
\newcommand{\Case}[1]{Case\,#1}

 \def\qi{q_\mathrm i}
 
 \newcommand{\Rp}[1]{\mathcal R^{({#1})}}
 \def\days{\,\text{d}}
 
 \def\kyr{\,\text{kyr}}
 \def\erg{\,\mathrm{erg}}
 \def\ergs{\,\mathrm{erg}\,\mathrm{s}^{-1}}
 \def\Mcsm{M_\mathrm{CSM}}
 \def\Mej{M_\mathrm{ej}}
 \def\Mni{M_\mathrm{Ni}}
 \def\Eexp{E_\mathrm{exp}}
 \def\vej{v_\mathrm{ej}}

 \def\Rin{R_\mathrm{in}}
 \def\Rout{R_\mathrm{out}}
 \def\Eexp{E_\mathrm{kin,ej}}
 \def\Mshell{M}
 \def\Vshell{v}
 \def\cm{\,\mathrm{cm}}
 \newcommand{\Type}[1]{\text{Type}\,\text{#1}}
 
 \newcommand{\SN}[1]{\object{SN#1}}

 \newcommand{\multilinecomment}[1]{}
\newcommand{\yr}{\,\mathrm{yr}}
 \newcommand{\de}{\mathrm{d}}

\def\simle{\mathrel{\hbox{\rlap{\hbox{\lower4pt\hbox{$\sim$}}}\hbox{$<$}}}}
\def\simgr{\mathrel{\hbox{\rlap{\hbox{\lower4pt\hbox{$\sim$}}}\hbox{$>$}}}}

\newcommand{\EDIT}[1]{#1}
\newcommand{\REM}[1]{}

\usepackage{soul}
\defcitealias{WuFuller_IbcLMT}{WF22}
\defcitealias{Ercolino_widebinary_RSG}{Paper I}
\def\WF22{\citetalias{WuFuller_IbcLMT}}

\begin{document}

   \title{Mass-transferring binary stars as progenitors of interacting hydrogen-free supernovae }
    
   %\subtitle{}

   \author{A. Ercolino
          \inst{1}
          \and H. Jin \inst{1} \and
          N. Langer\inst{1,2} \and L. Dessart \inst{3}}

   \institute{Argelander Institut für Astronomie,
              Auf dem Hügel 71, DE-53121 Bonn, Germany\\
              \email{aercolino@astro.uni-bonn.de}
         \and
          Max-Planck-Institut für Radioastronomie, Auf dem Hügel 69, DE-53121 Bonn, Germany
          \and
Institut d'Astrophysique de Paris, CNRS-Sorbonne Universit\'e, 98 bis boulevard Arago, F-75014 Paris, France
}

   \date{Received Month Day, 2024; accepted Month Day, 2024}

% \abstract{}{}{}{}{} 
% 5 {} token are mandatory

 % THE OLD ABSTRACT IS IN backup abs.tex

  \abstract
   {Stripped-envelope supernovae (SNe) are hydrogen-poor
   transients produced at the end of the life of massive stars that
   have previously lost most or all of their hydrogen-rich envelope. The progenitors of most stripped-envelope SNe are thought to be donor stars in mass transferring binary systems, which were stripped of their hydrogen-rich envelopes some $10^6$\,yr before core collapse.
   A subset of the stripped-envelope SNe exhibit spectral and photometric features indicative of early, intense interactions between their ejecta and nearby circumstellar material (CSM), occurring within days or weeks of the explosion. 
   }
   {We examine whether Roche lobe overflow during, or shortly before, core collapse in massive binary systems can produce the CSM inferred from the observations of interacting H-poor SNe. %We explore to what extent these observations can be explained by massive binary system in which mass transfer induced by Roche lobe overflow is ongoing at, or shortly before, the time when the iron-core collapse occurs in the mass donor.
   }
   {We select 44 models from a comprehensive grid of detailed binary evolution models which are representative of the subset in which the mass donors are hydrogen-free and explode while transferring mass to a main-sequence companion.
   We characterize the properties of the pre-SN stellar models and of the material surrounding the binary at the time of the SN. %We computed single and binary stellar evolutionary models with the MESA code, selecting a range of initial masses and orbital configurations so as to result in the HeS undergoing another phase of mass transfer in the last $30\kyr$ before core collapse (CC). 
   }
{We find that in these models, mass transfer starts less than $\sim20\kyr$ before, and often continues until the core collapse of the donor star. Up to $0.8\Msun$ of hydrogen-free material are removed from the donor star during this phase, of which a large fraction may be lost from the binary system and produce He-rich circum-binary material. We explore plausible assumptions for its spatial distribution at the time of explosion. When assuming that the CSM accumulates in a circumbinary disk, we find qualitative agreement with the supernova and CSM properties inferred from observed \Type Ibn SNe, and to a lesser extent with constraints from \Type Icn SNe. Considering the birth probabilities of our mass transferring stripped envelope SN progenitor models, we find that they may produce up to $\sim$10\% of all stripped envelope supernovae.}
{The generic binary channel proposed in this work can qualitatively account for the observed key properties and the observed rate of interacting H-poor SNe. Models for the evolution of the circumbinary material and for the spectral evolution of exploding progenitors from this channel are needed to further test its significance.
}

   \maketitle
%
%-------------------------------------------------------------------

\section{Introduction}
%\LUC{Style can be personal but a good guide is to always try to be as concise as possible. There is a tension here in connecting Ibn to Ibc in the sense that Ibn never look like Ibc, neither at discovery nor at late times. So, I am not sure Ibn are a subgroup of Ibc. They are, however, a subgroup of stripped envelope SNe so that might be a better way to presenting them.}
%\ANDREA{I think I addressed this in the abstract and introduction - you might want to take a look at it anyway}

%--------------------------------------------------------------------
Many massive stars end their lives as supernovae (SNe), bright transients which are {observable even in distant} galaxies.  Past surveys (PTF, \citealt{PTF}, and iPTF \citealt{iPTF}), and especially ongoing ones (e.g., ATLAS, \citealt{ATLAS}, ASAS-SN, \citealt{ASASN},  ZTF, \citealt{ZTF}) have dramatically increased the number of observed SNe over the last few decades, which led to the discovery of new, particular SN types. This is expected to continue in the coming years and decades, particularly thanks to the beginning of new surveys such as LSST \citep{LSST}. 

These surveys are providing growing evidence that the light of as much as $\sim\,10\%$ of all core-collapse (CC)SNe is affected, sometimes even dominated, by the interaction of the SN ejecta with circumstellar material (CSM) surrounding the exploding star \citep[e.g.,][]{Perley2020_ZTF_SNdemographics}. The relatively slow-moving CSM can produce narrow-line features in the SN spectrum, which prompts the classification of such transients as \Type IIn (with narrow H-emission lines), \Type Ibn (He), \Type Icn (C/O) and even \Type Ien \citep[S/Si, ][]{Schulze2024_narrowSSi_lines} SNe. In some instances, the interaction power is so strong as to result in superluminous (SL)SNe \citep{Chatzopoulos2012_generalizedlightcurvefitting} and can also explain more exotic events like some Fast Blue Optical Transients (FBOTs, e.g., AT-2018cow, \citealt{Margutti19_AT2018cow, Pastorello2022_CSM_of_Ibn, Ho2023_FBOT_extended_or_interaction}). The observed presence of CSM hence offers a valuable diagnostic for probing the late stages of stellar evolution. %\textcolor{red}{[RETHINK] The presence of this material is a crucial tool to understand the final phases of stellar evolution.}

Recent studies have investigated the characteristics of the CSM in interacting H-poor SNe, through explosion modeling \citep[e.g.,][]{Dessart2022_Ibn, Takei24_IbcnLCmodelling}, or by analyzing the light-curves of observed events \citep[e.g., ][]{Chatzopoulos2012_generalizedlightcurvefitting, Chatzopoulos2013_analiticalmodel}. For \Type Ibn and Icn SNe, not all events of the same class are associated with the same progenitor and CSM structure \citep[e.g.,][]{TurattoPastorello14_varietyIbn, Pellegrino2022_DiversitySNIcn}. Some, like the \Type Ibn SN\,2019uo, \citep{Gangopadhyay2020_2019uo}, and the \Type Icn SN\,2022ann, \citep{Davis23_SN2022ann} are highly energetic explosions associated to massive WR stars. Others, such as ultra-stripped (US) \Type Ibn SN\,2019dge \citep{Yao2020_2019dge}, are much less energetic and likely originate from low-mass progenitors. %, in particular those identified as ultra-stripped (US)SNe \citep[e.g., the \type Ibn SN\,2019dge,][]{Yao2020_2019dge}. 
These SNe prompt questions about the progenitors and the CSM required to explain the observations, as canonical single-star models struggle to reproduce the inferred CSM for these SNe. 

Some works suggest that single stars can undergo strong phases of mass loss due to mechanisms such as thermonuclear flashes in the core \citep{Woosley1980_TypeI_flashes, Woosley_Heger_2015_SiFlash}, wave-driven envelope excitations \citep{Quataert_Shiode_wavedriven_winds, WuFuller21_wavedrivenoutburst}, and LBV-type eruptions \citep{SmithArnett2014_Hydroinstab_Turb_preSN}. If triggered shortly before CC, these mechanisms not only produce the CSM, but also explain the observed pre-SN outbursts associated with some interacting SNe. 

Most massive stars are born as members of close binary systems \citep{Sana_massive_stars_binaries}, of which many undergo mass transfer to their companion star through Roche lobe overflow (RLOF) during their evolution\citep{Podsiadlowski_massive_star_binary_interaction_1992}. In most interacting binaries, this happens as a consequence of the post main-sequence expansion of the donor star, long before a \EDIT{SN} explosion occurs in the binary. However, some donor stars expand strongly after core helium exhaustion, which can trigger a late RLOF that continues until the mass donor's core collapses. For non-conservative mass transfer, this situation can lead to a dense CSM surrounding the binary system at the time of the \EDIT{SN} explosion. This is found in binary evolution models containing H-rich \citep{Matsuoka_Sawada_BinaryInteraction_IIP_Progenitors, Ercolino_widebinary_RSG} or H-poor donor stars \citep[][ referred to hereafter as \WF22]{WuFuller_IbcLMT}. This binary channel enables stars of the same initial mass to follow various evolutionary paths, depending on their initial orbital configuration, which may relate to the variety of \Type II SNe progenitors \citep[e.g., in wide binaries:][]{Ouchi2017_IIb_RSG_progenitors, Matsuoka_Sawada_BinaryInteraction_IIP_Progenitors, Ercolino_widebinary_RSG, Dessart24_IIP_to_Ib_fuErk24}, as well as \Type IIb \citep{Claeys_b, Sravan_b, Yoon_IIb_Ib} and \Type Ibc SNe \citep{Yoon2010_Ibc, Dessart2011_SESNe_from_CCofWR}. 
% from lower mass progenitors than predicted from single-star evolution. In the context of interacting SNe, \cite{Ercolino_widebinary_RSG} showed that different initial orbital configurations alone can provide diverse CSM and progenitor properties for \Type IIn SNe and SLSNe-II. 

In stripped-envelope SNe, the exploding star is a naked He-star (HeS). It has been established that low-mass HeS ($\simle3.5\Msun$) expand strongly after core-helium burning (e.g., \citealt{Paczynski1971_HeStars_firstGiant, Habets1986_HeS_RLOF, WLW1995_preSN_HeS, WL99_closebinaries_BH&SNe, Yoon2010_Ibc, KleiserFullerKasen18_HeG_RapidlyFading_Ibc, Woosley2019_Hestars, Laplace2020_Stripped_Envelope_SNe}, \WF22). In binaries, this expansion can trigger mass transfer. If some of the transferred mass escapes the system, it could form the CSM near the binary, which may interact with the ejecta during the first SN explosion that occurs in the binary.%, and it may develop interaction features with the SN ejecta.

In this paper, we investigate how interacting \Type Ibc SNe can arise in binaries undergoing multiple phases of stable mass transfer. Unlike many studies that focus on binaries with compact object companions, such as a black hole (BH, e.g. \citealt{Jiang2023_HeS+BH_GWprog}) or a neutron star (NS, e.g. \citealt{Tauris2013_USSNe_Ic,Tauris2015_USSNe, JiangTauris2021_USSNeBinary_until_CC, Jiang2023_HeS+BH_GWprog, Guo2024_ECSNe_caseX}, \WF22), this work examines the more abundant systems where the exploding star is the initially more massive star in the binary, orbiting a main-sequence (MS) companion. Although such binaries have been modeled before, e.g. \citep{Habets1986_HeS_RLOF, Laplace2020_Stripped_Envelope_SNe} and \WF22, we use a comprehensive approach here with the aim of covering the diversity of pre-SN CSM structures and to explore the frequency with which they occur in the universe.

 The paper is structured as follows. We begin by detailing the physics of the stellar evolutionary models (Sect.\,\ref{sec:methods}) and the parameter space for pre-SN mass transfer (Sect.\,\ref{sec:ps}). We then analyze binary evolution models and their pre-SN properties (Sect.\,\ref{sec:res}) followed by the impact of the CSM on the SN (Sect.,\ref{sec:SN}) before comparing between the model predictions with observed H-poor interacting SNe (Sect.\,\ref{sec:SN:compare}). Finally, we discuss key physical processes and uncertainties that affect our results (Sect.\,\ref{sec:disc}) before summarizing the results (Sect.\,\ref{sec:conclusions}).

\section{Method and assumptions}\label{sec:methods}
In this work, we study binary models that develop into HeS+MS systems undergoing mass transfer shortly before CC (which we refer to as \Case BC RLOF, cf. Sect.\,\ref{sec:ps:binary_grid}). These systems are potential progenitors of interacting H-poor SNe, assuming that some of the transferred mass escapes to form the CSM.   

Three series of stellar evolutionary models run with MESA \citep[r10398,][]{MESA_I, MESA_II, MESA_III, MESA_IV, MESA_V} are used. The first is a \EDIT{comprehensive} large-scale grid of previously calculated detailed binary evolutionary models (Jin et al., in prep.)  which, due to the numerical settings used, do not produce many mass-transferring HeS+MS binaries (cf., Sect.\,\ref{sec:ps}). To expand the coverage of systems undergoing \Case BC RLOF, we produce two other sets of models. The first is a set of detailed single HeS models, used to diagnose models in the \EDIT{binary grid} where the stripped primary (i.e., the initially more massive star) is expected to expand and trigger \case BC RLOF. By comparing the \EDIT{binary grid} with the single HeS models, we select and run a representative set of models \EDIT{in the binary grid} up to CC, using improved physics and numerical settings to better tackle their evolution. Throughout this work, we refer to the HeS primary in a binary system as a binary-stripped HeS, in contrast to a single HeS model.

The three sets of models share key settings as described in \cite{Ercolino_widebinary_RSG} and \cite{Jin2024_Boron}, including the initial metallicity \citep[$Z=Z_\odot$, with $Z_\odot=0.154$ from][]{Zsun_Asplund2021}, and wind mass-loss prescription. In the following, we present the differences in the physics assumptions between the sets.

\subsection{Initialization and rotation}\label{sec:methods:init}
In both the \EDIT{binary grid} and our binary models, each star is initialized at ZAMS with an initial rotation set to $20\%$ of its breakup velocity \citep{VLT_FLAMES_X}.

The single HeS models are initialized following \citet{DR1,DR2}, where a MS star is evolved with artificial full mixing and no winds until core H depletion. Afterwards, artificial mixing is disabled and the winds are introduced, allowing the model to evolve as a HeS.
Rotation is neglected in single HeS models as, in the \EDIT{binary grid}, the binary-stripped HeS models rotate very slowly during core He burning.

\subsection{Nuclear network and termination}\label{sec:methods:network}
The \EDIT{binary grid} adopts the \texttt{stern} nuclear network \citep{HLW2000, Jin2024_Boron}, and the models are terminated after the end of core C-burning, (Jin et al., in prep.; however some models are also terminated beforehand, see Sect.\,\ref{sec:methods:RLOF}). 

We adopt the \texttt{approx21} network in the single and binary-stripped HeS models, as it includes the bare minimum number of isotopes to evolve the models until CC.  As our parameter space focuses on low-mass cores (cf. Sect.\,\ref{sec:ps}), the nuclear network adopted here is suboptimal \citep[cf.,][]{Farmer2016_preSNmodels_network}. The next smallest viable nuclear network in MESA, which contains 75 isotopes, increases the computation time, hindering the feasibility of running a grid of models. 

\subsection{Overshooting and non-adiabatic convection}\label{sec:methods:overshooting}
Overshooting is included in the convective H-burning core based on the initial mass, following \cite{Hastings_BeFractions}. After core H-burning, overshooting is not included in any of the convective boundaries until the end of core C burning. 
After core C burning, when the models in the \EDIT{binary grid} are terminated, our single and binary-stripped HeS models introduce small-scale over and undershooting across all convective boundaries, extending only $0.008H_P$ from the boundary (with $H_P$ being the local pressure scale height at the convective boundary). This reduces computation difficulties, especially as successive core burning phases ignite off-center \EDIT{(see the example model in Sect.\,\ref{sec:res:example_model})}, causing convergence issues. 
Given the number of convective regions that develop as heavier elements are burnt, the adoption of convective over and undershooting may yield considerable differences in the final core structure.

Lastly, the \EDIT{binary grid} adopts MLT++ \citep{MESA_III} to handle convectively unstable regions with inefficient energy transport (i.e., subsurface convective regions) by artificially reducing the local superadiabaticity. This can lead to smaller radii in models with convective layers near the surface, as well as density inversions. We therefore employ the standard MLT in both our single- and binary-stripped HeS models, rather than MLT++.

\subsection{Roche lobe overflow}\label{sec:methods:RLOF}
The \EDIT{binary grid} of models adopts different mass transfer schemes depending on the primary star's evolutionary phase. If the primary is still a MS star, the mass transfer scheme is the `contact' scheme \citep{Pablo16_merginmassiveBG_contactScheme}. After the MS, the mass transfer scheme adopted is that of \citet{Kolb_scheme} to handle RLOF from a convective envelope. 

Since we run binary models that begin RLOF after the MS (cf. Sect.\,\ref{sec:ps}), we adopt the scheme from \cite{Kolb_scheme}, where we also consider the radiation pressure of the donor star in the mass-transfer calculation, as done in \WF22. Our treatment is the same as that in \cite{Ercolino_widebinary_RSG}. We assume that the transferred mass also carries angular momentum and that accretion is halted when the accretor reaches critical rotation \citep{Petrovic2005_WRO_RLOF_constraint}. In the models from the \EDIT{binary grid}, this typically results in a small amount of mass being accreted, around $\simle 0.3\Msun$ during \case B RLOF.  The mass that fails to be accreted is assumed to be lost from the binary system, carrying away the specific angular momentum of the accreting star. We assume that \case BC RLOF is completely inefficient (see  Sect.\,\ref{sec:res:example_model}).

 {In the \EDIT{binary grid}, a model is terminated when unstable mass transfer and therefore common envelope (CE) evolution ensue. Unstable RLOF is assumed to occur if the model exhibits inverse mass transfer, RLOF at ZAMS, Darwin instability \citep{Darwin1879_StabilityBinary}, or if it becomes an overcontact binary outflowing material from the L2 point}. An additional stopping condition in the \EDIT{binary grid} occurs when a model reaches a mass transfer rate of $10^{-1}\msoy$. We do not apply this condition in our binary models. We additionally consider mass transfer to turn unstable when the donor star reaches the volume-equivalent radius associated to its outer Lagrangian point ($\sim\,1.3R_\mathrm{RL,1}$, \citealt{Pablo_Kolb}). This can lead to outer Lagrangian-point outflows \citep[OLOF, ][]{Pavlovskii_Ivanova_2015_MT_from_Giants, Ercolino_widebinary_RSG}. These conditions provide a broad parameter space to investigate models (but see Sect.\,\ref{sec:disc:MT_stability} for a discussion using different assumptions).

\section{Parameter Space}\label{sec:ps}
The key ingredient for an interacting H-poor SN is the presence of a dense, H-poor CSM. Assuming that the CSM forms from the material not accreted by the companion star during RLOF shortly before CC, we search for H-poor donors that will undergo \case BC RLOF. 

 Since \EDIT{binary grid} underestimates the parameter space for models undergoing \case BC RLOF (cf. Sect.\,\ref{sec:ps:binary_grid}), we aim to identify the systems that are expected to do so. We first investigate the radial expansion of single HeS models in their later evolutionary stages (Sect.\,\ref{sec:ps:single_he}). We then compare the single HeS models with the binary-stripped HeS models in the \EDIT{binary grid} at the end of core He burning to assess which models are expected to undergo RLOF before CC (Sect.\,\ref{sec:ps:binary_grid}). This enables us to select the initial parameters of the models to recompute (Sect.\,\ref{sec:ps:grid}).

\subsection{Single HeS models}\label{sec:ps:single_he}

\begin{table*}[]
\caption{Properties of the single HeS models.}

\centering 
\resizebox{\textwidth}{!}{
\begin{tabular}{lcccccccccccccr}
\hline\hline
$M_\mathrm i$ & $M_\mathrm{He,conv}^\mathrm{max}$ & $M_\mathrm{He-dep}$  & $M_\mathrm{CO,max}$ & $\Delta t_\mathrm{C-end}$ & $R_\mathrm{max}^\mathrm{20kyr}$ & $R_\mathrm{max}^\mathrm{10kyr}$ & $R_\mathrm{max}^\mathrm{5kyr}$ & $R_\mathrm{max}^\mathrm{1kyr}$ & $R_\mathrm{max}^\mathrm{1yr}$ & $R_\mathrm{max}$& $\log \rho_\mathrm{c}$ & $\log T_\mathrm{c}$ & $\log T_\mathrm{max}$ & End \\
$\Msun$ &  $\Msun$ & $\Msun$  & $\Msun$ & kyr & $\Rsun$& $\Rsun$ & $\Rsun$ & $\Rsun$ & $\Rsun$ & $\Rsun$ & $\mathrm{g}\,\mathrm{cm}^{-3}$ & K & K & of run\\
\hline
(1) & (2) & (3) & (4) & (5) & (6) & (7) & (8) & (9) & (10) & (11) & (12) & (13) & (14) & (15) \\ \hline
2.99 &  1.16 & 2.49 & 1.38 & 31.8 & 23.4 & 133  & 152  & 195  & 499  &4254  & 8.93 & 8.77 & 9.11  & Ne$^\mathrm{EC}$*\\
3.16               &  1.26 & 2.61 & 1.43 & 27.0 & 11.0 & 36.0 & 123  & 191  & 191  & 237  & 9.07 & 9.00 & 9.49 &  Ne/O*  \\
3.35               &  1.36 & 2.73 & 1.52 & 23.1 & 5.37 & 15.3 & 37.7 & 171  & 185  & 185  & 8.03 & 9.03 & 9.44  & O\\
3.55               &  1.48 & 2.86 & 1.61 & 19.6 & 2.64 & 8.90 & 17.4 & 74.6 & 166  & 225  & 8.89 & 9.34 & 9.56  & Si*\\
3.76               &  1.61 & 2.99 & 1.70 & 16.6 & 1.79 & 6.39 & 10.6 & 21.0 & 57.4 & 173  & 8.81 & 9.22 & 9.58  & Si*\\
3.98               &  1.75 & 3.13 & 1.80 & 14.0 & 1.41 & 4.28 & 7.55 & 12.9 & 20.4 & 20.4 & 7.94 & 9.23 & 9.57  & Si\\
4.22               &  1.90 & 3.28 & 1.91 & 12.0 & 1.17 & 2.63 & 5.97 & 9.54 & 12.2 & 12.2 & 9.53 & 9.84 & 9.84  & Si\\
4.47 &  2.06 & 3.43 & 2.02 & 10.0 & 0.99 & 1.91 & 4.61 & 7.69 & 9.14 & 9.14 & 10.2 & 10.0 & 10.0  & CC\\
4.73 &  2.24 & 3.59 & 2.14 & 8.4  & 0.84 & 1.51 & 3.19 & 6.50 & 7.50 & 7.50 & 10.2 & 10.0 & 10.0  & CC\\
5.01 &  2.43 & 3.75 & 2.28 & 7.1  & 0.72 & 1.24 & 2.27 & 5.48 & 6.36 & 6.36 & 10.1 & 10.0 & 10.0  & CC\\
5.31 &  2.64 & 3.92 & 2.41 & 6.0  & 0.62 & 1.05 & 1.78 & 4.52 & 5.51 & 5.67 & 10.1 & 9.99 & 9.99  & CC\\
\end{tabular}}
\tablefoot{The columns show the initial mass (1), the maximum extent of the convective He-burning core (2), the mass at core helium depletion (3), the maximum mass of the CO-core (4), the time between core C-ignition and the end of the run (5), the maximum radius exhibited between core He-depletion and $20\kyr$ before the end of the run (6), $10\kyr$ (7), $5\kyr$ (8), $1\kyr$ (9), $1\yr$ (10), and the absolute maximum value (11). Columns 12-14 show the central density, central and maximum temperature in the model at the 6end of the run respectively. Column 15 highlights the burning phase at the end of the run  labeled according to the nuclei being burnt, if it didn't reach CC because of numerics. (*) Models have experienced a strong expansion of the envelope before terminating (see Sect.\,\ref{sec:ps:single_he}). (EC) The lowest-mass model is expected to end as an EC-SN.   }
\label{tab:data_sHeS}
\end{table*}

\begin{figure}
    \includegraphics[width=1\linewidth]{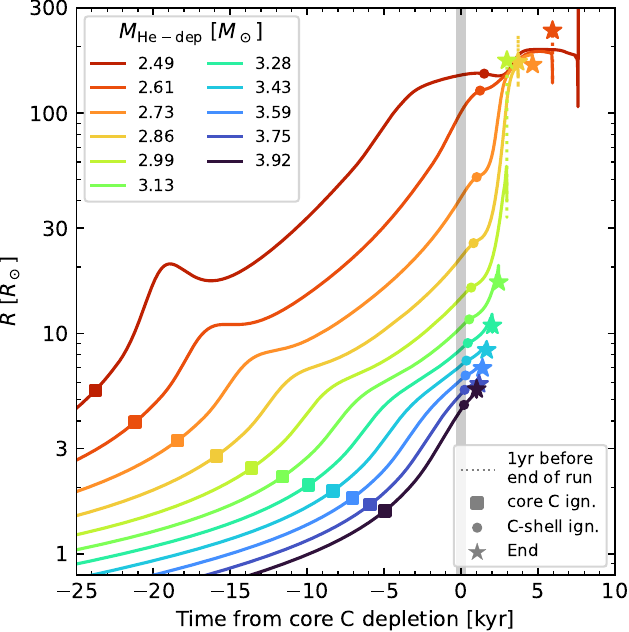}
    \caption{Evolution of the radius of single HeS models as a function of time during and after core carbon burning, with $t=0$ set at core carbon depletion. The square markers indicate core carbon ignition, circles  indicate the time when the first carbon shell develops, and stars mark the end of the calculation. For the lowest mass model, no endpoint is shown (see Sect.\,\ref{sec:ps:single_he}).}
    \label{fig:single_he_rad}
\end{figure}

We compute single HeS models with initial masses between $2.99$ and $5.31\Msun$ and focus on their radius evolution (cf. Table\,\ref{tab:data_sHeS} and Fig.\,\ref{fig:single_he_rad}).

During core He burning, these stars have radii below $\sim\,1\Rsun$. Stellar winds reduce their mass over time,  resulting in a decrease of the convective He-burning core mass, which affects the final CO-core mass. 
We label the models by their mass at core He-depletion ($M_\text{He-dep}$), prefixed with  `He'. For example, model He2.49 is the lowest-mass the model, with an initial mass $M_i=2.99\Msun$ and $M_\mathrm{He-dep}=2.49\Msun$.  

After core He-burning, shell He-burning drives the expansion of the He-rich envelope, which is more significant for lower mass models (cf., Fig.\,\ref{fig:single_he_rad}).
The expansion continues during C-burning. In the lowest mass models, the expansion temporarily stalls (models He2.49, and He2.61), due to the more significant decrease of the power of the He-burning shell. The stellar expansion continues beyond core-C exhaustion. In the $\sim\,2-3\kyr$ that follow the ignition of the first C-burning shell, the radius grows significantly, with the lowest mass models reaching $\sim\,200\Rsun$. More massive models (with $M_\mathrm{He-dep}>3.59\Msun$) explode before they can expand significantly. Following C-shell burning, the He-burning shell becomes convective for all models. In the least massive models, which developed the largest radii (i.e., models from He2.49 to He2.99), the outer \EDIT{\REM{the}} envelope also turns convective.

After Ne and O ignite in the core, the envelope contracts significantly in smaller-mass models (cf. Fig.\,\ref{fig:single_he_rad}) but remains largely unaffected in higher-mass ones. Models He2.86 and He2.99 exhibit ingestion of He inside the CO-core shortly before core-Si ignition, causing the envelope to expand. \EDIT{In the final stages before the end of the run, model He2.86 also exhibits a numerical oscillation of the envelope}. In models He2.61 and He2.49, the surface and the He-burning convective regions merge, leading to significant radius expansion. 

The key results of the simulations are shown in Table\,\ref{tab:data_sHeS}. All models {, except the least massive, are expected to end their evolution as CC-SNe, even though many terminated shortly before CC}.
In particular, the models' central density and temperature increase towards \EDIT{\REM{towards}} the end of the run, resembling the CC progenitors in \cite{Tauris2015_USSNe}. The lowest-mass model in the set (He2.49) evolves with an almost constant central temperature (below that achieved during core C burning) and increasing densities, resembling the electron-capture (EC) SN progenitor in \cite{Tauris2015_USSNe}. HeS models with \EDIT{similar masses to} that of model He2.49 are expected to explode as EC-SNe, while those at lower masses will turn into a WD.

\begin{figure}
    \includegraphics[width=\linewidth]{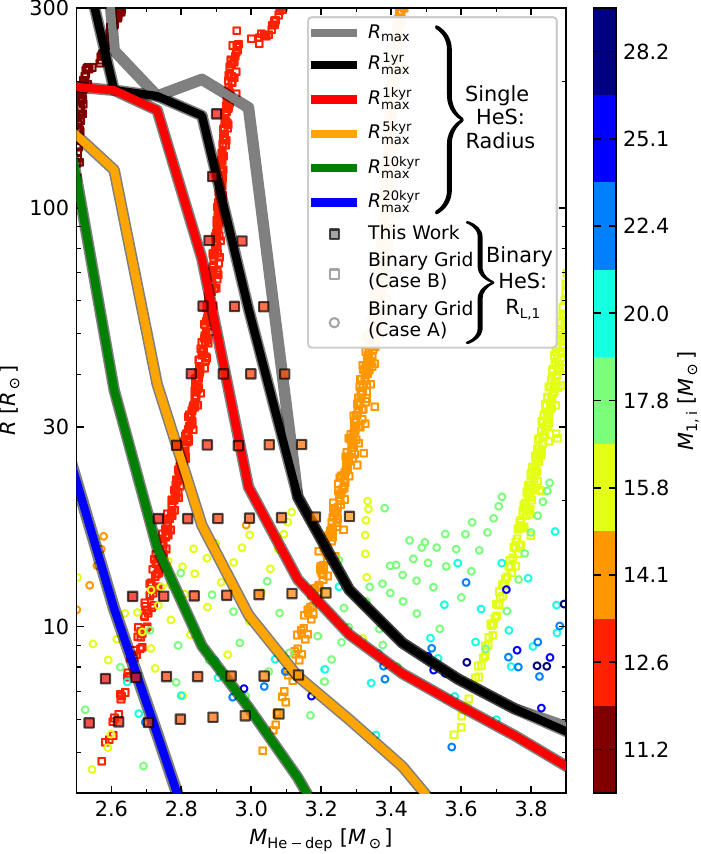}
    \caption{{ Mass-radius diagram for single HeS models (showing the stellar radius, lines) and binary-stripped HeS models (showing the Roche lobe radius, symbols) as a function of the mass at core He-depletion $M_\mathrm{He-dep}$. For single HeS models, the radius at a time $t$ before the end of the run ($R_\mathrm{max}^{t}$) is plotted for different $t$ (see legend). Binary-stripped HeS models are shown for different initial primary masses $M_\mathrm{1,i}$ (see colorbar), and different markers are used for \Case A (circles) and \Case B systems (squares). The binary-stripped HeS models are distinguished between those from the \EDIT{binary grid from Jin et al. (in prep., empty markers}), and this work (\EDIT{filled markers with black outline}).}}
    \label{fig:R_vs_RL}
\end{figure}

These models enable mapping the radius evolution as a function of mass of the HeS. We build a mass-radius diagram (cf. Fig.\,\ref{fig:R_vs_RL}) showing how long before CC a single HeS of mass $M_\mathrm{He-dep}$ reaches a radius $R$. 

\subsection{Comparing HeS models to the binary grid}\label{sec:ps:binary_grid}

 \begin{figure}
   \includegraphics[width=\linewidth]{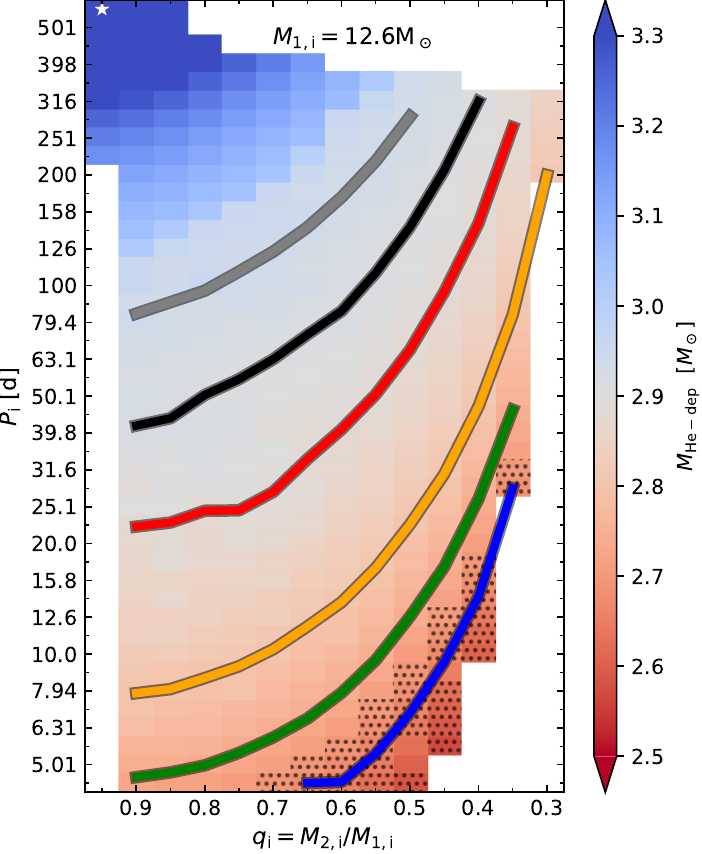}
    \caption{The $\log P_\text{i}-\qi$ diagram for the \EDIT{binary} grid \EDIT{of models from Jin et al. (in prep.)}, with $M_\mathrm{1,i}=12.6\Msun$,  {where each pixel represents one detailed binary evolution model}. The color coding represents the mass of the primary star at core helium depletion.  The solid lines cover the models where the radius of a single HeS of the same mass 
    (cf. Fig.\,\ref{fig:R_vs_RL}) matches their Roche lobe radius at core He depletion.  White pixels are models that terminated before core-carbon ignition (see Sect.\,\ref{sec:methods:RLOF}). The dotted patches correspond to the models which exhibited \Case BC RLOF in the \EDIT{binary grid}. The star marker in the top-left corner corresponds to the lowest-period model shown in \citet{Ercolino_widebinary_RSG}.}
    \label{fig:logPq_B}
\end{figure}

Using the mass-radius map from single HeS models (Sect.\,\ref{sec:ps:single_he}), we identify the binary-stripped HeS models \EDIT{in the binary grid that are expected} to undergo \case BC RLOF. This is done by comparing their Roche lobe radius ($R_\mathrm{RL,1}$) with the expected radius evolution of a single HeS model with the same $M_\mathrm{He-dep}$.
\EDIT{\REM{We focus exclusively on models undergoing Case\,B RLOF as the first phase of mass-transfer. We briefly discuss Case A systems in Appendix\,A}}.

 \EDIT{In the models in the binary grid} where \case B is the first mass-transfer phase, the donor does not lose a fixed amount of mass or the entire H-rich envelope, as often assumed in rapid-binary evolution codes \citep[e.g., ComBinE;][]{Kruckow18_COMBINE}. Rather, the amount of mass-loss depends on the initial mass ratio ($\qi$) and orbital period  ($P_\mathrm i$), leading to varying degrees of stripping for primaries of same initial mass ($M_\mathrm{1,i}$). Moreover, during core He burning, core growth depends on the mass of the leftover H-rich envelope \citep[e.g.,][]{Ercolino_widebinary_RSG}. If the H-envelope mass is $\simle 1\Msun$, the long timescales for core-He burning allow stellar winds to shed it away entirely (though wind mass-loss rates are uncertain, see Sect.\,\ref{sec:disc:winds}).  {These effects lead to the differences of up to $1\,\Msun$ in $M_\mathrm{He-dep}$ in stars of the same $M_\mathrm{1,i}$ found in binaries (see Fig.\,\ref{fig:logPq_B}).} 

All \Case B models in the \EDIT{binary grid} exhibit $R_\mathrm{RL,1}\simgr5\Rsun$ during core He burning, which is larger than the maximum radius expected from single HeS models in that phase (cf. Sect.\,\ref{sec:ps:single_he}). This implies that RLOF can only occur after core He depletion. Following core He depletion, the timescales ensure that wind mass loss from either star does not significantly alter the orbital configuration, leaving $R_\mathrm{RL,1}$ unchanged until the onset of another phase of RLOF or CC. 

To identify models that will undergo \case BC RLOF, we compare $R_\mathrm{RL,1}$ at core He-depletion with the maximum radius of a single HeS model of the same mass $M_\text{He-dep}$. Figure\,\ref{fig:logPq_B} shows that there is a much larger parameter space for systems undergoing \case BC RLOF, compared to \EDIT{that seen in the binary grid}. This difference likely stems from the adoption of MLT++ in the \EDIT{binary grid} and the termination of the calculation at core C depletion (cf. Fig.\,\ref{fig:single_he_rad}).
\EDIT{The \Case B m}odels expected to undergo \Case BC RLOF are concentrated at low initial $M_\text{1,i}$, low $\qi$ and low $P_\text i$.

We find \EDIT{limits} in the initial mass for the parameter space \EDIT{of \Case B models} where the primary undergoes CC following \case BC RLOF. Binary-stripped HeS models produce CCSNe only for $M_\text{1,i}\geq11.2\Msun$. On the other hand, models with $M_\text{1,i}> 15.8\Msun$ develop $M_\mathrm{He-dep}>3.6\Msun$ and do not expand enough to fill their Roche lobes again. 

%\section{Parameter space from \Case A systems}\label{sec:ps:caseA}

%\begin{figure}
%    \includegraphics[width=\linewidth]{Plots/logPq_compact_M_he_core_1.200.pdf}
%\caption{The same diagrams shown in Fig.\,\ref{fig:logPq_B} but for \case A systems with $M_\mathrm{1,i}=15.8\Msun$. }
%    \label{fig:logPq_A}
%\end{figure}

\EDIT{\Case A models also contribute to the population of HeS undergoing \Case BC RLOF. In this case, however, the mass-loss of the donor star during the MS} will \EDIT{result} in less massive cores, \EDIT{shifting} the parameter space for CC progenitors to higher initial masses compared to \case B systems (cf. Fig.\,\ref{fig:R_vs_RL}). 
The orbit following \case AB RLOF is wide enough to allow RLOF once again only after core He depletion. The major difference compared to the \Case B models is the mass of the companion, which here can accrete a few solar masses (compared to $\simle 0.5\Msun$ for \Case B systems) thanks to the tides that spin down the accretor during RLOF. We also find more \EDIT{pronounced} differences in $M_\mathrm{He-dep}$ for models of the same initial mass but different $\qi$ and $P_\mathrm{i}$ compared to \Case B models.

\EDIT{\REM{In conclusion}} We expect to find that \Case A systems can undergo \Case BC RLOF and reach CC for $14.1\leq M_\mathrm{1,i}<28.2\Msun$. These higher initial masses compared to those expected in \Case B systems will result in a less significant contribution of \case A systems to the population of systems undergoing \case BC RLOF due to the IMF \citep{Salpeter_IMF_55}.

\subsection{Initial conditions for the representative set of models to rerun}\label{sec:ps:grid}
Assuming that the CSM forms from the unaccreted mass during \case BC RLOF, this defines a clear parameter space to study interacting H-poor SN progenitors (Fig.\,\ref{fig:R_vs_RL}). 
 
Models occupying similar points in mass-radius diagram (Fig.\,\ref{fig:R_vs_RL}) can originate from different $\qi$, $P_\mathrm{i}$ and $M_\mathrm{1,i}$, resulting in significantly different companion masses ($M_2$). We argue that the effect of different $M_2$ have a secondary impact on the amount of mass removed during \case BC RLOF (see Appendix\,\ref{sec:appendix:Mhe_Rl_q}).

 Under these assumptions, the parameter space for interacting H-poor SNe is defined by two variables: $M_\text{He-dep}$ and $R_\mathrm{RL,1}$. We will work with \Case B systems, which are favored by the IMF (see \EDIT{Sect.\,\ref{sec:ps:binary_grid}}). \case B systems also offer a convenient degeneracy in the grid (see Figs.\,\ref{fig:R_vs_RL} and \ref{fig:logPq_B}). For a given initial mass $M_\text{1,i}$, there is a one-to-one correlation between $R_\mathrm{RL,1}$ and $M_\text{He-dep}$ enabling the tracing of distinct progenitor systems with different $(\qi,P_\text i)$. 

We rerun a subset of these models at $\qi=0.50$. This choice reflects that systems with lower $\qi$ are more likely to experience unstable mass transfer during \Case B RLOF, while those at higher $\qi$ have a narrower range of $P_\text i$ for which systems undergo \case BC RLOF (cf., Fig.\,\ref{fig:logPq_B}). We chose initial masses of $12.3\Msun\leq M_\mathrm{1,i}\leq 14.1\Msun$ (in logarithmic steps of 0.01, instead of 0.05 from the \EDIT{binary grid}). The initial orbital period $P_\mathrm{i}$ is selected adaptively to only include binaries that are expected to undergo \case BC RLOF with $P_\mathrm{i}>5\days$ (cf. Table\,\ref{tab:data}).

\begin{table*}[h]
\caption{Key parameters for the binary-models rerun in this work.}
\centering 
\resizebox{\textwidth}{!}{
\begin{tabular}{lc|cc|ccccc|cc|cccc|ccccccr}
\hline\hline
\multicolumn{2}{r}{} & \multicolumn{2}{c}{\Case B RLOF} &  \multicolumn{5}{c}{Core He burning}&  \multicolumn{2}{c}{Core He-dep} & \multicolumn{4}{c}{\Case BC RLOF} & \multicolumn{7}{c}{End} \\

$M_\mathrm{1,i}$ & $P_\mathrm{i}$ & $\Delta M_\mathrm{1}$ &  $\Delta M_\mathrm{2}$ &  $\Delta t$ & $M_\mathrm{conv}^\mathrm{max}$ &  $L$ & $\mathcal{L}$ & $T_\mathrm{eff}$ & $M_\mathrm{1}$  & $R_\mathrm{RL,1} $ & $\Delta t$ & $\Delta M_1$ & $\dot M_\mathrm{max}^{(-4)}$ & $\Delta t_\mathrm{max}$ &  $M_\mathrm{1}^\dagger$ & $M_\mathrm{CO}^\dagger$ & $\left< Z \right>_\mathrm{env}^\dagger$ & $\log \rho_\mathrm{c}$ & $\log T_\mathrm{c}$ & $\log T_\mathrm{max}$ & End \\
$\Msun$ & d & $\Msun$ &  $\Msun$ &  Myr & $\Msun$ & k$\Lsun$ & k$\mathcal{L}_\odot$ & kK  & $\Msun$ & $\Rsun$ & $\kyr$ & $\Msun$ &   & $\kyr$ &  $\Msun$ & $\Msun$   &  & $\mathrm{g}\ \mathrm{cm}^{-3}$ & K & K & of run \\ \hline
(1) & (2) & (3) & (4) & (5) & (6) & (7) & (8) & (9) & (10) & (11) & (12) & (13) & (14) & (15) & (16) & (17) & (18) & (19) & (20) & (21) & (22)  \\ \hline
12.3 &    5.01 &    8.61 &    0.26 &    1.72 &     1.3 &    8.66 &    3.02 &      80.0 &    2.53 &    5.88 &    26.1 &    0.79 &    1.18 &     4.19 &    1.62 &    1.44 &    0.14 &    7.68 &   8.68 &  9.27  & Ne$^\mathrm{EC}$ \\
12.3 &    6.31 &    8.57 &    0.21 &    1.67 &    1.32 &     9.30 &    3.16 &    80.1 &    2.58 &    7.51 &    23.2 &    0.78 &    1.35 &     3.61 &    1.68 &    1.46 &    0.19 &   8.71 &   9.00 &  9.42 &  Ne/O/Si*\\
12.3 &      10.0 &     8.50 &    0.18 &     1.60 &    1.36 &    10.5 &    3.43 &    79.9 &    2.66 &    11.8 &    13.7 &    0.74 &    1.95 &     2.64 &    1.81 &    1.50 &    0.19 & 8.26 &   9.07 &  9.51 & Ne/O/Si \\
12.3 &    15.8 &    8.45 &    0.15 &    1.53 &    1.38 &    11.5 &    3.64 &    79.3 &    2.73 &    18.1 &    8.44 &    0.70 &    2.85 &     1.95 &    1.92 &    1.53 &    0.16 & 8.26 &   9.11 &  9.43 & Ne/O \\
12.3 &    25.1 &    8.42 &    0.15 &    1.49 &    1.40 &    12.2 &    3.79 &    78.7 &    2.78 &    27.1 &    4.98 &    0.57 &    3.10 &     1.15 &    2.10 &    1.55 &    0.12 & 8.21 &   9.12 &  9.39 & O \\
12.3 &    39.8 &    8.39 &    0.16 &    1.45 &    1.41 &    12.8 &    3.91 &    78.1 &    2.83 &    40.1 &    2.22 &    0.41 &    2.92 &     0.67 &    2.30 &    1.57 &    0.10 & 8.83 &   9.21 &  9.58 &   Si* \\
12.3 &    63.1 &    8.36 &    0.16 &    1.43 &    1.42 &    13.2 &    4.01 &    77.6 &    2.86 &    58.2 &    1.52 &    0.28 &    2.85 &     0.02 &    2.46 &    1.58 &    0.09 & 8.79 &   9.34 &  9.55  & Si \\
12.3 &     100 &    8.35 &    0.16 &    1.43 &    1.44 &    13.6 &    4.08 &    77.2 &    2.88 &    83.5 &    1.13 &    0.18 &    2.96 &     0.02 &    2.58 &    1.59 &    0.08 & 8.84 &   9.34 &  9.55 &  Si \\
12.3 &     158 &    8.33 &    0.17 &    1.43 &    1.46 &    13.8 &    4.13 &    77.1 &    2.89 &     119 &    0.78 &    0.10 &    1.75 &     0.15 &    2.67 &    1.61 &    0.07 & 8.68 &   9.35 &  9.55  & Si \\
12.3 &     251 &    8.32 &    0.16 &    1.43 &    1.47 &    14.0 &    4.18 &    76.8 &    2.90 &     168 &       / &       / &       / &        / &    2.78 &    1.62 &    0.06 & 7.80 &   9.07 &  9.50  & Ne/O/Si \\ 
\hline
12.6 &    5.01 &    8.75 &    0.26 &    1.64 &    1.37 &    9.59 &    3.22 &   81.2 &    2.62 &    5.92 &   22.5 &    0.78 &    1.28 &    3.27 &    1.72 &    1.48 &    0.20 &    8.55 &   9.03 &  9.44 & Ne/O/Si* \\
12.6 &    6.31 &    8.71 &    0.20 &    1.59 &    1.40 &   10.3  &    3.38 &   81.4 &    2.67 &    7.56 &   19.5 &    0.76 &    1.49 &    2.73 &    1.79 &    1.51 &    0.20 &    7.93 &   9.01 & 10.2  & Ne/O/Si \\
12.6 &    10.0 &    8.64 &    0.19 &    1.52 &    1.44 &   11.6  &    3.65 &   81.2 &    2.75 &   11.8  &   10.4 &    0.71 &    2.30 &    1.84 &    1.93 &    1.56 &    0.18 &    8.85 &   9.21 &  9.57 &  Si* \\
12.6 &    15.9 &    8.59 &    0.15 &    1.47 &    1.46 &   12.6  &    3.86 &   80.6 &    2.82 &   18.1  &  5.70  &    0.54 &    2.79 &    0.94 &    2.16 &    1.59 &    0.13 &    8.85 &   9.33 &  9.56 &  Si* \\
12.6 &    25.1 &    8.55 &    0.13 &    1.42 &    1.48 &   13.4  &    4.02 &   80.0 &    2.87 &   27.1  &  2.24  &    0.31 &    2.57 &    0.37 &    2.45 &    1.61 &    0.09 &    8.70 &   9.23 &  9.57&   Si* \\
12.6 &   39.8  &    8.52 &    0.16 &    1.39 &    1.49 &   14.0  &    4.15 &   79.4 &    2.92 &   40.1 &   1.16  &    0.16 &    2.26 &    0.04 &    2.64 &    1.63 &    0.08 &    8.63 &   9.34 &  9.56 & Si* \\
   12.6 & 63.1 &    8.49 &    0.17 &    1.37 &    1.50 &   14.5 &    4.25  &   79.0 &    2.95 &   58.1 &    0.65 &    0.06 &    1.65 &    0.04 &    2.77 &    1.64 &    0.07 &    8.78 &   9.37 &  9.55 & Si* \\
   12.6 &  100 &    8.48 &    0.17 &    1.36 &    1.51 &   14.9 &    4.32  &   78.6 &    2.97 &   83.3 &    0.28 &    0.01 &    1.34 &    0.02 &    2.84 &    1.65 &    0.07 &    8.70 &   9.24 &  9.57 &  Si* \\
\hline
   12.9 &    5.01&    8.89 &    0.32 &    1.56 &    1.45 &   10.6 &    3.43 &   82.5 &    2.70 &   5.90 &   19.26 &    0.77 &    1.38 &    2.50 &    1.81 &    1.53 &    0.21 & 9.09 &   9.31 &  9.51 &  Si* \\
   12.9 &    6.31&    8.84 &    0.21 &    1.52 &    1.49 &   11.4 &    3.60 &   82.6 &    2.76 &   7.59 &   15.35 &    0.74 &    1.70 &    1.99 &    1.90 &    1.56 &    0.20 & 8.57 &   9.26 &  9.38 &    Si \\
   12.9 &   10.0 &    8.77 &    0.20 &    1.46 &    1.52 &   12.7 &    3.87 &   82.5 &    2.83 &   11.8 &    7.66 &    0.61 &    2.49 &    1.06 &    2.11 &    1.62 &    0.14 & 8.76 &   9.21 &  9.57 &  Si* \\
   12.9 &   15.9 &    8.72 &    0.16 &    1.41 &    1.55 &   13.8 &    4.09 &   82.1 &    2.90 &   18.1 &    3.47 &    0.29 &    2.53 &    0.02 &    2.50 &    1.65 &    0.10 & 8.85 &   9.45 &  9.98 & Si* \\
   12.9 &   25.1 &    8.68 &    0.16 &    1.37 &    1.56 &   14.7 &    4.26 &   81.5 &    2.96 &   27.0 &    1.04 &    0.10 &    1.87 &    0.03 &    2.75 &    1.68 &    0.08 & 9.02 &   9.40 & 10.0&   Si* \\
   12.9 &   39.8 &    8.65 &    0.16 &    1.34 &    1.58 &   15.3 &    4.38 &   81.0 &    3.00 &   40.1 &    0.40 &    0.02 &    1.04 &    0.03 &    2.86 &    1.69 &    0.08 & 8.56 &   9.43 &  9.57 &   Si*\\
   12.9 &   63.1 &    8.62 &    0.17 &    1.32 &    1.59 &   15.9 &    4.49 &   80.5 &    3.03 &   58.0 &   / &    / &    / &   / &    2.92 &    1.71 &    0.07 & 8.46 &   9.44 &  9.56 &  Si \\
\hline
   13.2 &    5.01 &    9.02 &    0.29 &    1.49 &    1.53 &   11.7 &    3.67 &   83.7 &    2.79 &    6.01 &   16.1 &    0.76 &    1.54 &    1.77 &    1.92 &    1.59 &    0.21 &   8.90  &   9.33 &  9.54&   Si* \\
   13.2 &    6.31 &    8.98 &    0.26 &    1.45 &    1.56 &   12.5 &    3.83 &   83.8 &    2.84 &    7.59 &   11.4 &    0.71 &    1.96 &    1.29 &    2.02 &    1.62 &    0.18 &   8.84  &   9.35 &  9.55 &   Si* \\
   13.2 &    10.0 &    8.90 &    0.20 &    1.39 &    1.61 &   14.0 &    4.11 &   83.8 &    2.93 &   11.9 &    5.30 &    0.41 &    2.40 &    0.33 &    2.41 &    1.68 &    0.10 &   8.41  &   9.42 &  9.57 &   Si \\
   13.2 &    15.9 &    8.84 &    0.16 &    1.35 &    1.64 &   15.1 &    4.33 &   83.4 &    2.99 &   18.1 &    1.35 &    0.10 &    1.79 &    0.02 &    2.77 &    1.72 &    0.08 &   8.30  &   9.53 &  9.57 &   Si \\
   13.2 &    25.1 &    8.80 &    0.14 &    1.32 &    1.66 &   16.1 &    4.52 &   82.9 &    3.05 &   27.2 &    0.26 &    0.01 &    0.57 &    0.03 &    2.93 &    1.75 &    0.08 &   8.71  &   9.52 &  9.55 &  Si* \\
   13.2 &    39.8 &    8.77 &    0.16 &    1.29 &    1.67 &   16.8 &    4.65 &   82.4 &    3.09 &   40.1 &    / &    / &   / &    / &    2.98 &    1.76 &    0.08 &   8.27  &   9.53 &  9.56 &  Si \\
\hline
   13.5 &    5.01 &    9.15 &    0.29 &    1.42 &    1.63 &   13.0 &    3.91 &   84.9 &    2.89 &    6.07 &   13.0 &    0.72 &    1.75 &    1.12 &    2.05 &    1.66 &    0.17 &   8.77 &   9.38 &  9.55 &  Si* \\
   13.5 &    6.31 &    9.11 &    0.27 &    1.38 &    1.66 &   13.8 &    4.08 &   85.1 &    2.94 &    7.62 &   8.57 &    0.60 &    2.15 &    0.60 &    2.23 &    1.69 &    0.13 &   9.00 &   9.37 &  9.43&  Si* \\
   13.5 &   10.0  &    9.03 &    0.21 &    1.33 &    1.71 &   15.3 &    4.36 &   85.1 &    3.02 &   11.9 &    3.41 &    0.18 &    2.08 &    0.02 &    2.72 &    1.74 &    0.09 &   8.73 &   9.24 &  9.57 &   Si* \\
   13.5 &   15.9  &    8.97 &    0.16 &    1.29 &    1.73 &   16.5 &    4.58 &   84.8 &    3.08 &   18.2 &    0.50 &    0.02 &    0.76 &    0.01 &    2.95 &    1.78 &    0.09 &   8.74 &   9.50 &  9.55 &   Si* \\
   13.5 &   25.1  &    8.92 &    0.14 &    1.26 &    1.75 &   17.6 &    4.77 &   84.3 &    3.14 &   27.2 &    / &   / &    / &    / &    3.03 &    1.81 &    0.08 &   7.86 &   9.54 &  9.57 &  Si \\ 
\hline
   13.8 &    5.01 &    9.28 &    0.31 &    1.36 &    1.72 &   14.3 &    4.15 &   86.1 &    2.98 &    6.12 &   10.2 &    0.63 &    1.94 &    0.55 &    2.24 &    1.72 &    0.15 &  8.44 &   9.35 &  9.56 &  Si* \\
   13.8 &    6.31 &    9.24 &    0.28 &    1.32 &    1.75 &   15.2 &    4.33 &   86.3 &    3.03 &    7.61 &    6.38 &    0.40 &    2.36 &    0.02 &    2.52 &    1.76 &    0.12 &  8.22 &   9.53 &  9.56 &  Si \\
   13.8 &   10.0 &     9.16 &    0.21 &    1.28 &    1.80 &   16.7 &    4.61 &   86.4 &    3.11 &   12.0 &    1.66 &    0.05 &    1.14 &    0.01 &    2.95 &    1.81 &    0.09 &  7.86 &   9.51 &  9.54 &    Si \\
   13.8 &   15.9 &     9.10 &    0.17 &    1.24 &    1.84 &   18.1 &    4.84 &   86.1 &    3.18 &   18.3 &   / &   / &   / &    / &    3.07 &    1.84 &    0.09 &  10.1 &  10.0 & 10.0& CC \\ 
\hline
   14.1 &    5.01 &    9.41 &    0.32 &    1.30 &    1.83 &   15.7 &    4.41 &   87.3 &    3.08 &    6.18 &    7.56 &    0.44 &    2.14 &    0.02 &    2.53 &    1.79 &    0.12 &   7.88 &   9.54 &  9.57 &  Si \\
   14.1 &    6.31 &    9.37 &    0.31 &    1.26 &    1.86 &   16.7 &    4.59 &   87.6 &    3.13 &    7.66 &    4.51 &    0.21 &    1.76 &    0.01 &    2.81 &    1.82 &    0.10 &   7.83 &   9.52 &  9.55 &   Si \\
   14.1 &   10.0 &    9.28 &    0.23 &    1.22 &    1.91 &   18.3 &    4.87 &   87.7 &    3.21 &   12.0 &    0.43 &    0.01 &    0.41 &    0.01 &    3.09 &    1.88 &    0.09 &   10.2 &  10.0 & 10.0 & CC \\
   14.1 &   15.9 &    9.22 &    0.17 &    1.19 &    1.94 &   19.7 &    5.11 &   87.4 &    3.28 &   18.3 &    / &    / &    / &    / &    3.17 &    1.92 &    0.09 &   10.2 &  10.0 & 10.0  & CC \\
\end{tabular}
}
\tablefoot{The rows are grouped by models with the same initial primary mass $M_{\rm 1}$. The columns are divided in different groups, for different evolutionary phases. Columns\,1 and 2 show the initial conditions: the initial primary mass (Col.\,1) and initial orbital period (2). Columns\,3 and 4 show the amount of mass shed by the primary (3) and accreted by the secondary (4) during \case B RLOF. Columns\,5-9 show parameters during core He burning: the time spent in this phase (5),  the maximum extent in mass of the convective core (6), the average luminosity, spectroscopic luminosity and effective temperature of the primary (7-9, with a standard deviation of about $\sim\,10\%$). Columns\,10-11 show the parameters at core He depletion: the mass of the primary (10) and and the Roche lobe radius (11). Columns\,12-15 show parameters associated to \Case BC RLOF: the time betwen the end of the run and the onset of \case BC RLOF (12), the amount of mass shed by the primary (13), the maximum mass transfer rate in units of $10^{-4}\msoy$ (14) and the time before the end of the run at which the maximum mass transfer rate was reached (15). Columns 16-22 show the parameters at the end of the run: the final mass of the primary (16), its CO-core mass (17), the average metal abundance in the envelope (18), central density and temperature (19-20), the maximum temperature (21) and the burning phase at the end of the run  labeled according to the nuclei being burnt, if it didn't reach CC (22). (*) Models have experienced \Case X RLOF (see Sect.\,\ref{sec:overv:caseX}). (EC) The least-massive model is expected to end as an EC-SN.  ($\dagger$) The data is taken at 1 year before the end of the run to filter out the effects of \case X RLOF.}
\label{tab:data}
\end{table*}

\section{Results}\label{sec:res}

\begin{figure*}
\centering
\includegraphics[width=\linewidth]{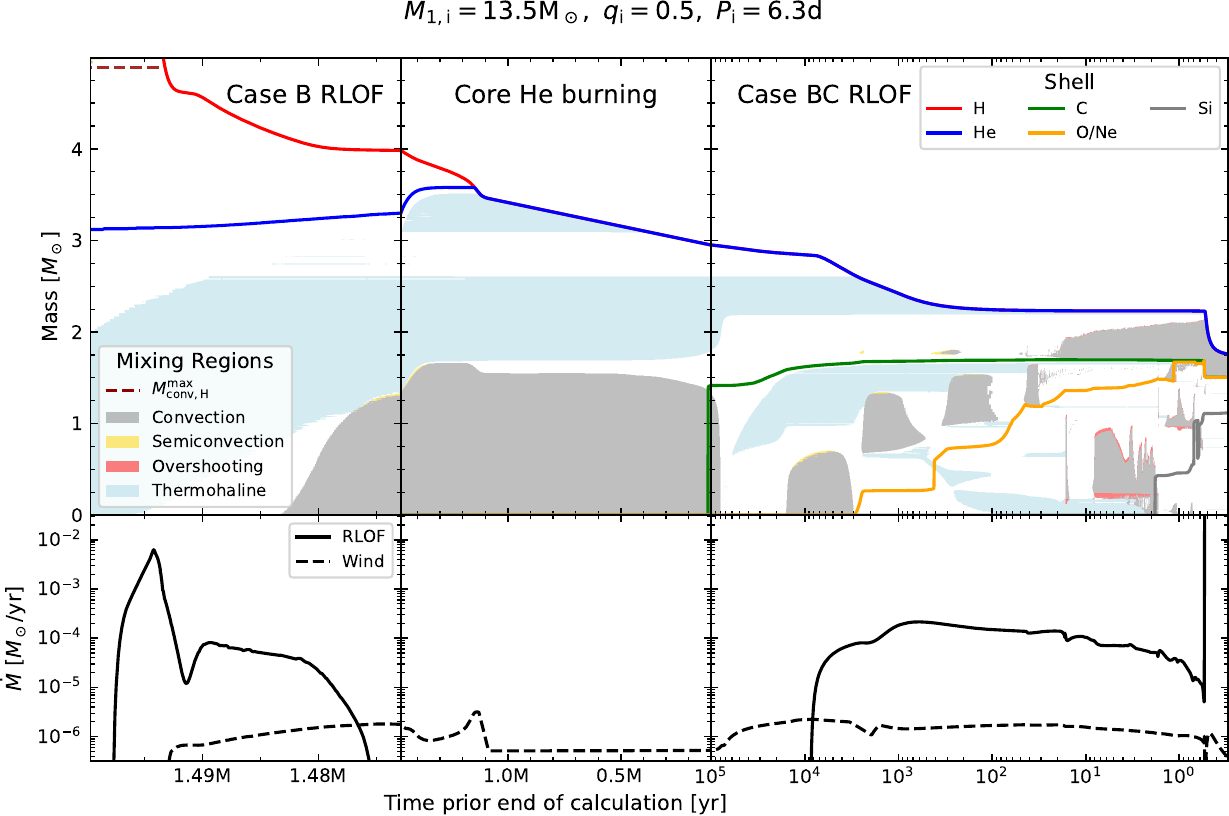}

\caption{Time evolution of the primary star in model $M_\text{1,i}=13.5\Msun$, $\qi=0.50$ and  $P_\text i = 6.3\,\text{d}$ from the onset of \case B RLOF (left), during core He burning (center), and until the end of the run (right).  {\textbf{Top}}: Kippenhahn diagram in which the outermost boundaries are colored corresponding to the most abundant element just below
(see legend on the right) and colored patches based on the dominating mixing mechanism (see legend on the left). A dashed red line shows the maximum extent of the convective H-burning core during the MS.  {\textbf{Bottom}}: Mass transfer rate (solid) and wind mass loss rate from the primary (dashed). }
    \label{fig:example_run}
\end{figure*}

In this section, we present the results of the rerun binary models. In  Sect.\,\ref{sec:res:example_model} we will focus on one specific model, followed by Sect.\,\ref{sec:res:overview} with an overview across all the models.  {We discuss the generic features of the SN progenitors in Sect.\,\ref{sec:SN:type} with a brief discussion on the post-SN evolution of these systems in Sect.\,\ref{sec:SN:postSN}.}

\subsection{Example Model}\label{sec:res:example_model}

    Here, we examine the evolution of a single model, with $M_\text{1,i}=13.5\Msun$, $\qi=0.50$ and $P_\text i = 6.3\,\text{d}$ (Fig.\,\ref{fig:example_run}).

\subsubsection{\Case B mass transfer and subsequent stripping}\label{sec:res:example_mode:caseB}
Both stars evolve effectively as single stars until after the primary reaches TAMS. The primary then expands, filling its Roche lobe and initiating mass transfer. During this phase, most of the H-rich envelope is quickly stripped until nuclear-processed material from the core is exposed. This material, initially part of the convective H-burning core, remains H-rich but is He- and N-enhanced (cf. Fig.\,\ref{fig:example_run}). At this point, $8.3\Msun$ of the envelope have been removed, and mass transfer becomes less intense, removing an additional $0.77\Msun$. Mass transfer stops as the primary star ignites He, leaving behind a $0.67\Msun$ envelope and a $3.31\Msun$ He-core.

During core He-burning, the He-core initially grows to $3.57\Msun$ due to the H-burning shell.
Stellar winds also contribute to the erosion of the envelope, which is lost $340\kyr$ after \case B RLOF. At this point, the primary star is a HeS. Over the remaining phase of core He burning ($\sim\,10^6\yr$), the primary star sheds $0.59\Msun$ of its He-rich envelope via winds, reaching core He-depletion with $M_\mathrm{He-dep}=2.94\Msun$.

The secondary star spins up to critical rotation during RLOF due to the accretion of angular momentum. The time before the CC of the donor star is insufficient for a significant slow down. The secondary is therefore expected to appear as a Be star, both before and after the CC of the primary. 

\subsubsection{\case BC mass transfer}
Comparison the single HeS models of similar mass shows that the primary star will expand to fill its Roche lobe ($7.6\Rsun$) roughly $10\kyr$ before collapse (cf. Fig.\,\ref{fig:single_he_rad} and Fig.\,\ref{fig:R_vs_RL}).

Figure\,\ref{fig:example_run} shows that $8.7\kyr$ before CC, the primary began \case BC RLOF, reaching a maximum rate of $\sim\,2\times 10^{-4}\msoy$ as the first C-burning shell turns off. By the end of core-O burning, RLOF removed about $0.71\Msun$ from the He-rich envelope, which now has a mass of $0.52\Msun$ with an average metal abundance $\left<Z\right>=0.13$ (mostly C, N and O).   

\subsubsection{\Case X mass transfer}\label{sec:res:example_mod:caseX}

\begin{figure}
\centering
\includegraphics[width=1\linewidth]{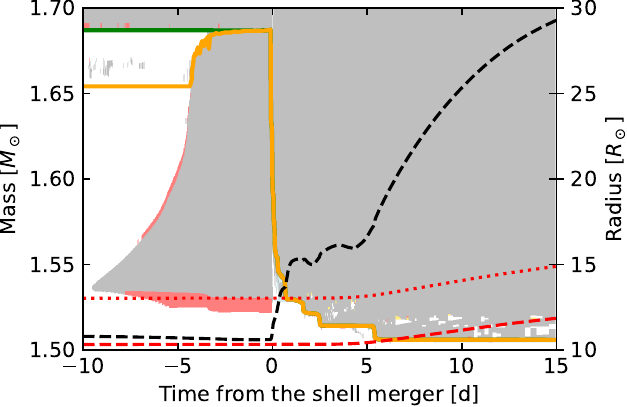}

\caption{ A zoom in on the Kippenhahn diagram close to the \EDIT{ edge of the CO-core where the Ne-burning shell ignites at a mass coordinate of $1.54\Msun$ and develops a convective region just above it which then merges with the He-burning shell after about $10\days$.} The solid lines, and patch colors are the same as in Fig.\,\ref{fig:example_run}. The dashed lines correspond to the radius (black, dashed), the Roche lobe radius (red, dashed) and the volume-equivalent radius of the outer Lagrangian point (red, dotted).}
    \label{fig:shell_merger_ex}
\end{figure}

Following \EDIT{radiative} Si-burning, a \EDIT{\REM{thin}} Ne-burning shell ignites at the outer edge of the CO-core\EDIT{, at around a mass coordinate of $1.54\Msun$), developing a thin convective region}. \EDIT{This region then} merges with the He-rich convective shell above (cf. Fig.\,\ref{fig:shell_merger_ex}), leading to ingestion of He in hotter and denser layers. He is now burnt at higher densities and temperatures, increasing the shell's luminosity, while metal-rich material is dredged-up, raising the envelope's opacity. The envelope becomes  fully convective, and the radius expands significantly from $10$ to $56\Rsun$ in $\sim\,20\,\text{d}$ (cf. Fig.\,\ref{fig:shell_merger_ex}). Additionally,  $^{14}\mathrm{N}$ is dredged down into the burning shell, triggering the chain $^{14}\mathrm{N}(\alpha,\gamma)\,^{18}\mathrm{O}(\alpha,\gamma)\,^{22}\mathrm{Ne}(\alpha,n)\,^{25}\mathrm{Mg}$ which releases free neutrons and likely initiates s-process nucleosynthesis. 

The abrupt expansion triggers a new phase of mass transfer, which we will refer to as \Case X RLOF\footnote{This is referred to as `late-stage' mass transfer in \WF22 or \Case BBB RLOF in \citealt{Tauris2013_USSNe_Ic}. We use a different nomenclature to highlight this phenomenon's uncertain and extreme nature.}. The resulting overflow ($R/R_\mathrm{RL,1}\simeq 3$)  is enough to engulf the companion, thus initiating a CE-phase. The model sheds $0.60\Msun$ during \case X RLOF,  though this is a lower-bound estimate as mass transfer was ongoing at  rates of $\simgr0.1\msoy$ when the model terminated. The SN-progenitor would retain a low-mass ($<0.13\Msun$), metal rich $(\left<Z\right>>0.5)$ envelope. With an orbital period of $8\,\text{d}$  before \case X RLOF, the remaining time left to CC suggests the primary will likely explode during CE-evolution. 

\subsection{Overview}\label{sec:res:overview}

The binary models rerun in this work evolve similarly to the model in Sect.\,\ref{sec:res:example_model}. Here, we now discuss the quantitative differences and trends, in the model grid. Each model is labeled  `B' followed by its initial primary mass, \EDIT{ a `p' and its} initial orbital period. For example, B13.5\EDIT{p6.3} refers to the model with $M_\mathrm{1,i}=13.5\Msun$ \EDIT{and} $P_\mathrm{i}=6.3\text{d}$, discussed in Sect.\,\ref{sec:res:example_model}.

\subsubsection{\Case B RLOF and envelope loss}\label{sec:overv:caseB}

All the models in this set undergo \Case B RLOF as the first mass transfer event, removing about $\sim\,90\%$ of the primary's envelope mass and leaving behind between $0.55$ and $0.83\Msun$ of the H-rich envelope.
This phase of mass transfer is stable in all models per our stability criterion (cf. Sect.\,\ref{sec:methods:RLOF}). 

After the onset of core He-burning, the remaining H-rich envelope is lost via shell burning and winds within $20\%$ to $70\%$ of the core He-burning lifetime. 
By the end of core He-burning, models with the same initial masses evolve into HeSs of varying masses  (cf Table.\,\ref{tab:data}).
Losing the H-rich envelope after a significant fraction of the core He-burning phase introduces some differences in the internal structure between single HeS and binary-stripped HeS models, discussed in Appendix\,\ref{sec:appendix:A_BETTER_PARAMETER}.   

\begin{figure}
\centering
\includegraphics[width=\linewidth]{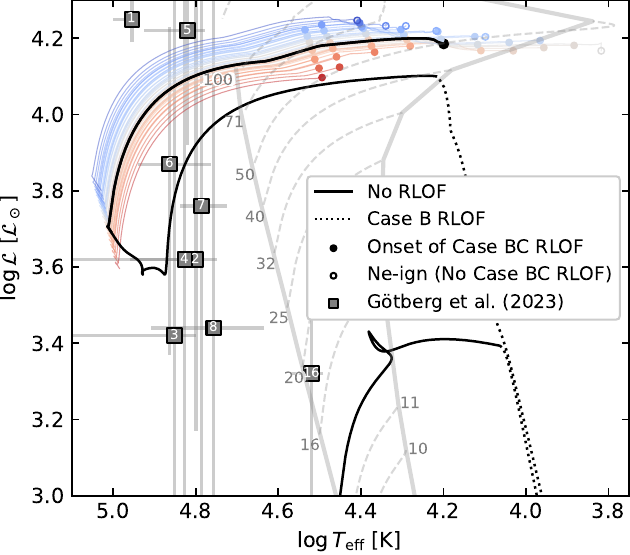}
\caption{The spectroscopic HR-diagram of the binary-stripped models explored in this work, color-coded by the mass at core He-depletion $M_\mathrm{He-dep}$ (cf., \EDIT{Fig.\,\ref{fig:logPq_B}}). Each track is plotted from the end of core He-burning until the onset of Case BC RLOF (filled circle). For those that do not undergo RLOF, the plot is shown until the point found at 20yr before the end of the run (empty circle). The track of model B13.5\EDIT{p6.3} is highlighted in black and shown from ZAMS. The He-stars in the SMC and LMC analyzed by and \cite{Gotberg2023_stripped_stars_in_binaries} are also shown (black squares), and numbered as in \cite{Drout2023_Observed_stripped_HeStars_inBinaries}
}
\label{fig:HR}
\end{figure}

\EDIT{\REM{Observationally}} The models align in the sHR diagram as a function of their core mass (cf. Fig.\,\ref{fig:HR}). 
The phase of core He-burning, while long-lived ($\sim\,1.2-1.8\,\text{Myr}$), is challenging to observe, particularily in the optical, due to the strong flux from the higher-mass MS companion. Recent UV observations by \cite{Drout2023_Observed_stripped_HeStars_inBinaries} reveal HeS companions to otherwise apparently single MS stars exhibiting UV-excess. Some of their observed stars may evolve similarly to our models, depending on the mass and distance of the MS companion.

\subsubsection{\case BC RLOF}\label{sec:overv:caseC}

All but the widest models in the set undergo \Case BC RLOF after core He depletion. Although the majority of the models crashed before reaching CC (see Sect.\,\ref{sec:SN:type}), the bulk of the mass transferred during \case BC has been captured due to the limited mass transfer rates  and the short time remaining before CC. This phase of mass transfer is stable for all models.

\begin{figure}
\centering
\includegraphics[width=\linewidth]{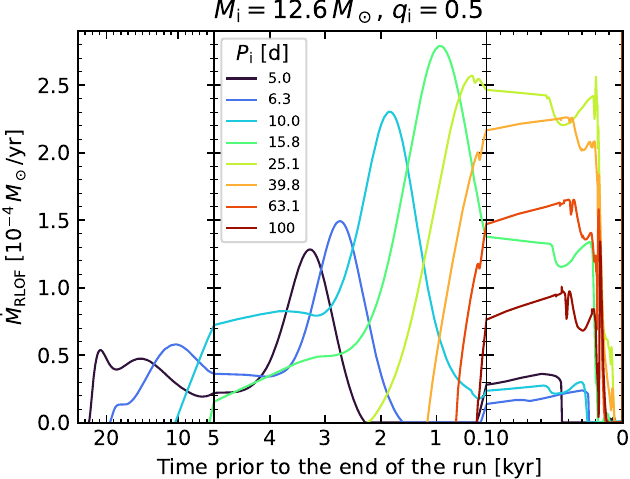}
\caption{Mass loss as a function of time prior to the end of the run for the set of models with $M_\mathrm{1,i}=12.6\Msun$ with different initial orbital period, marked by different colors. The three different panels highlight different time scales. 
}
    \label{fig:mdot_preCC}
\end{figure}

\begin{figure}
\centering
\includegraphics[width=\linewidth]{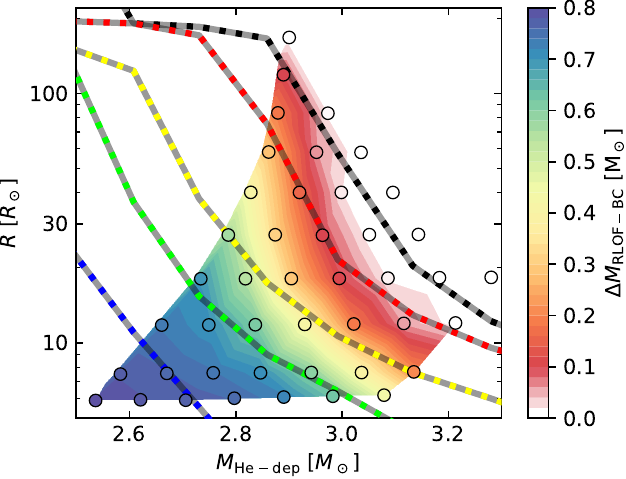}
\vspace{0.1em}
\includegraphics[width=1\linewidth]{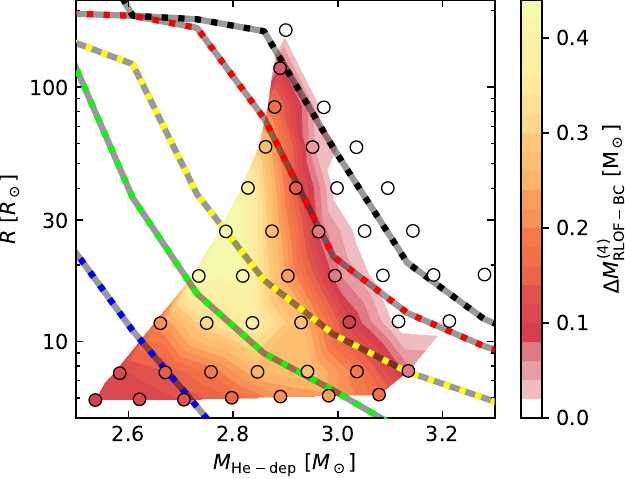}

\caption{ {Maps showing the $\Delta M_\mathrm{RLOF-BC}$ (top) and the amount lost after $\dot M_\mathrm{RLOF}$ exceeds $10^{-4}\msoy$ (i.e., $\Delta M_\mathrm{RLOF-BC}^{(4)}$, bottom) in the $M_\mathrm{He-dep}-R_\mathrm{RL,1}$ diagram. The scatter highlight the models, and the map is linearly interpolated between them}. The dashed curves correspond to the radii of single HeS models as a function of mass at different times before the end of the run ($20\kyr$ in blue, $10\kyr$ in green, $5\kyr$ in yellow, $1\kyr$ in red and $1\yr$ in black, cf. Fig.\,\ref{fig:R_vs_RL})}
\label{fig:tri_deltaMC_hecore}
\end{figure}

The time evolution of the mass transfer rate (cf. Fig.\,\ref{fig:mdot_preCC}) depends on when RLOF begins. Generally, two phases can be distinguished. A weaker phase ($\dot M_\mathrm{RLOF}<10^{-4}\msoy$) occurs $>5\kyr$ before CC, and is seen only in the tight systems starting RLOF much earlier. A stronger phase follows, as the envelope expands after the first C-burning shell ignites (between $6\kyr$ and $1\kyr$ before CC, depending on mass, cf. Fig.\,\ref{fig:single_he_rad}), leading to mass-transfer rates of up to $3\times 10^{-4}\msoy$. For the systems starting RLOF only in their last $5\kyr$, $\dot M_\mathrm{RLOF}$ reaches high rates in the final $\sim\,100\yr$ before CC. In contrast, those already undergoing RLOF at this time show significantly lower rates closer to CC, and in some cases even detach. For all but the most massive models, mass transfer will terminate $\sim\,20-40\yr$ before CC, following the ignition of Ne.

Figure\,\ref{fig:tri_deltaMC_hecore} shows that the mass-loss during \case BC ($\Delta M_\mathrm{RLOF-BC}$) scales with how early the star is expected to fill its Roche lobe, determined by comparing the models' Roche lobe radius with the radius of single HeS models of the same initial mass. Trends in $\Delta M_\mathrm{RLOF-BC}$ align more closely to the single HeS radius evolution when compared using a variable other than $M_\mathrm{He-dep}$ (cf.\, Appendix \ref{sec:appendix:A_BETTER_PARAMETER}). Mass lost during \case BC RLOF depends on \EDIT{how} early the model began RLOF, independently of the HeS mass (cf. Fig.\,\ref{fig:tri_deltaMC_hecore} and Fig\,\ref{fig:triangle_deltaMc}). \EDIT{I}f RLOF begins $20\kyr$, $10\kyr$, $5\kyr$ and $1\kyr$ before CC, it removes about $0.8\Msun$, $0.6\Msun$, $0.4\Msun$ and $0.1\Msun$ respectively. The stripping only affects regions of the envelope that are still rich in N.

\subsubsection{\Case X RLOF}\label{sec:overv:caseX}
This phase of mass transfer, previously outlined in Sect.\,\ref{sec:res:example_mod:caseX}, affects only half of the models, all of which fail to reach CC.  \Case X RLOF typically reaches mass loss rates of $10^{-2}-1\msoy$, removing $0.2-1\Msun$ of material. These values are lower-limits, as the models terminate before CC \EDIT{of the donor}. \case X RLOF is always unstable as all models undergo OLOF and, in some cases, the HeS expands enough to engulf the secondary star.

\EDIT{The onset of} \Case X RLOF \EDIT{is linked to the evolution of the donor star}. \EDIT{It typically occurs after} the onset of radiative Si-burning \EDIT{and the ingestion of He in the outer core of the donor star}. This ingestion leads to the increase in the luminosity of the \EDIT{donor} star as well as the envelope opacity, causing the sudden expansion (cf., Sect.\,\ref{sec:res:example_mod:caseX}). However, there are some models \EDIT{where the donor star} exhibit\EDIT{s} He-ingestion without expanding. The reason as to why helium is ingested in the first place remains uncertain. 

Convective over- and undershooting likely play a key role in this phenomenon, (see \EDIT{\REM{Appendix} Sect.\,} \ref{sec:appendix:shell_merger_overshooting}). These methods lack physical backing and were introduced to help convergence in the calculations. Consequently, we limit our discussion of the effects of \Case X on the SN in detail to a more qualitative analysis (see \WF22 for a detailed analysis of this phase of RLOF in HeS+NS binaries). To exclude the effects of He-ingestion and \case X RLOF, we use the pre-SN data at $1\yr$ before the model termination in subsequent sections.

\subsection{Termination and Explosion Properties}\label{sec:SN:type}
 {The majority of our models develop a Si-rich core, and only three (B13.8\EDIT{15.8}, B14.1\EDIT{p10.0}, and B14.1\EDIT{p15.8}) reach the point of CC. These models provide estimates for the time between CC and specific burning phases: $22-10\yr$ after core Ne-burning, $10-1\yr$ after core O-burning, and about $10\text{d}$ after core Si-burning. }

 {Models B12.3\EDIT{p5.01-B12.3p251}, B12.3\EDIT{p251}, B12.6\EDIT{p5.01 and B12.6p6.31} terminate while burning Ne/O off-center. Of these, B12.3\EDIT{p5.01} (the most-stripped model) exhibits central density and temperature evolution akin to that of a ECSN progenitor \citep[cf. Sect.\,\ref{sec:ps:single_he} and ][]{Tauris2015_USSNe}. All other models \EDIT{develop} higher temperatures, indicating that they will eventually undergo CC.}

\begin{figure}
\centering
\includegraphics[width=01\linewidth]{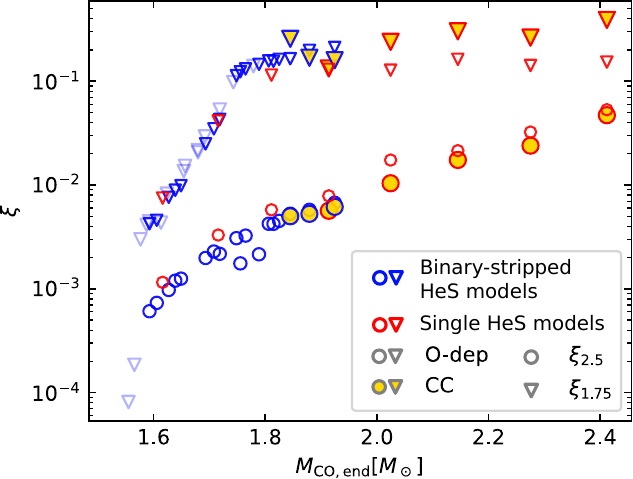}

\caption{Scatter plot showing the compactness parameter $\xi_m=(m/\Msun)/ (R(m)/1000\,\mathrm{km})$ for $m=1.75\Msun$ (triangle marker) and $2.5\Msun$ (circle marker) as a function of the final CO-core. The results of both the single (red) and binary-stripped HeS models (blue) are shown. The compactness is evaluated at core O-depletion for all models (empty markers), including at CC for those that reached it (gold fill). Models with $M_\mathrm{end}<2.5\Msun$ are shown with lighter colors.}

    \label{fig:compactness}
\end{figure}

We apply the prescriptions from \cite{MHLC16} and \cite{MM2020} to the models reaching CC, and obtain explosion energies of $\Eexp=5.8-7.0\times 10^{50}\erg$, nickel-masses of $M_\mathrm{Ni}=0.03-0.04\Msun$ and NS masses of $M_\mathrm{ns}=1.34-1.37\Msun$.
These values resemble those reported in \cite{Ertl2020_HeS_exp} based on the HeS models from \cite{Woosley2019_Hestars}. We investigate the compactness parameter $\xi_m$ \citep[][]{OConnor2011_compactness_parameter, Sukhbold14_compactness_preSN} of our models at the end of the run (cf. Fig.\,\ref{fig:compactness}), and find that $\xi_{2.5}$ decreases with decreasing $M_\mathrm{CO,end}$. This trend aligns with previous studies \citep[e.g.,][]{Ertl2020_HeS_exp, DR1, DR2} where also $\Eexp$ also decreases to as little as $1.0-1.2\times10^{50}\erg$. 
Since all these values are derived from recipes based on 1D-explosion models, they may be uncertain, and we also consider other values for the explosion energy and nickel mass below.

\begin{figure}
\centering
\includegraphics[width=1\linewidth]{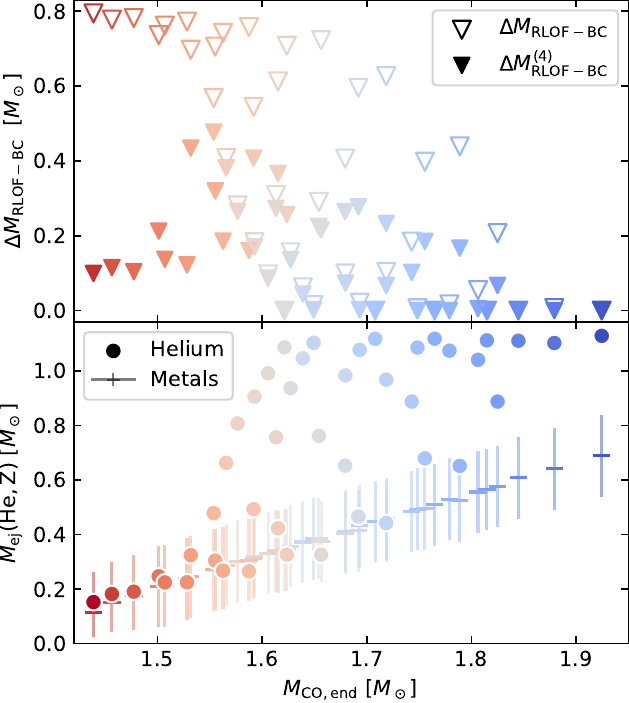}

\caption{Properties of the primaries as a function of the final CO-core mass of the primary star. The markers are colored as a function of $M_\text{He-dep}$, as in \EDIT{Fig.\,\ref{fig:logPq_B}}.  {\textbf{Top:}} Amount of mass ejected during \case BC RLOF. The empty markers represent the total mass lost via mass transfer, while the filled markers represent the mass shed during mass transfer after $\dot M_\mathrm{RLOF}>10^{-4}\msoy$ for the first time.  {\textbf{Bottom}:} The amount of helium (circles) and metals (error bars) found in the ejecta. The uncertainty in the metal content derives from assuming $M_\mathrm{ns}$ to be within $0.15\Msun$ of the fiducial value (horizontal marker).
}
    \label{fig:deltaMC_Mhe_Mz}
\end{figure}

In the following sections, we adopt a mass $M_\text{ns}=1.35\Msun$ for the NS, and discuss the effects of the adoption of other values between $1.20\Msun$ and $1.50\Msun$. The ejecta mass of the resulting SNe is given by $M_\text{ej}=M_\text{end}-M_\text{ns}$. The composition of the ejecta is also noted, and it is distinguished between He and metals ($\Mej(\mathrm{He})$ and $\Mej(\mathrm{Z})$ respectively), as shown in Fig.\,\ref{fig:deltaMC_Mhe_Mz}.

Except for the most heavily stripped models, the ejecta is He-rich (Fig.\,\ref{fig:deltaMC_Mhe_Mz}), and we can therefore assume that these will explode as \Type Ib SNe.  {However, simulations of the explosion of low-mass HeS models show peculiar nebular spectra, which do not at present have any observational counterparts }\citep{Dessart21_nebularSN_HeS, Dessart23_nebularSN_HeS_longer}. One possible explanation is that they instead lead to \Type Ibn SNe \citep{Dessart2022_Ibn}. 
The average He content in the ejecta varies significantly depending on the CO-core mass and the degree of stripping. The least massive models at CC ($M_\mathrm{end}<2.1\Msun$) have comparable amounts of He and metals in the ejecta, within a reasonable deviation of $M_\mathrm{ns}$ from its fiducial value. The transition to a \Type Ic SN also depends on the details of the explosion  \citep[e.g. the degree of mixing of Ni in the He-rich regions, as well as asymmetries in the explosion itself, would affect the excitation of HeI lines,][]{Lucy1991_He_excitation_in_SN_spectra, Dessart2012_IbcSNe, WoosleySukhboldKasen2021_IbcSNe}. More stripped models, with low He and high metal abundances in their ejecta, are more likely to explode as \Type Ic SNe. 

While we will solely focus on the CSM interaction features of the SN in later sections, we also outline the effects of \EDIT{\REM{the newly-born NS on the SN and}}  
the ejecta-companion interaction in \EDIT{\REM{Appendix} Sect.}\,\ref{sec:SN:compare:inter:secondary}. 

\subsection{Post-SN evolution of the system}\label{sec:SN:postSN}
Following the SN, intense mass-loss and natal-kicks on the newly-born NS, \citep{Burrows95_CCSNe_breakingsymmetry, Janka217_tugboatNS} affect the orbital evolution of the system, increasing its eccentricity and even unbinding it. The overall low-mass of the SN progenitors may result in a lower natal kick \citep{Janka217_tugboatNS}. Combined with the relatively short pre-explosion orbital periods, this increases the probability of these systems remaining bound, potentially forming a Be-X-ray binaries (cf. Sect.\,\ref{sec:res:example_mode:caseB}). 

The secondary is expected to evolve much like a single star until it is close to filling its Roche lobe. Here, mass transfer will likely turn unstable, triggering a CE-phase. If the CE is ejected, a tight WD+NS binary may form. Conversely, if the CE is not ejected, the NS will merge with the companion forming a Thorne–$\dot{\mathrm{Z}}$ytkow object.  These results can also be extended to systems with more massive secondaries than those presented here (cf. Appendix\,\ref{sec:appendix:Mhe_Rl_q}). In such cases,  {the CE-phase may produce a tight HeS+NS binary, where the HeS is massive enough to undergo CC. The resulting SN explosion could exhibit interacting features if \case BC RLOF occurs between the HeS and its NS companion (e.g., \WF22). If the natal kick of the newly-formed NS does not disrupt the binary, the system forms a double-NS system, which would eventually merge and emit gravitational waves \citep{Qin24_caseBB_HeS+NS_resultsin_GW}.}

\section{Supernova-CSM interaction}\label{sec:SN}

All SN-progenitor models that undergo \case BC RLOF explode with a significant amount of H-poor material likely remaining close to the binary system. 
Here, we discuss the CSM properties from our models (Sect.\,\ref{sec:SN:csmMass} and \ref{sec:SN:interact:R}), their potential effects on the SN light curve (Sect.\,\ref{sec:SN:compare:inter:cumulative}-\ref{sec:SN:interaction:CBD_lightcurve}) and a discussion on the number of interacting H-poor SNe expected from binary models (Sect.\,\ref{sec:SN:pop}). 
We do not pursue a discussion of the effect of the CSM on the spectral evolution \citep[but see][]{Dessart2022_Ibn}, as this depends on strongly non-linear physics and requires detailed radiative-transfer modeling.

\subsection{CSM mass}\label{sec:SN:csmMass}

\begin{figure}
\centering
\includegraphics[width=\linewidth]{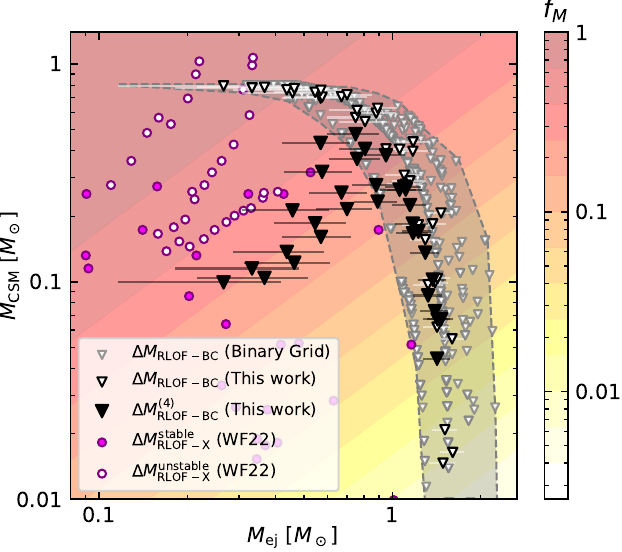}
\caption{The CSM in our models at the time of the SN, according to various estimates, versus the SN ejecta mass. The CSM mass given by $\Delta M_\mathrm{RLOF-BC}$ (white, with black outline) and $\Delta M_\mathrm{RLOF-BC}^{(4)}$ (black) is shown. CSM mass estimates for models in the \EDIT{binary grid} are given as smaller white triangles with gray rim, and the region occupied by these models is shaded in gray. The results from the models in \WF22  \EDIT{undergoing \case X RLOF are shown as purple dots and distinguished between the case in which mass transfer is assumed to be stable (filled) and unstable (empty)}. The background colormap represents the estimated maximum conversion fraction of kinetic energy into radiation via CSM interaction ($f_M$, Eq.\,\eqref{eq:interaction})
}
    \label{fig:Mej_Mcsm_Theo}
\end{figure}

\begin{figure*}
\centering
\includegraphics[width=0.96\linewidth]{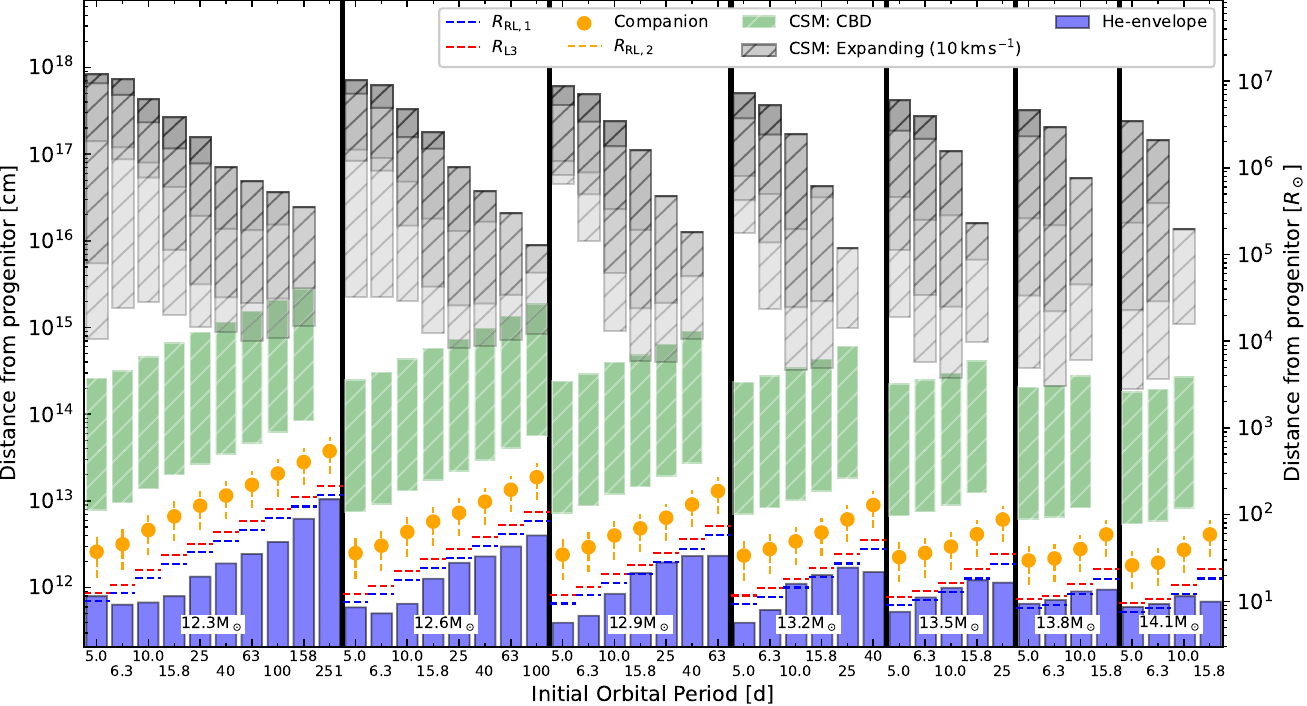}

\caption{Bar chart of the spatial distribution of material from the SN progenitor models $1\yr$ prior to the end of the run for the models run in this work. Color bars highlight the envelope (blue filled bar) and the regions where the CSM would be found (hatched bars) if it were distributed along a CBD (green) or if it were expanding isotropically, at a constant speed of $10\kms$ (gray). In the latter case, the minimum radius is set as the point where the cumulative mass of the CSM from the progenitor star reaches $0.001\Msun$, and increases in contrast when it reaches $0.01$, $0.1$ and $0.5\Msun$. The Roche lobe radius of the progenitor star is shown (blue hatched-line) as well as the volume-equivalent radius of the outer Lagrangian point (red hatched-line). Finally the companion's location is also shown (yellow marker) alongside its Roche lobe (vertial yellow hatched-line).  The models are grouped by different initial masses (indicated in the bottom of each box).
}
    \label{fig:Rbars}
\end{figure*}

The upper limit on the CSM mass ($\Mcsm$) in our models is given by $\Delta M_\mathrm{RLOF-BC}$. Some of this mass may escape the system and therefore not affect the SN. Most observations do not track the latet-ime evolution of the SN, often missing out on the prolonged interaction phase \citep[e.g., \Type IIn SNe, ][]{Smith17_IIn2005ip_latetime_interaction}. This restricts inferred CSM properties to the closer, denser regions where interaction power dominates the observed light-curve.

Our models constrain the total mass lost from the binary but are less predictive of the circumbinary density distribution. To exploit the time-evolution of the mass outflow from the binary model, we compute not only its total amount but separately also the amount of mass that was lost during the stronger phase of mass transfer in the last $\simle3\kyr$  before CC (cf. Fig.\,\ref{fig:mdot_preCC}), as
\begin{equation}\label{eq:deltaMRLOF4}
    \Delta M_\mathrm{RLOF-BC}^\mathrm{(4)}=\int_{t_4}^{t_\mathrm{CC}}\dot M_\mathrm{RLOF}  \, \mathrm{d}t,
\end{equation}
where $t_4$ is the time during \case BC RLOF such that $\dot M_\mathrm{RLOF}$ exceeds $10^{-4}\msoy$ for the first time. This quantity is not sensitive to the adopted threshold (see Fig.\,\ref{fig:mdot_preCC}). Unlike $\Delta M_\mathrm{RLOF-BC}$, $\Delta M_\mathrm{RLOF-BC}^\mathrm{(4)}$ does not correlate with the time the star fills its Roche lobe. Systems that undergo RLOF exclusively within $10\kyr$ before CC achieve the highest values of $\Delta M_\mathrm{RLOF-BC}^{(4)}$ (cf. Fig.\,\ref{fig:triangle_deltaMc}, and Sect.\,\ref{sec:overv:caseC}), and those that begin RLOF even later ($\simle 3\kyr$ before CC) lose all their mass at high rates, resulting in $\Delta M_\mathrm{RLOF-BC}^{(4)}\simeq \Delta M_\mathrm{RLOF-BC}$ (cf. Fig.\,\ref{fig:mdot_preCC}).

 In Fig.\,\ref{fig:Mej_Mcsm_Theo}, we find a strong anticorrelation between $\Mej$ and $\Mcsm$ when assuming $\Mcsm=\Delta M_\mathrm{RLOF-BC}$, which is also present when we extrapolate $\Delta M_\mathrm{RLOF-BC}$ in the whole \EDIT{binary grid}. This extrapolation is done by mapping each model in the \EDIT{binary grid} onto Fig.\,\ref{fig:triangle_deltaMc} and assuming $M_\mathrm{ns}=1.35\Msun$.  This anticorrelation is driven by the fact that the lower-mass models expand more and therefore lose more mass (cf. Sect.\,\ref{sec:overv:caseC} and Fig.\,\ref{fig:tri_deltaMC_hecore}) compared to higher-mass models. The combinations of $\Mcsm$ and $\Mej$ result in different effects on the light-curve (background color, Fig.\,\ref{fig:Mej_Mcsm_Theo}) which will be discussed in Sect.\,\ref{sec:SN:compare:inter:cumulative}.
 
 When assuming $\Mcsm=\Delta M_\mathrm{RLOF-BC}^\mathrm{(4)}$, trends differ only for $\Mej<1\Msun$, where $\Mcsm$ decreases with decreasing $\Mej$ (cf. Fig.\,\ref{fig:Mej_Mcsm_Theo}). This is because, even though lower mass models generally lose more mass, the absolute amount lost in the last $5\kyr$ is proportionally much less than in other systems (cf. Fig.\,\ref{fig:mdot_preCC} and the bottom panel in Fig.\,\ref{fig:tri_deltaMC_hecore}).

\subsection{CSM extension and shape}\label{sec:SN:interact:R}

As we do not model the CSM density structure, we estimate its spatial extension analytically under simple assumptions. For this, we consider two approaches.

In the first approach,
we assume that the CSM consists of a steady spherical outflow with constant velocity $v_\mathrm{CSM}$, from which we can infer its extent.
 For $v_\mathrm{CSM}=10\kms$, the CSM would extend to distances of $10^{15}\sim\,10^{18}\mathrm{cm}$ (cf. Fig.\,\ref{fig:Rbars}). These distances correlate with the mass of the CSM, with the more extended CSM containing more mass (cf. Sect.,\ref{sec:overv:caseC}).  Most of our mass donors detach from their Roche lobe before CC (except those with $M_\mathrm{He-dep}>2.90\Msun$ and $R_\mathrm{RL}<30\Rsun$), usually by core Ne-ignition. This leaves a cavity, extending to $\sim\,6\times 10^{14}\,\mathrm{cm}$, filled by donor's stellar wind. \EDIT{This cavity implies that the interaction between the ejecta and the CSM will be delayed by several days, assuming $v_\mathrm{ej}\sim 9\,000\kms$}. In models that do not detach, mass transfer in the last $\sim 20\yr$ deposits $\sim 0.001\Msun$ of material in that region (Fig.\,\ref{fig:Rbars}). 

Alternatively, material that leaves the binary may accumulate in a circumbinary disk (CBD), if channeled through the outer Lagrangian point of the accretor \citep[e.g.,][]{Lu2023_L2_outflow_during_MT}. Whether or not the material remains bound depends on its velocity. Given the high mass-ratio of the systems during \case BC RLOF, material escaping from the outer Lagrangian point with a small perturbation may remain gravitationally bound, feeding the disk. The inner disk radius is expected to be comparable to two to three times the semi-major axis of the binary \citep{ArtymowiczLubow94_Binary_Disk, Pejcha16_CBD_LumRedNovae}, corresponding to $10^{13}-10^{14}\,\mathrm{cm}$ in our models. The outermost disk radius is limited by photoevaporation from the radiation emitted by the stars in the inner binary \citep{Hollenbach94_photoevaporation_of_disks}. For a post-CE binary with $M_1=M_2=10\Msun$ in a tight orbit with $a=30\Rsun$ and a CSM of $1\Msun$, \cite{TunaMetzger2023_longtermevolutionCBD_postCE} find that the CBD extends up to $3\,000\sim\,10\,000\Rsun$ (cf. their Eq.\,29 and Fig.\,5), i.e., about $100-300$ times the orbital separation. Adopting the conservative value of $100a$ as the outermost edge of the disk, we estimate CBDs in our models to extend between $6\times 10^{12}\cm$ and $4\times10^{15}\cm$ (cf. Fig.\,\ref{fig:Rbars}).

We can extend this consideration to the \EDIT{whole binary grid \REM{of binary models}}. \EDIT{The f}inal orbital separation for systems undergoing \Case BC RLOF \EDIT{is} derived by estimating $\Delta M_\mathrm{RLOF-BC}$ (see Sect.\,\ref{sec:SN:csmMass}) and integrating Eq.\,(5) in \citet{Willcox2023_stabilityRLOF_am} for each candidate model. The resulting CBDs \EDIT{across all models in the binary grid} are expected to \EDIT{be found} between $3\times10^{12}\,\mathrm{cm}$ and $10^{16}\,\mathrm{cm}$.  \EDIT{The inner boundary of the disk can be between $3\times 10^{12}\cm$ to $10^{14}\cm$, implying that ejecta-CSM interaction is always delayed, from several hours to up to just above a day, assuming $\vej\sim9\,000\kms$.}

\subsection{Cumulative effect on the light-curve}
\label{sec:SN:compare:inter:cumulative}

As the expanding ejecta collide with the CSM, a fraction of its kinetic energy will be converted into radiation,  thus affecting the SN light curve. \Type Ibc SNe show a total radiated energy of $\simle 10^{49}\erg$ \citep{Nicholl2015_IbcTemplate}. For an explosion energy $\Eexp$ of $6\times 10^{50}\erg$ (cf. Sect.\,\ref{sec:SN:type}), CSM interaction could dominate the typical \Type Ibc light curve more than $\sim 2\%$ of the kinetic energy is converted into light. {For $\Eexp=\,10^{50}\erg$, more than $10\%$ of $\Eexp$ would need to be converted.}

Assuming an inelastic collision, the cumulative interaction power can be estimated as \citep{DRAD18_GRB_SLSN_Ic}
\begin{equation}\label{eq:interaction}
 E_\mathrm{rad} = \eta \Eexp f_Mf_v^2,
\end{equation}
where $\eta$ is the efficiency with which kinetic energy is converted into radiation, and the two multiplicative factors represent the mass ($f_M=\frac{\Mcsm}{\Mej+\Mcsm}$) and velocity  ($f_v=\frac{\vej-v_\mathrm{CSM}}{\vej}$) contrasts between the ejecta and the CSM. Assuming that the CSM is slow-moving ($f_v\sim\,1$), and that the conversion is highly efficient ($\eta\sim\,1$), the fraction of kinetic energy converted into radiation equals $f_M$, which is only a function of the ejecta mass and the CSM mass.

The value of $f_M$ increases with decreasing ejecta mass and increasing CSM mass (cf. Fig.\,\ref{fig:Mej_Mcsm_Theo}), peaking in the models with the most stripping ($\sim\,0.75$ if $M_\mathrm{CSM}=\Delta M_\mathrm{RLOF-BC}$, and $\sim\,0.45$ if $\Mcsm=\Delta M_\mathrm{RLOF-BC}^{(4)}$). Models with less stripping during \case BC RLOF, can still convert as much as $10\%$ of $\Eexp$ into radiation if $M_\mathrm{CSM}>0.1\Msun$. For both cases, interaction would convert a significant fraction of the SN kinetic energy into radiation, strongly impacting the SN light curve. \EDIT{It is worth noting that these values of $f_M$ are upper limits, as they assume that all the mass remains close enough to the progenitor to affect the light-curve.}

The low mass-loss rates in our models imply low CSM densities, meaning the SN radiation \EDIT{may} not thermalize and therefore not result in an optically-luminous transient. Instead, most of the flux would come out in the UV or as X-rays \citep{Dessart2022_interactingII_xray_emission}. Consequently, we focus on the case in which the CSM is distributed along a CBD, where higher densities are achievable. 
We consider isotropically expanding ejecta \EDIT{impacting} a CSM confined within a half-opening angle $\theta$ from the orbital plane.  The effective ejecta mass $M_\mathrm{ej,eff}$, that is the part of the ejecta that will impact the CSM, is smaller than the total amount $\Mej$ and is determined by the solid angle that the CSM subtends:
\begin{equation}\label{eq:Mej_eff}
    \frac{M_\mathrm{ej,eff}}{\Mej} = \frac{\int_0^{2\pi} \de \varphi \int_{-\theta}^\theta \cos\alpha\de\alpha}{4\pi}=\sin\theta.
\end{equation}
Recalculating the energy lost due to the inelastic collision, we obtain a similar form to Eq.\,\eqref{eq:interaction}, with  
\begin{equation}\label{eq:fM_theta}
f_M(\theta)=\sin\theta\left(1+\frac{\Mej\sin\theta}{\Mcsm}\right)^{-1}.
\end{equation} 

For $\Mej\sim\Mcsm$ and $\theta\ll1$, then $f_M(\theta)\sim\,\theta$. The cumulative interaction-power will exceed the typical radiated energy in \Type Ibc SNe as long as $\theta>1^\circ$ for $\Eexp=6\times 10^ {50}\erg$, or $\theta>5.5^\circ$ for $\Eexp=10^{50}\erg$. Even for very thin disks, interaction power can dominate the cumulative radiation.  In thin disks, high densities may trap radiation, causing it to expand the material rather then escape as radiation. Consequently, our estimates represent upper-limits on radiated energy.

Below, we consider at which time the light curve might be affected the most, based on different assumptions on the CSM structure.

\subsection{Interaction light-curve for a constantly expanding CSM}\label{sec:SN:compare:inter:lightcurve}

\begin{figure*}
\centering
\includegraphics[width=\linewidth]{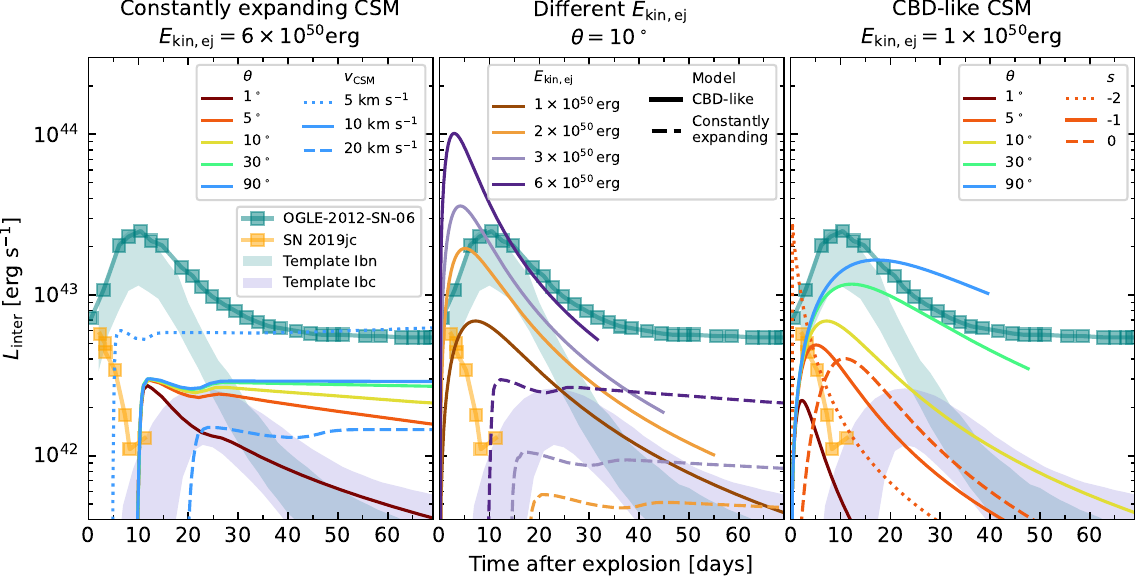}
\caption{Interaction-powered light curves for model B12.3\EDIT{p25.1} assuming a constantly-expanding CSM (left panel), a CBD-like CSM (right panel), and a comparison between the two with different explosion energies (center panel). In each panel, two observations are shown, namely the \Type Ibn SN OGLE-2012-SN-006 \citep[][ light-blue scatter]{Pastorello2015_massivestars_herichCSM_OGLE2012SN006}, and the \Type Icn SN\,2019jc \citep[][ orange scatter]{Pellegrino2022_DiversitySNIcn} as well as the template light curves for \Type Ibc SNe \citep[][in violet, where the peak is located at the same time as in the light-curve of model he4 from \citealt{Dessart2020_Ibc}]{Nicholl2015_IbcTemplate}  and \Type Ibn SNe \citep[][in blue, where the peak is arbitrarily placed at $10\days$ ]{Hosseinzadeh2017_Ibn_lightcurve}. 
 {Left}: the CSM is assumed to constantly expanding at $v_\mathrm{CSM}=10\kms$, with different opening angles (with different colors) and with $\Eexp=6\times 10^{50}\erg$. For $\theta=90^\circ$, models with $v_\mathrm{CSM}$ of $5\kms$ (dotted line),  and $20\kms$ (dashed line) are also shown.  {Center}: interaction power from the CBD-like CSM (solid, with $\Rin=10^{13}\cm$, $\Rout=10^{15}\cm$ and $s=-1$) and constantly expanding CSM (dashed, with $v_\mathrm{CSM}=10\kms$) are shown with different $E_\mathrm{kin,ej}$ (different colors), assuming an opening angle $\theta=10^\circ$.  {Right}: the CSM is assumed to be CBD-like, with $
\Rin=10^{13}\cm$, $\Rout=10^{15}\cm$, $s=-1$ and different opening angles (different colors). For $\theta=5^\circ$, the cases with $s=0$ (dashed) and $s=-2$ (dotted) are also shown.
}
    \label{fig:interaction}
\end{figure*}

We present simplified light curve models, assuming interaction as the sole source of SN luminosity. We neglect diffusion processes and the spatial extension of the ejecta and the swept-up CSM, and focus on the conversion rate of kinetic energy into escaping radiation.
The equations are detailed in Appendix\,\ref{sec:appendix:L_interaction_vCSM}.

We use model B12.3\EDIT{p25.1} as our fiducial model, as it shows the strongest mass-loss rate close to CC of all models ($3.1\times 10^{-4}\msoy$). The results are shown in Fig.\,\ref{fig:interaction}. We also assume that the CSM is isotropically distributed. 

The interaction luminosity ($L_\mathrm{inter}$) becomes significant only after a few days, due to the presence of the cavity in the CSM (Sect.\,\ref{sec:SN:interact:R}) which would otherwise be filled by winds which are presently ignored. The light-curve also exhibits a plateau after the onset of the interaction, which is qualitatively similar to that obtained in detailed radiative-transfer models in \cite{Dessart2022_Ibn} for a wind-like CSM. This plateu, which is not usually present in most observed \Type Ibn SNe with the exception of OGLE-2012-SN-006 \citep{Pastorello2015_massivestars_herichCSM_OGLE2012SN006}, arises from the roughly constant mass-loss rate. The ejecta therefore sweep about the same amount of mass per unit time, producing a steady energy conversion rate.  The absence of diffusion processes in the model is also gives rise to this peculiar signal, as its inclusion would have altered the luminosity evolution at early times, where the CSM is denser and more optically thick. 

The effect of varying $v_\mathrm{CSM}$  are shown in Fig.\,\ref{fig:interaction}. Higher $v_\mathrm{CSM}$ increases the lag time before the onset of interaction and decreases $L_\mathrm{inter}$ (cf., Eq.\,\ref{eq:interaction_luminosity}) and vice-versa.  The interaction persists for as long as several years, as the ejecta sweeps material that has travelled since the beginning of \case BC RLOF. However, the optical signal will remain weak and likely become undetectable much earlier than that, as only the innermost CSM is optically thick.

An aspherical CSM distributed along a torus, confined near the orbital plane within a half-opening angle $\theta$, presents a more interesting scenario. Here, only a fraction of the ejecta interacts with the CSM ($M_\mathrm{ej,eff}$, cf. Sect\,\ref{sec:SN:compare:inter:cumulative}), which decreases with smaller $\theta$.  At the onset of interaction $L_\mathrm{inter}$ remains unaffected, as the same amount of mass would be swept up independently of $\theta$. However, smaller $\theta$ cause a stronger deceleration of the effective ejecta,  reducing $L_\mathrm{inter}$ at later times (see Eq.\,\ref{eq:interaction_luminosity}). Significant effects arise only for very small opening angles ($< 10^\circ$).

 {Lower $\Eexp$ affect the light-curve in two ways, due to the slower expanding ejecta: reduced luminosity and delayed onset of interaction  (cf. Fig.\,\ref{fig:interaction}). As in the model $L_\mathrm{inter}\sim\,\Eexp^{3/2}$, the overall luminosity decreases strongly with decreasing $\Eexp$. The onset of interaction is instead delayed by a factor $\vej\sim\,\Eexp^{1/2}$. For $\Eexp\leq 10^{50}\erg$, the interaction luminosity drops to $\simle 2\times 10^{41}\ergs$, and would remain largely undetectable. 

\subsection{Interaction light-curve for a CBD-like CSM}\label{sec:SN:interaction:CBD_lightcurve}
We now model the interaction assuming the CSM is a bound CBD (such that $v_\mathrm{CSM}=0$) rather than spherically symmetric (cf. Sect.\,\ref{sec:SN:interact:R}). This approach introduces additional parameters and assumptions -- such as the half-opening angle $\theta$, the CBD density profile and its radial extent --  which are not constrained by our models. We will assume that $\rho(r)\sim\,r^{s}$, with $s\leq0$, and that the CBD is confined between $\Rin=10^{13}\cm$ and $\Rout=10^{15}\cm$.

As a proof of concept, we present a simplified light curve model (see Appendix\,\ref{sec:appendix:L_interaction:CBD}), assuming $\Eexp=10^{50}\erg$ (Fig.\,\ref{fig:interaction}). The effect of varying $\theta$ is more pronounced than in the case of a constantly expanding CSM.  {For thinner disks, $L_\mathrm{inter}$ is weaker and dims faster, as less ejecta interacts with the CBD.} The characteristic the bell-shape feature, typical of many \Type Ibn SNe \citep[cf.][]{Hosseinzadeh2017_Ibn_lightcurve}, is qualitatively reproduced. Varying $s$ also has a considerable effect, as a steeper density profile (i.e., lower $s$) confine the CBD closer to the binary, resulting in a earlier interaction and higher initial $L_\mathrm{inter}$ (Fig.\,\ref{fig:interaction}).  Even when adopting such a low $\Eexp$, the interaction power is still significant even for $\theta=5^\circ$ where $L_\mathrm{inter}$ peaks at $>2\times10^{42}\ergs$. Unlike the constantly-expanding CSM scenario, interaction terminates after $\sim\,70\days$, as the entire CBD is swept away.

 {Adopting $\Eexp>10^{50}\erg$ results in a brighter and faster-evolving light-curve, similarily to the case of a constantly expanding CSM (cf.\,Sect.\,\ref{sec:SN:compare:inter:lightcurve}, and Fig.\,\ref{fig:interaction}), as the peak brightness scales with $\Eexp^{3/2}$ and the peak time with $\Eexp^{1/2}$. Models with $\theta\geq 10^\circ$ are capable of exceeding the typical luminosity of \Type Ibc SNe, even with $\Eexp<10^{50}\erg$. }
This model can also qualitatively reproduce the behavior of \Type Icn SN\,2019jc, especially in the case of thin, confined CBD and higher $\Eexp$.

This experiment, while very crude, demonstrates that a disk-like CSM can significantly shape the light-curve, particularily for thin disks, which qualitatively mash some observed \Type Ibn and \Type Icn light curves. This model however neglects radiation transport, which would smear the interaction power on longer timescales, or even reduce the radiated energy if some is used to expand the material in the CBD. The interaction's asymmetry introduces viewing angle effects. In the case of edge-on observations, the light-curve may be additionally smeared-out \citep[see][in the context of \Type II SNe]{Vlasis16_asymmetricmodels_IIn, Suzuki2019_HrichSN2DDiskCSM}. Finally, a disk-like structure may conflict with the narrow-line emission spectra \citep{Smith17_SNHandbook_InteractingSNe}.

\subsection{Predicting the number of interacting H-poor supernovae}\label{sec:SN:pop}
Since we infer the mass-loss during \case BC RLOF from each model in the \EDIT{comprehensive binary grid of models from Jin et al. (in prep.,} cf. Sect.\,\ref{sec:SN:csmMass} and Fig.\,\ref{fig:Mej_Mcsm_Theo}), we provide population estimates for the models we predict will develop into an interacting SN. As shown throughout Sects.\,\ref{sec:SN:compare:inter:cumulative}-\ref{sec:SN:interaction:CBD_lightcurve}, a significant fraction of the kinetic energy of the ejecta can be converted into radiation when $M_\mathrm{CSM}\simgr0.1\Msun$. We assume that an interacting H-free \EDIT{SN} may likely develop in those systems where the \EDIT{strpped-envelope} donor star underwent \Case BC RLOF. 

\EDIT{We perform a population synthesis calculation, assigning an initial probability distribution to each model in the binary grid from Jin et al.\, (in prep.) assuming that all stars are born in binaries with a \citet{Salpeter_IMF_55} initial mass-function for the initially more massive star and a flat initial $\log P_\mathrm{i}$ and $q_\mathrm{i}$ distribution}. We also assume that \EDIT{the binaries} that undergo unstable RLOF will merge (see Sect.\,\ref{sec:methods:RLOF}), and that the merger product will evolve like a single star of the same mass as the sum of the two component stars. \EDIT{We identify SN progenitors undergoing \Case BC RLOF by adopting the method shown in Sect.\,\ref{sec:SN:csmMass}, and obtain that up to $\sim 12\%$ of all stripped-envelope SN progenitors will develop into an interacting H-free SN.} The observed fraction of Type Ibn SNe reported in \cite{Perley2020_ZTF_SNdemographics} is $9.2\%$, although this is a magnitude-limited sample and therefore \EDIT{may overestimate the true population of these events.} This similarity suggests that binary mass transfer might be a suitable channel to create the majority of interacting H-free SNe.

\section{Comparing observed interacting H-poor SNe with the models}\label{sec:SN:compare}

To date, there is a growing number of H-poor SNe that exhibit features compatible with that of interaction with a nearby CSM. The light curve and the spectra provide insight into both the explosion parameters of the SN ($\Mej$, $\vej$, $\Mni$ and $\Eexp$) and the properties of the CSM ($\Mcsm$, $R_\mathrm{CSM}$), albeit under a series of assumptions (see \texttt{MOSFiT}, \citealt{MOSFIT}, and also \citealt{Chatzopoulos2012_generalizedlightcurvefitting, Chatzopoulos2013_analiticalmodel, Villar17_lightcurves}). 
These models typically assumed isotropic CSM distributions around the progenitor, with a density profiles of the form $\rho(r)\sim\,r^{s}$ starting at a distance $R_\mathrm{CSM}$. Many works usually keep the parameter $s$ fixed to either $0$ (a constant distribution) or $-2$ (wind-like) or explore both, yielding significantly different light-curve fitting parameters \citep[e.g., ][]{BenAmi2023_SN2019kbj}. More recent theoretical works have produced \Type Ibn SNe by exploding a low-mass He-star model while surrounded by a massive CSM, yielding interesting matches to the light curves and spectra of events like SN\,2006jc, SN\,2018bcc and SN\,2011hw \citep{Dessart2022_Ibn} 

 There is a notable lack of systematic analysis of H-poor interacting SNe. Many studies incorporate additional physical processes, such as Ni-decay or magnetar spin-down, which complicate the interpretation of the effects of CSM interaction. Moreover, the discussion in Sect.\,\ref{sec:SN:interact:R} tells us that the CSM may be distributed aspherically, violating the assumption of isotropy used in standard light-curve fitting methods. These modeling uncertainties inevitably affect the interpretation of observations.
 
 In this work, we compare our model predictions with published observationally-inferred properties of H-poor interacting SNe, without making claims about any specific observed event. It should be emphasized that we use literature values without assessing their accuracy or robustness. Some of the values, especially from older studies, may be inaccurate.  

\begin{figure*}
\centering
\includegraphics[width=1\linewidth]{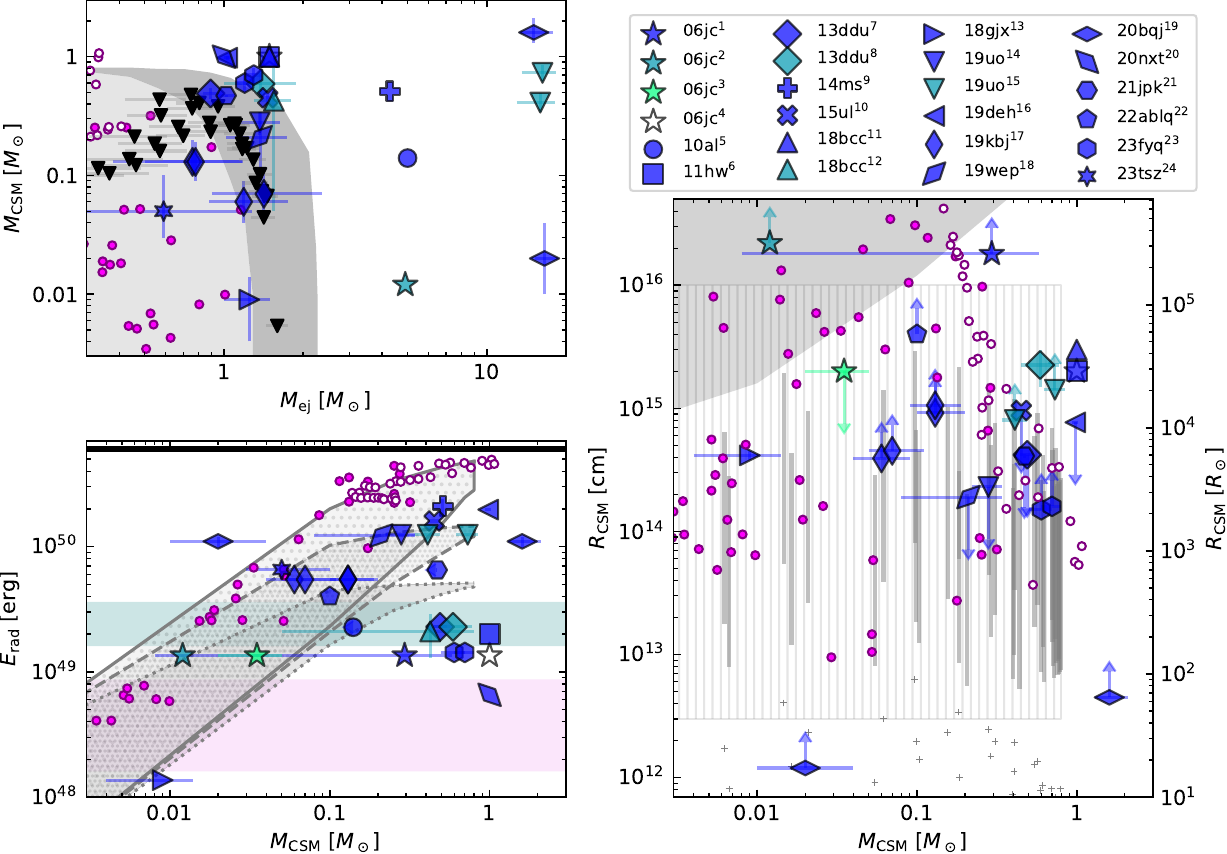}
\caption{Comparison between model data and parameters inferred from observed \Type Ibn SNe (blue-scale colored markers), with different markers associated to different SNe, and different colors for the same SN highlighting different references. Estimates inferred from the models of \WF22 are also included \EDIT{and are shown as in Fig.\,\ref{fig:Mej_Mcsm_Theo}}.  \textbf{Top Left:} Scatter plot showing the amount of mass of the CSM against the ejecta mass in the models as in Fig.\,\ref{fig:Mej_Mcsm_Theo}. As the values of $M_\mathrm{CSM}$ from the models is an upper limit, a light-gray region is added where models can still be expected.  \textbf{Right:} Scatter plot showing the extent of the CSM against its mass. The position of CBD-like CSM from our models is shown with a gray line for $\Mcsm=\Delta M_\mathrm{RLOF-BC}$. The region where CBDs from the \EDIT{models in the binary grid are expected to be found} is shown with a vertically-hatched region. The radii of our theoretical model are also shown (cross scatter).  A gray patch is drawn where to expect the CSM if it were to move constantly at a speed of $10\kms$.  \textbf{Bottom Left:} Scatter plot showing the energy released by interaction (Eq.\,\eqref{eq:interaction}) as a function of CSM mass for the models in this work and \WF22 assuming $\Eexp=6\times10^{50}\erg$ (shown with a black horizontal line), and the total light-curve energy of the observed SNe. The shaded regions represent the typical $E_\mathrm{rad}$ of \Type Ibc \citep[light-red, ][]{Nicholl2015_IbcTemplate} and \Type Ibn SNe \citep[light-blue, ][]{Hosseinzadeh2017_Ibn_lightcurve}. The values expected from our models and \EDIT{those in the binary grid} are summarized in gray regions, which also show the expected $E_\mathrm{rad}$ if $\Mcsm$ is lower than the values inferred. Each gray region assumes that the CSM is distributed  spherically (solid line, light-hatch), along a CBD with $\theta=15^\circ$ (dashed line, dense hatch) and $5^\circ$ (dotted line, gray fill) of the orbital plane (cf. Eq.\,\eqref{eq:fM_theta}) .\\
 {References:} \object{SN2006jc}: (1) \citealt{Mattila2008_SN2006jc, Pastorello2007_2006jc_preSNoutburst}
(2) \citealt{Tominaga2008_2006jc, Pastorello2007_2006jc_preSNoutburst}
(3) \citealt{Chugai2009_2006jc, Pastorello2007_2006jc_preSNoutburst};
(4) \citealt{Dessart2022_Ibn, Pastorello2007_2006jc_preSNoutburst}
\object{SN2010al}:
(5) \citealt{Chugai2022_2010al}; 
\object{SN2011hw}: (6) \citealt{Dessart2022_Ibn, Pastorello2015_massivestars_herichCSM_2010al_2011hw};
\object{LSQ13ddu}: (7) \citealt{Pastorello2022_CSM_of_Ibn, Clark2020_LSQ13ddu}
(8) \citealt{Clark2020_LSQ13ddu, Bathauer2022_2014C_plus_interactingSNe}; 
\object{ASASSN-15ms}:
(9) \citealt{Vallely2018_ASASSN-14ms}; 
\object{iPTF15ul}:
(10) \citealt{Pastorello2022_CSM_of_Ibn, Hosseinzadeh2017_Ibn_lightcurve}; 
\object{SN2018bcc}:
(11) \citealt{Dessart2022_Ibn, Karamehmetoglu2021_2018bcc}; 
(12) \citealt{Karamehmetoglu2021_2018bcc}; 
\object{SN2018gjx}:
(13) \citealt{Prentice2020_2018gjx}; 
\object{SN2019uo}:
(14) \citealt{Pastorello2022_CSM_of_Ibn, Gangopadhyay2020_2019uo}
(15) \citealt{Gangopadhyay2020_2019uo}; 
\object{SN2019deh}:
(16) \citealt{Pastorello2022_CSM_of_Ibn}; 
\object{SN2019kbj}:
(17) \citealt{BenAmi2023_SN2019kbj}; 
\object{SN2019wep}:
(18) \citealt{Pastorello2022_CSM_of_Ibn, Gangopadhyay22_2019wep}; 
\object{SN2020bqj}:
(19) \citealt{Kool2021_SN2020bqj_Ibn_longplateau}; 
\object{SN2020nxt}:
(20) \citealt{Qinan2024_2020nxt}; 
\object{SN2021jpk}: 
(21) \citealt{Pastorello2022_CSM_of_Ibn};
\object{SN2022ablq}:
(22) \citealt{Pellegrino24_2022ablq}; 
\object{SN2023fyq}:
(23) \citealt{Dong2024_2023fyq_precursor_eqCSM}; 
\object{SN2023tsz}:
(24) \citealt{Warwick24_SN2023tsz}
}
    \label{fig:Mej_Mcsm_Ibn}
\end{figure*}

We searched the literature for H-poor interacting SNe and included only those for which $M_\mathrm{CSM}$  values are available.  {In the following, we discuss \Type Ibn SNe (Sect.\,\ref{sec:SN:obs:Ibn}) and \Type Icn/Ic-CSM SNe (Sect.\,\ref{sec:SN:obs:Icn})\EDIT{.\REM{, while in Appendix\,G we discuss also other transients like USSNe and SLSNe.}} 

\subsection{\Type Ibn SNe}\label{sec:SN:obs:Ibn}
The sample of \Type Ibn SNe includes H-poor and He-rich transients that showcase narrow He-emission lines. The sample spans a wide range of $\Mej$  (between $0.6$ and $20\Msun$), and $\Mcsm$ (between $0.01$ and $2\Msun$). Most  SNe in the sample cluster around $\Mej\sim\,1\Msun$ and $\Mcsm$ between $\sim\,0.1$ and $1\Msun$. Despite their large uncertainties, these values align with the parameter space predicted by our models (cf. top-left panel of Fig.\,\ref{fig:Mej_Mcsm_Ibn}), regardless of whether $\Mcsm$ corresponds to $\Delta M_\mathrm{RLOF-BC}$ or $\Delta M_\mathrm{RLOF-BC}^{(4)}$. Some SNe fall outside of the predicted $\Mcsm-\Mej$ region (SN\,2006jc, SN\,2010al, ASASSN-14ms, SN\,2020bqj, SN\,2019uo). These outliers, characterized by high $\Mej$, likely originate from a different evolutionary channel.

The typical inferred CSM extension for \Type Ibn SNe exceeds $10^{14}\,\mathrm{cm}$ (cf., Fig.\,\ref{fig:Mej_Mcsm_Ibn}, right panel), consistent with the CBD-like CSM expected from the models. Notable exceptions include SN\,2020bqj, whose inferred CSM extent is similar to the radii of our SN-progenitor models, and SN\,2006jc ($>10^{16}\,\mathrm{cm}$), which appears to be instead in agreement with the case of a slow-moving CSM.

The majority of the \Type Ibn SNe collected deviate from the \Type Ibn template of \cite{Hosseinzadeh2017_Ibn_lightcurve} as reflected in their the total radiated energy in the light-curve $E_\mathrm{rad}$  (cf., Fig.\,\ref{fig:Mej_Mcsm_Ibn}, bottom-left panel). Most SNe from \cite{Hosseinzadeh2017_Ibn_lightcurve} lack $\Mcsm$ estimates and were therefore excluded from our sample. The collected SNe show a large spread of $E_\mathrm{rad}$, with most having a higher value than inferred from the template \Type Ibn SNe, and only a few with significantly lower values (SN\,2018gjx and SN\,2020nxt), which may be the effect of a shorter observation campaign or poor multi-band coverage. 

The radiated energy from the models scales with the explosion energy $\Eexp$ and decreases with very small half-opening angles (cf. Eq.\,\ref{eq:interaction}-\ref{eq:fM_theta}).  The observations with the higher $E_\mathrm{rad}$ and $M_\mathrm{CSM}$ align with our models assuming $\Eexp=6\times 10^{50}\erg$ and a spherically distributed-CSM (cf., Fig.\,\ref{fig:Mej_Mcsm_Ibn}, bottom-left panel), although, they can still be explained assuming a lower $\Eexp$ or a CBD-like CSM with $\theta\geq15^\circ$.  {Dimmer events are consistent with significantly lower explosion energies ($\Eexp\sim\,10^{50}\erg$), expected in the models with the stronger mass-loss and hence smaller final core masses (Sect.\,\ref{sec:SN:type}), as well as a thinner CBD-like CSM, with $\theta\simle 5^\circ$.} 

The progenitor and CSM properties inferred from the observed \Type Ibn SN we collected \EDIT{appear to be consistent with} the CBD-like CSM sorrunding the binary models explored in this work. However, challenges remain regarding narrow-lines observations. As the narrow lines would originate within the disk, which is likely engulfed by the freely-expanding ejecta in the other directions, the narrow-line component would likely be thermalized and therefore unobservable \citep{Smith17_SNHandbook_InteractingSNe}.

\subsection{\Type Icn and Ic-CSM SNe}\label{sec:SN:obs:Icn}

\begin{figure*}
\centering
\includegraphics[width=1\linewidth]{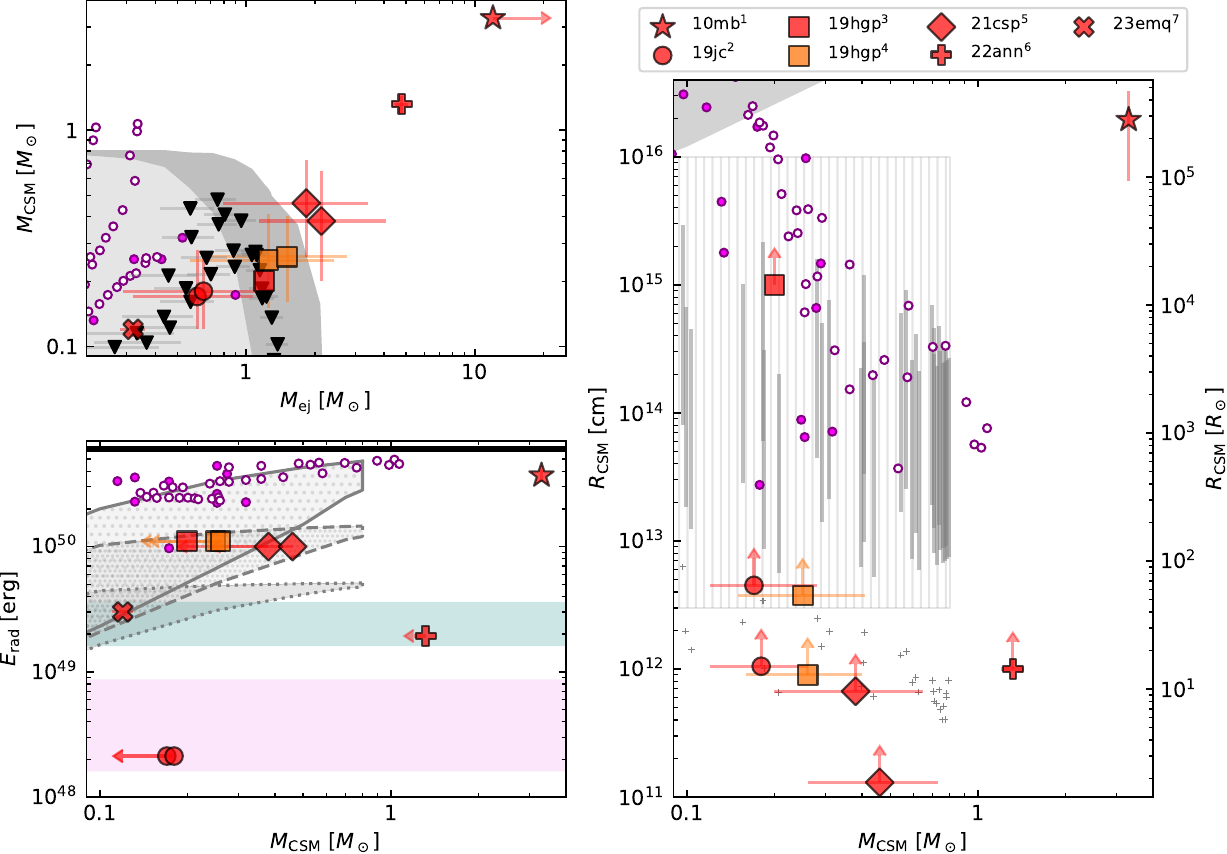}
\caption{The same as Fig.\,\ref{fig:Mej_Mcsm_Ibn} but for \Type Icn SNe and the \Type Ic-CSM SN\,2010mb. \\
 {References:} 
\object{SN2010mb}: (1) \citealt{BenAmi2014_2010mb, Bathauer2022_2014C_plus_interactingSNe}; 
\object{SN2019jc}:
(2) \citealt{Pellegrino2022_DiversitySNIcn}; 
\object{SN2019hgp}:
(3) \citealt{GalYam2022_SN2019hgp_Icn};
(4) \citealt{Pellegrino2022_DiversitySNIcn, GalYam2022_SN2019hgp_Icn};
\object{SN2021csp}:
(5) \citealt{Pellegrino2022_DiversitySNIcn, Perley2022_SN2021csp_Icn};
\object{SN2022ann}:
(6) \citealt{Davis23_SN2022ann}; 
\object{SN2023emq}:
(7) \citealt{Pursiainen23_SN2023emq}.\\
%\EDIT{Note: The data regarding $M_\mathrm{CSM}$ in \cite{Nagao23_SN2021ckj} is extrapolated by integrating the CSM density law they provided between $10\Rsun$ and $10^{15}\cm$.}
}
    \label{fig:Mej_Mcsm_Icn}
\end{figure*}

The observed \Type Icn/Ic-CSM SNe feature a diverse sample of events, from the very stripped SN\,2023emq ($\Mej\sim\,0.3\Msun$) to the massive Wolf-Rayet progenitor of SN\,2010mb ($\Mej>10\Msun$). Interestingly, the sample shows a possible correlation between $\Mej$ and $\Mcsm$, with $\Mej\sim\,4\Mcsm$ (cf., Fig.\,\ref{fig:Mej_Mcsm_Icn}, upper-left panel), though this be a result of the low number of observed events. SNe with $\Mej\simle 1.5\Msun$ align with our models if $\Mcsm=\Delta M_\mathrm{RLOF-BC}^{(4)}$ (and also $\Delta M_\mathrm{RLOF-BC}$ for those that also have $\Mej\simgr1\Msun$). The ones with higher ejecta and CSM masses  (i.e., SN\,2010mb, SN\,2021ckj, and SN\,2022ann) cannot be explained by our models. 

The CSM radii in \Type Icn SNe are more uncertain than \Type Ibn SNe.  For example the CSM is inferred to extend from close to the progenitor  \citep[$\sim\,1\Rsun$ \EDIT{in} SN\,2021csp,][]{Pellegrino2022_DiversitySNIcn} to $10^{15}\,\mathrm{cm}$ (SN\,2019hgp, \citealt{GalYam2022_SN2019hgp_Icn}). For some of these SNe, the adoption of different assumptions in the light-curve modeling changes the value of $R_\mathrm{CSM}$ by about one order of magnitude (e.g., SN\,2019jc, SN\,2019hgp ad SN\,2021csp; cf. Fig.\,\ref{fig:Mej_Mcsm_Icn}, right panel), while $\Mej$ and $\Mcsm$ are not affected by more than a factor of two. The inferred CSM extension in these SNe is compatible with either an extended envelope or a CBD. The slow-moving CSM model can, instead, be ruled out.

The radiated energy of these SNe varies significantly. Brighter events, (e.g., SN\,2019hgp, SN\,2021csp), align with our models assuming a spherically-symmetric CSM , as well as a CBD-like structure with $\theta>20^\circ$.  
\EDIT{The} underluminous event \EDIT{\REM{SN\,2021ckj and}} SN\,2019jc \citep{Pellegrino2022_DiversitySNIcn} \EDIT{is} also consistent with our models when considering lower $\Eexp$ and a CBD-like CSM with small opening angles.

Similar to the collected \Type Ibn SNe, some \Type Icn SNe are compatible with the binary channel proposed and a CBD-like CSM structure proposed in this work.  However, explaining their spectral evolution remains challenging,
especially as the CSM of all our models is He-rich. While there are models with very low He in the ejecta ($\simle 0.2\Msun$, cf. Fig.\,\ref{fig:deltaMC_Mhe_Mz}), the metal-rich lines would likely appear broad rather than narrow.

\subsubsection*{Summary}

Our models indicate that \Case BC RLOF can produce progenitors and CSM structures broadly consistent with those inferred for a subset of the observed \Type Ibn, especially when considering a CSM distributed along a CBD. While our models struggle to fully account for \Type Icn/Ic-CSM SNe, the similarities in bulk properties (especially $\Mcsm, \Mej$ and $R_\mathrm{CSM}$) and light-curves (cf. Sect.\,\ref{sec:SN:interaction:CBD_lightcurve}) warrant further investigation of the spectral evolution in progenitors like those presented in this work.

Caution is warranted when interpreting these results. The CSM mass predicted by our models represents an upper limit, while observational estimates are biased toward a spherically symmetric distribution, which our models disfavor. These discrepancies highlight the need for further investigation of the CSM structure and the interaction features it produces, particularly in the case of a CBD-like configuration.

\subsection{Comparisons with \Case X RLOF}\label{sec:SN:caseX}
If the progenitor underwent \case X RLOF, the SN would occur while the system is embedded in a H-poor, He- and metal-rich CE (cf. Sect.\,\ref{sec:res:example_mod:caseX} and \ref{sec:overv:caseX}), which would likely form the CSM. The CE's impact on the light curve would be significant (as $f_M\sim\,1$), likely turning the SN into a SLSN.  The CSM structure will be qualitatively different from that generated during \Case BC RLOF, due to its more dynamic nature and the shorter timescale before CC.

\Case X RLOF has been investigated systematically in \WF22 \EDIT{for HeSs with NS companions, assuming that it provides either a stable or an unstable phase of mass transfer.} The timescales \EDIT{\REM{and mass transfer rates}} for \case X RLOF differ significantly \EDIT{between their models and our own}: in \WF22, \case X RLOF begins even decades before CC, whereas in our models it is confined to only a few weeks or months.  \EDIT{We compare their results with the observed \Type Ibn and Icn SNe (cf. Figs.\,\ref{fig:Mej_Mcsm_Ibn}-\ref{fig:Mej_Mcsm_Icn}). Their stable \case X scenario predicts $\Mcsm$ of up to $\sim0.3\Msun$ and $0.1\Msun \simle \Mej \simle 1\Msun$, which is consistent with several observed \Type Ibn and Icn SNe with $\Mej\simle1\Msun$ and $\Mcsm<0.3\Msun$, but falls short to explain the bulk of the \Type Ibn observations with higher $\Mcsm$. The unstable \Case X scenario, which yields $\Mcsm$ of up to $\sim1\Msun$ and $0.1\Msun < \Mej < 0.4\Msun$, appears incompatible with most observations, due to the very low predicted ejecta masses in that case.} 
 
\EDIT{\WF22 provide values for $R_\mathrm{CSM}$ assuming $\vej$ equal to the orbital velocity of the L2 point, usually between $100$ and $500\kms$. As a result,} their constantly expanding CSM \EDIT{\REM{(assuming $v_\mathrm{CSM}=200\kms$)}} is comparable in size to that of a CBD-like CSM from \Case BC RLOF in our models (cf. Figs.\,\ref{fig:Mej_Mcsm_Ibn}-\ref{fig:Mej_Mcsm_Icn}) and also many observations of \Type Ibn SNe \EDIT{(as also discussed in their Sect.\,3.5)}. 
\EDIT{On the other hand}, The CSM geometry may be more complicated than that expected from \EDIT{our models during} \case BC RLOF, \EDIT{especially if \case X RLOF is unstable}.   
  
The numerical settings used in \WF22 are similar to our own, suggesting that their models may also be affected by numerical issues that give rise to this phase of RLOF (see  \EDIT{\REM{Appendix} Sect}\,\ref{sec:appendix:shell_merger_overshooting} for a discussion).

\section{Discussion}\label{sec:disc}
The results from the models show a series of features that are highly dependent on the assumptions made. We will discuss the main uncertainties affecting our result and provide comparisons to previous works.

\subsection{Winds and metallicity}\label{sec:disc:winds}
The evolution of the models discussed is strongly influenced by the adopted wind scheme. For HeS mas-loss rates, our models use the prescription from \citet[][]{Yoon_wind} \citep[cf. ][]{Jin2024_Boron}, which combines empirical rates of observed Wolf-Rayet stars with higher luminosities than those found in our models \citep[$\log L/\Lsun>4.90$ from ][ against the maximum luminosity before \case BC RLOF in our models of $\log L/\Lsun=4.75$]{NugisLamers2000, Hamann2006_Potsdam, Hainich_WRwinds, TSK_wind}.
This raises the question about the validity of our choice of wind recipes. 

Recent studies like \cite{Vink_WR} argue for lower mass-loss rates for low-mass HeSs. \cite{Gilkis_Wind} show that this wind mass-loss rate actually prevents the complete loss of the H-rich envelope if adopted when $X_s$ drops below 0.4. This contrasts with the higher rates predicted by \citealt{NugisLamers2000} , which our models adopt. Additionally, \cite{Gotberg2023_stripped_stars_in_binaries} suggest that the winds of some observed HeSs observed in the LMC and SMC \citep{Drout2023_Observed_stripped_HeStars_inBinaries} are comparable to, or even lower than, those described in \cite{Vink_WR}. 

Significantly weaker winds following \case B RLOF, as would occur with lower initial metallicity, shift the parameter space for \case BC RLOF towards systems with lower initial masses (cf. Sect.\,\ref{sec:ps}), because the core remains more massive. A key consequence of weaker winds is that the H-rich envelope may not be completely removed by core He-depletion. This thin H-rich envelope \EDIT{\REM{which}} may be extended enough to also increase the probability of \Case BC RLOF \citep[][]{Yoon_IIb_Ib}. \EDIT{The shift to lower initial masses and the larger radii increase the population of interacting \EDIT{SNe}, which are however expected to} appear as a \Type IIb SN with narrow H-lines if \Case BC RLOF fails to strip the remaining H-rich envelope.

Binary models at lower metallicities also fail to lose the H-rich envelope completely \citep{Yoon_IIb_Ib, Laplace2020_Stripped_Envelope_SNe}, due to the reduced winds. In some cases, RLOF may occur before core He-exhaustion \citep{ Klencki_partialstripping_2022}, which can be followed again by \Case BC RLOF to further strip down the envelope. 

\EDIT{The shift to lower metallicities offers, on the one hand, similar effects as those expected when decreasing the wind, namely that the parameter space for \Case BC systems would shift towards lower initial masses. However, there are other metallicity-related effects, such as the changes to the radius evolution, that may reduce the parameter space for \Case BC systems. Such effects require detailed calculations to assess the metallicity-dependence of \Case BC RLOF and hence the population of interacting \EDIT{SNe}.} 

\subsection{Mass transfer}
\subsubsection{Mass accretion and angular-momentum loss}
The amount of mass accreted by the companion star during mass transfer and the amount of angular momentum carried away by unaccreted material are critical parameters influencing the evolution of mass-transferring binaries \citep{Podsiadlowski_massive_star_binary_interaction_1992}. We report the key effects on our models, and we refer to \cite{Ercolino_widebinary_RSG} (and references therein) for a more detailed discussion.

Angular momentum losses may exceed those accounted for in our models due to additional mechanisms. For example, outflows from the outer Lagrangian point of the accretor, that may occur when mass transfer rates exceed $10^{-4}\msoy$ \citep[][]{Lu2023_L2_outflow_during_MT}, can carry away significant angular-momentum. Furthermore, unaccreted material remaining in the vicinity of the system can exchange angular momentum with the inner binary. In the presence of a CBD, additional processes such as mass reaccretion, angular momentum exchanges and eccentricity pumping may also occur \citep[e.g.,][]{Wei2023_fate_of_postCCEbinaries, Valli24_CBD_binaryevolution}. Higher angular momentum losses would tighten the orbit during RLOF and therefore remove more mass from the donor star. 

For \case B RLOF, increase angular momentum losses would result in smaller $M_\mathrm{He-dep}$ and $R_\mathrm{RL,1}$, which would increase the parameter space for \case BC RLOF. However, this would also make mass transfer more prone to becoming unstable \citep{Willcox2023_stabilityRLOF_am}. 
In the context of \case BC RLOF, these effects will likely lead to more mass-loss and therefore larger $M_\mathrm{CSM}$. In this case however, the stability of mass transfer would likely remain unaffected due to the significantly more massive companion and contained mass-loss.

Our models may also underestimate the amount of mass accreted by the secondary \citep{ Vinciguerra20_BeXray_Accretion_efficiency}. Increase in mass transfer efficiency would remove more mass from the donor \cite[e.g.][]{Claeys_b}, expanding the parameter space for \case BC RLOF to occur as well as increasing the likelihood of mass transfer turning unstable \citep[][]{Braun_Langer_95,SchuermannLanger2024_StabilitySecAccrete}. Increased accretion efficiency would reduce the amount of CSM as more of the transferred mass would be deposited on the companion star, rather than lost from the system.

\subsubsection{Stability of Mass transfer}\label{sec:disc:MT_stability}

Even though the computation of mass transfer proceeds unimpeded, special care is required to ensure the results have physical significance. This includes verifying whether the underlying assumptions hold during the calculations \citep{Temmink23_MT_stability} or determining if the system enters an unstable phase of mass transfer. The criteria for identifying unstable mass transfer remain debated and we discuss a few in the context of \case B and \case BC RLOF below. We will not discuss the effects on \case X RLOF here (but see Sect.\,\ref{sec:overv:caseX}). 

Our calculations assume that unaccreted material is ejected as a fast wind from the secondary star \citep{Soberman97_StabilityCriteriaMassTransfer_abg}. However, there is a limited understanding of how the unaccreted material escapes the inner binary. If it is not ejected efficiently, it may exert drag and extract orbital angular momentum, driving the system into an unstable phase of RLOF.  

  \begin{figure}
    \includegraphics[width=1\linewidth]{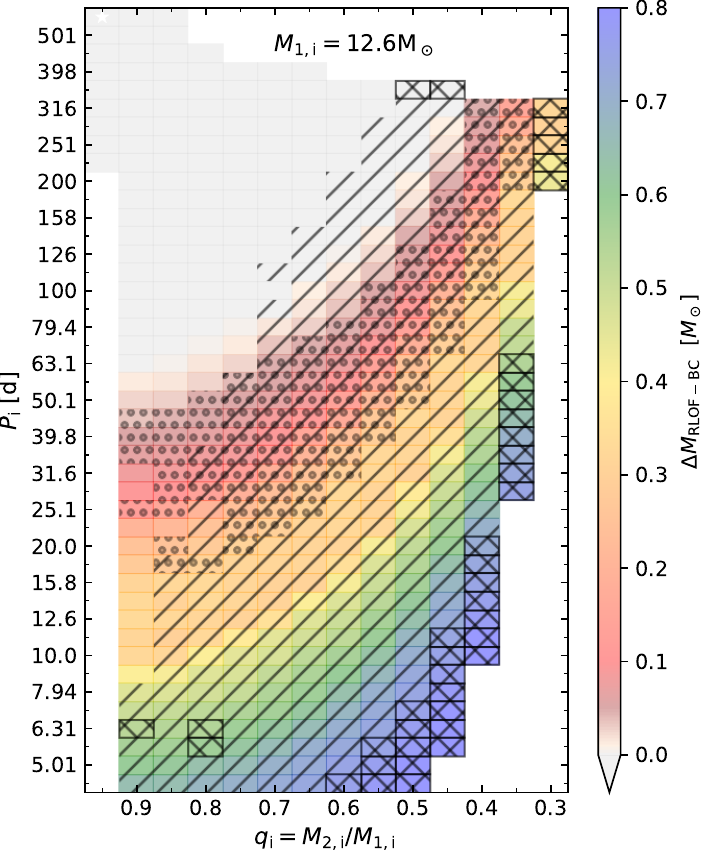}
\caption{The $\log P_\mathrm{i}-\qi$ diagram of the the \case B binary models in the \EDIT{binary grid from Jin et al. (in prep.)} with $M_\mathrm{1,i}=12.6\Msun$, color-coded based on the amount of mass expected to be shed during \case BC RLOF. The hatching indicates models that are flagged as having undergone unstable \case B RLOF according to the criterion from \cite{PABLO_LMC_GRID} (single hatch), \cite{Pauli_MasterThesis} (cross hatch) and \cite{Pavlovskii_Ivanova_2015_MT_from_Giants} (circle hatches).}
    \label{fig:logPq_MTS}
\end{figure}

\cite{PABLO_LMC_GRID} proposes that the unaccreted material can be ejected via a radiation-driven wind, which defines an upper limit on the mass transfer rate $\dot M_\mathrm{high}$. This rate represents the threshold above which the combined luminosity of the two stars is not able to drive the material away radiatively. The un-accreted material is assumed to lie on the outer edges of the accretion disk, at a distance $R_\mathrm{RL,2}$ from the secondary star, and the limit is given by
\begin{equation}\label{eq:pablo_Mdot_high}
\dot M_\mathrm{high} = (L_1+L_2)\frac{R_\mathrm{RL,2}^2}{GM_2},
\end{equation}
where $M_2$ is the mass of the secondary, and $L_1$ and $L_2$ the luminosities of the primary and secondary stars respectively. If mass transfer is inefficient and $\dot M_\mathrm{RLOF}>\dot M_\mathrm{high}$, the system cannot remove the unaccreted material from the inner binary, leading to unstable RLOF. Fig.\,\ref{fig:logPq_MTS} illustrates a slice of the parameter space of \Case BC systems at $M_\mathrm{1,i}=12.6\Msun$, revealing that many systems but those with high $P_\mathrm{i}$ or $\qi$ would undergo CE evolution during \case B RLOF. This is because during \case B RLOF the donor star temporarily dims as it is thrown off of thermal equilibrium. Unless the binary is wide or the companion is very luminous, this instability criterion is met. In our models, where $\qi=0.5$, are always flagged as unstable according to this criterion. 

\citet{Pauli_MasterThesis} argues that during the short timescale in which the primary star dims, the material can still be pushed away from the inner binary to a finite distance from the system before the primary resettles to a higher luminosity. This may subsequently enable the radiation-driven ejection of the material. They argue that in Eq.\,\eqref{eq:pablo_Mdot_high}, $L_1$ should be replaced by the maximum luminosity reached by the donor during mass transfer. This makes the criterion less stringent than that of \citet{PABLO_LMC_GRID}, classifying only the models with the lowest $\qi$ and $P_\mathrm{i}$ as unstable during \case B RLOF. In our rerun models, this affects only the models with $P_\mathrm{i}<10\days$.

Figure\,\ref{fig:logPq_MTS} highlights the systems expected to undergo \case BC RLOF. Using the stricter criterion from \cite{PABLO_LMC_GRID} significantly reduces the parameter space for the binaries that survive \Case B RLOF. For instance, in the grid at $M_\mathrm{1,i}=12.6\Msun$, most models predicted to experience \case BC RLOF (which give rise to H-poor intreacting SNe) would instead have undergone unstable \case B RLOF and likely merge (likely producing a \Type II SN progenitor). By contrast, the criterion from \cite{Pauli_LMC} only excludes a handful of systems. Both criteria flag more systems as stable with increasing $M_\mathrm{1,i}$ due to the higher luminosities. 

Another indicator of the onset of unstable RLOF is when the donor star swells to the point of undergoing OLOF \citep[][]{Pavlovskii_Ivanova_2015_MT_from_Giants, Ercolino_widebinary_RSG}. In this case, the results in the \EDIT{binary grid} differ from the binaries we rerun. Due to the inclusion of radiation pressure in the mass-transfer scheme (cf. Sect.\,\ref{sec:methods:RLOF}), the donor stars in our models do not expand as much as \EDIT{those} in the \EDIT{binary grid} during \case B mass transfer, and therefore do not exhibit OLOF, even in cases where OLOF is flagged in the models \EDIT{from the binary grid}. \Case BC RLOF is never flagged as unstable per the criterion from \citet{Pavlovskii_Ivanova_2015_MT_from_Giants} .

Our analysis demonstrates that applying different mass transfer stability criteria to the same 1D simulations significantly alters the predicted outcome, particularly for the first phase of mass transfer (i.e., \case B RLOF in these systems, and a similar discussion also holds for those undergoing \case A RLOF). Conversely, \Case BC RLOF is consistently stable across all the criteria mentioned, as the mass-loss rates are much lower, the donor does not expand significantly, and the mass ratio is inverted.

\subsection{Uncertainties in the latter evolutionary stages}\label{sec:disc:CSM:density}
\subsubsection{{Pre-supernova outbursts}}
Precursors to the SN explosion have been observed for some interacting SNe \citep[e.g.,][]{Strotjohann2021_interactionpoweredSN}. Among \Type Ibn SNe, pre-explosion outbursts have been observed for SN\,2006jc \citep{Pastorello2007_2006jc_preSNoutburst}, SN\,2019uo \citep{Strotjohann2021_interactionpoweredSN}, and SN\,2023fyq \citep{Brennan2024_SN2023fyq_precursor_activity, Dong2024_2023fyq_precursor_eqCSM}. It is not yet clear if these outbursts are characteristic of all \Type Ibn progenitors. In some cases, pre-explosion photometry rules out the presence of bright precursors, like SN\,2015G \citep{Shivvers2017_2015G}, SN\,2020nxt \citep{Qinan2024_2020nxt}, and SN\,2022ablq \citep{Pellegrino24_2022ablq}. 

These outbursts may drive intense mass-loss, providing an alternative mechanism to produce the nearby CSM observed in many \Type Ibn SNe \citep[][]{ Dessart2022_Ibn, MaedaMoriya2022_SIMSofIbn, Takei24_IbcnLCmodelling}. Simulations of SNe with progenitors similar to those in this work were carried out by \citep{Dessart2022_Ibn}, where they included a CSM formed by an eruption in the last $\simle 2\yr$ before CC. They find that an outburst-driven CSM leads to light-curve and spectral features compatible to observations of \Type Ibn events.

 {Lower-mass models can experience Si-flashes \citep{Woosley_Heger_2015_SiFlash, Woosley2019_Hestars} that release sufficient energy to expand, or even unbind, parts of the envelope. This expansion may also be triggered by Ne-flashes \citep{WuFuller2022_wavedriven_semidegenerate_radialexpansion}. In \citet{Woosley_Heger_2015_SiFlash}, these flashes impact models with $M_\mathrm{CO,end}<1.68\Msun$, which correspond to about half of the models investigated here. While our models lack the proper physics to account for this effect, we argue that the sudden expansion alone can trigger \Case X RLOF and create a close-by and dense CSM, perhaps even accompanied by an observable pre-SN outburst.
}

\subsubsection{\EDIT{He/CO shell merger and \case X RLOF}} \label{sec:appendix:shell_merger_overshooting}

One peculiar behavior that arises in some of our models (and also found in others in the literature, such as \citealt{Tauris2013_USSNe_Ic, Tauris2015_USSNe, Wei2023_fate_of_postCCEbinaries, Guo2024_ECSNe_caseX} and, more interestingly, \WF22) is the presence of a final mass transfer event, which we labeled as \case X RLOF, shortly before CC. As noted in Sect.\,\ref{sec:res:example_mod:caseX} and \ref{sec:overv:caseX}, this phase of mass transfer is commonly linked to the He-burning shell digging significantly inside the CO-core, which is also observed in more massive models in the literature \citep[e.g.,][]{WuFuller21_wavedrivenoutburst}. 

We argue that this phenomenon is likely caused by the use of overshooting in the model. To test this, we ran the model discussed in Sect.\,\ref{sec:res:example_model} from the time of core O-depletion, and deactivated overshooting. The following evolution in terms of the development of the subsequent shells is similar, with some slight differences. We will focus exclusively on the effects close to the \EDIT{outer edge of the} CO-core.

\begin{figure}
\centering
\includegraphics[width=\linewidth]{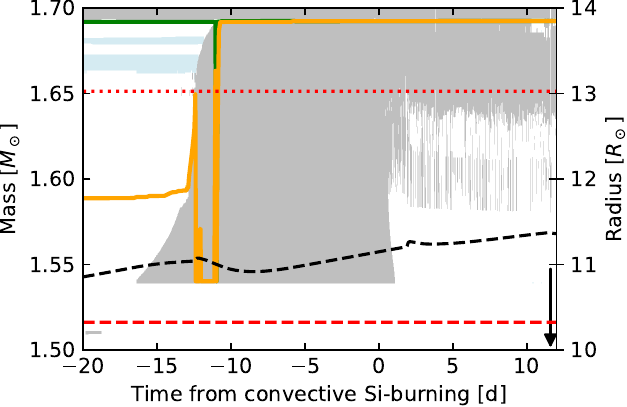}

\caption{The same Kippehnahn plot shown in Fig.\,\ref{fig:shell_merger_ex} for system B13.5\EDIT{p6.31} (Sect.\,\ref{sec:res:example_model}) and rerun from core O-depletion without any overshooting. The zero point in the $x$-axis is set to the onset of convective core Si-burning. The black arrow shows the moment when core Si-burning ends.   
}
    \label{fig:shell_merger_noos}
\end{figure}

As noted already in Sect.\,\ref{sec:res:example_mod:caseX}, Ne begins burning at a mass coordinate of $1.54\Msun$ roughly a month prior to the end of the run, and it develops a convective region on top of it. This region extends above the Ne/O rich \EDIT{region} (which extends up to $1.59\Msun$), \EDIT{thus including} the meshes that still retain residual C up to \EDIT{close to the interface with the He-burning shell}. Here, differently from the original model, He is not ingested inside this convective region, and the envelope's composition and extension remain unaltered (cf. Fig.\,\ref{fig:shell_merger_noos}). At the onset of convective Si-burning, this convective region dies off. 

This model showcases two main differences with the original one showed in Sect.\,\ref{sec:res:example_model}. Not only did the shell merger not occur, but the lifetime of the model also changed dramatically. The original model, following the onset of this external Ne-burning region, would have to wait for another $\sim\,200\text{d}$ to ignite Si convectively, as the core's internal structure is affected following the sudden mass-loss and core-erosion. Without these sudden changes to the stellar structure, convective Si burning begins only two weeks after the onset of the \EDIT{\REM{external}} Ne-burning shell \EDIT{close to the outer edge of the CO core}, and the model even reaches core-Si depletion, just 11 days after. 

In this section we show that the implementation of over- and undershooting is the direct cause for the anomalous behavior experienced in the model. We caution that it remains to be seen if these results also hold in the case of models with a different nuclear network, as the shell structure would change. It is however important to mention that the general phenomenon of shell mergers appears in 3D-hydro simulations \citep[e.g., between the C and Ne/O shells in][]{Yadav2020_shellmerger3D_CNeO}, and therefore warrants further study.

Regardless of the cause of \case X RLOF, it has already been present in previous models in the literature, though with different features and timescales, such as \WF22 and \cite{Guo2024_ECSNe_caseX} for HeS+NS systems. In their models, no shell merger is mentioned to occur, and yet a final phase of mass transfer ensues nonetheless about $\sim\,10-50\yr$ prior to core Si-ignition and the onset of an EC-SNe respectively.  Other models have exhibited such He/C shell-merger, like \cite{Habets1986_HeS_RLOF} and \cite{WLW1995_preSN_HeS}, without necessarily exhibiting this expansion phase. This may then indicate that the shell-merger we observed in the models is perhaps one of many symptoms that can cause \EDIT{the radial} expansion and thus \case X RLOF.

\subsection{CSM geometry surrounding a binary}\label{sec:disc:CSM:symm}
The structure of the CSM in our models depends on the mechanism that drives the ejection of unaccreted material. If the material is lost through isotropic re-emission \citep{Soberman97_StabilityCriteriaMassTransfer_abg}, it would distribute evenly in all directions. Alternatively, if it is lost through the accretor's outer Lagrangian point, it may accumulate near the orbital plane \citep[e.g.][]{Lu2023_L2_outflow_during_MT} and form a bound CBD. Such a disk could retain a significant fraction of the unaccreted mass \citep[at least $80\%$ of the mass lost during \case BC RLOF,][]{Pejcha16_CBD_LumRedNovae}. The presence of such massive CBD will also introduce a complex shock geometry, particularly at the boundary regions of the disk.

The recent work of \cite{TunaMetzger2023_longtermevolutionCBD_postCE} offers an analytical description of the CBD's evolution but for the case of a post-CE CBD. While the evolutionary history of disks in our models differ, their findings offer relevant insight.
They suggest that the CBD may persist for $\sim\,50\kyr$, with most of its mass lost through reaccretion onto the inner binary, while only a small fraction is lost by photoevaporation \citep{Hollenbach94_photoevaporation_of_disks}. In contrast, in our models, the CBD would be continuously supplied with mass by the inner binary. The extension of the CBD -- which is determined by the region where photo-evaporation dominates -- depends on parameters of the inner binary such as its mass and total luminosity, as well as composition and temperature of the disk \citep{TunaMetzger2023_longtermevolutionCBD_postCE}, which we have neglected here. 

 While we assume that the CSM is entirely contained within a CBD or freely expanding in all directions, the reality is likely more complex. It is reasonable to expect that, while the binary loses material to form a CBD, a fraction of the material can be lost through other mechanisms (e.g., isotropic re-emission, \citealt{Soberman97_StabilityCriteriaMassTransfer_abg}) resulting in a multicomponent CSM.   Thus, while the assumption that the RLOF-induced CSM impacts the SN is reasonable, significant uncertainties remain about its mass, extent, and geometry.

 {Low-mass HeS with high $L/M$ ratios are expected to be pulsationally unstable \citep{ SWC84_nonadiab_radpuls_heS_LM, GG90_pulsationHeS, GautschySaio95_pulsations_generic_HeS_NS_rev}. This is due to the expansion following core He depletion where the outer regions of the envelope experience HeII recombination and become locally super-Eddington,  producing a density inversion.  
 These pulsations could trigger episodic phases of mass transfer, as the star would periodically fill the Roche Lobe, therefore losing more mass and further increasing its $L/M$. These pulsations could therefore produce a more mass loaded CSM, similar to what is expected in the case of a wide RSG+MS mass-transferring binary \citep{Ercolino_widebinary_RSG}.}

\subsection{\EDIT{The effect of a close MS companion on the supernova}}\label{sec:SN:compare:inter:secondary}

The presence of a close-by MS star at the time of the explosion may affect the SN, as the ejecta rams with the companion star. While this has been predicted in the context of \Type Ia SNe \citep[see][for a review]{Liu_23_ReviewIa}, one of the first observations hinting at this process for CCSNe was recently reported for the \Type Ic \SN{2022jli} \citep{Moore2023_SN2022jli_Ic_Binary, Chen2024_SN2022jli_binary}. The peculiarity of this object lies in the $12.4\,\text{d}$ modulation of the light-curve. The extended observations presented in \citet{Chen2024_SN2022jli_binary} showcase an abrupt decline in the light-curve luminosity  (after $\sim\,300\days$), which they argue is compatible with a shortly-lived phase of mass transfer between the MS companion, thrown off of equilibrium by the impacting ejecta, and the newly-formed compact object gulping material during the passage at periastron. 

Since all of our SN progenitors are in orbit with a MS companion, this effect may be present for the tighter systems at CC. In our models, the companion typically subtends up to $1\%$ of the field of view. Assuming an explosion energy of $\sim\,6\times 10^{50}\erg$, the secondaries may capture as much as $\sim\,6\times 10^{48}\erg$ of energy, of which only $\sim\,10\%$ may be injected inside the stellar structure to drive its expansion \citep[cf., ][the remaining is instead injected onto the unbound material, which may be as much as $\simle0.01\Msun$]{Hiray2018_EjectaCompanionInteraction, Hirai2023_postSN_CEI}. If we apply these estimates to the empirical model fits in \cite{Ogata21_inflatedcompanion_afterSN}, the companion may swell to as much as $\sim\,1000\Rsun$, with a very short-lived expansion of at most a few years, which is of the same order of magnitude as the timescale for the luminosity drop in SN\,2022jli \citep{Chen2024_SN2022jli_binary}.

As the impact with the MS companion will unbind some of its H-rich envelope \citep{Liu15_CCejecta_on_MScompanion, Hiray2018_EjectaCompanionInteraction}, we expect that the SNe from our models should showcase narrow H-line emission at late times in the evolution of the SN.

\subsection{Comparison to previous works}

The mass transfer phase we focused on in our models (\case BC RLOF) results from the expansion of low-mass HeS, a phenomenon well-documented in the literature (\citealt{Paczynski1971_HeStars_firstGiant, Habets1986_HeS_RLOF, Tauris2013_USSNe_Ic, Tauris2015_USSNe, KleiserFullerKasen18_HeG_RapidlyFading_Ibc, Woosley2019_Hestars, Laplace2020_Stripped_Envelope_SNe}), which our models replicate qualitatively well.
Many previous studies of HeSs undergoing RLOF focused on systems with a compact object companion (NS or BH) following a phase of CE-evolution (e.g., \citealt{DewiPols2003_CaseBC_HeSNS,Tauris2013_USSNe_Ic,Tauris2015_USSNe,JiangTauris2021_USSNeBinary_until_CC,Jiang2023_HeS+BH_GWprog}, \WF22, \citealt{ Wei2023_fate_of_postCCEbinaries, Guo2024_ECSNe_caseX, Qin24_caseBB_HeS+NS_resultsin_GW}). In such HeS+NS binaries, the mass ratio during \case BC RLOF is typically $<0.5$, which is significantly lower than our systems. As a result, their orbit shrinks during RLOF, whereas ours expands, leading to more mass being transferred overall (see also Appendix\,\ref{sec:appendix:Mhe_Rl_q}). For the HeS+NS systems mentioned above, typical orbital periods are in the order of a few days or less, which can allow the HeSs to fill their Roche lobes during or immediately after core He burning, allowing for more overall stripping than in our models. Such a tight binary with instead a MS companion do not emerge in the \EDIT{binary} grid (Sect.\,\ref{sec:ps:binary_grid}). 

One work that modeled RLOF between a HeS and a MS companion was  \citet{Habets1986_HeS_RLOF}. In their work, RLOF occurs in a very tight orbit ($M_\mathrm{HeS}=2.5\Msun, R_\mathrm{RL,1}=1\Rsun, M_2=17\Msun$), \EDIT{in the last $\simle3\kyr$ before core Ne-burning, which removes} about $0.3\Msun$ of the HeS's envelope. This is much less than what we expect from our models (\EDIT{our models would undergo RLOF $\simgr 20\kyr$ before CC and therefore remove about $0.8\Msun$, cf. Fig.\,\ref{fig:tri_deltaMC_hecore} and Sect.\,\ref{sec:overv:caseC})}. \EDIT{Such discrepancy from our models' radius evolution is to be expected since they adopt older opacity tables and different assumptions for convective mixing, namely that it is instantaneous and that its boundary is defined with the Schwarzschild criterion}.

One of our main concerns from our method was the adoption of a limited nuclear network (cf. Sect.\,\ref{sec:methods:network}), especially during the final evolutionary phases. Prior studies in the literature that have adopted larger nuclear networks during \case BC RLOF show results broadly consistent to ours \citep[e.g.,][]{JiangTauris2021_USSNeBinary_until_CC, Jiang2023_HeS+BH_GWprog}, with some even showcasing \case X RLOF \citep{Wei2023_fate_of_postCCEbinaries, Guo2024_ECSNe_caseX}.  
In \citet{JiangTauris2021_USSNeBinary_until_CC, Jiang2023_HeS+BH_GWprog}, where they studied HeS+NS/BH binaries with an detailed nuclear network, mass transfer rates in the last months prior to collapse (i.e., \case X RLOF) were noted to be sensitive to some solver flags (their models, however, do not reach rates comparable to the extreme ones of our models). They were able to derive the SN-explosion properties from their least massive model, which yielded an explosion energy of $6\times 10^{50}\,\text{erg}$ and a remnant mass of $1.34\Msun$. These values are compatible with the ones adopted in this paper (cf. Sect.\,\ref{sec:SN:type}).

\section{Conclusions}\label{sec:conclusions}

In this work, we investigated a comprehensive set of binary evolution models. We focus on those in which the primary is stripped to become a helium star after core hydrogen depletion, and undergoes a second mass transfer phase with a main-sequence companion after core helium exhaustion up until close to core-collapse. The later phase of mass transfer is triggered in the systems where the helium-star primary has a mass of less than $3.4\Msun$ by the time of core He-depletion. During this phase, the amount of mass lost depends on how early mass transfer begins, which can remove as much as $0.8\Msun$ of the He-rich envelope for the models that started about $20\kyr$ before core-collapse. 

In our binary models, the main-sequence companion does not accrete significant amounts of the transferred material, which instead is lost from the system to form a nearby, H-free, circumbinary medium. Its typical mass of $0.4\Msun$ is large enough to give rise to significant interaction features, especially by extracting kinetic energy from the impacting supernova ejecta, which can be as high as $70\%$. Using a simplified model that neglects diffusion, we explore the impacts of different structures of the circumbinary material on the supernova light-curve. 
A circumbinary disk-like structure qualitatively reproduces the light-curve morphology of many \Type Ibn supernovae.

We find that our models produce masses and radii of the circumbinary material, as well as ejecta masses and integrated bolometric luminosities, which are compatible with those observed in many \Type Ibn supernovae.
However, these studies mostly assume a spherical distribution of the material, while our results show that a disk-like distribution is another reasonable possibility. 
Future studies on the evolution of the circumbinary material produced by binary mass-transfer will be crucial to address the structure and geometry of the material at core-collapse. This simultaneously calls for future research to explore how these large-scale asymmetries influence the light-curve and spectral evolution of interacting supernovae, and how these factors impact inferred properties from the observed data.

Finally, we developed a method to map the properties of a binary model at core He-depletion to the amount of mass being transferred during the later phase of mass-transfer. Applying this to a comprehensive grid of binary models, we find that $12\%$ of all stripped-envelope SN progenitors produced in binaries are expected to undergo mass-transfer until core-collapse, and can therefore be expected to become H-poor interacting supernovae, which is in broad agreement with observational rate estimates of \Type Ibn SNe \citep{Perley2020_ZTF_SNdemographics}. 
Future observations from ongoing surveys like ZTF, and the beginning of the science operations of LSST in 2025, will yield more significant constraints on the observed population of these supernovae.

\begin{acknowledgements}
\EDIT{The authors thank the referee Xiangcun Meng for the comments that improved the quality of the manuscript.} AE is grateful for the discussions with Lorenzo Roberti, David R. Aguilera-Dena, Thomas Tauris, Christoph Schürmann, Eva Laplace, Alexander Heger, and Ruggero Valli. \EDIT{The authors also thank Samantha Wu, Jim Fuller, and Ylva G\"otberg for comments on their data shown in this paper.}   AE acknowledges the support from the DFG through grant LA 587/22-1.

\end{acknowledgements}

\bibliographystyle{aa}
\bibliography{aanda}

\begin{appendix} 

\FloatBarrier

\section{The effect of the secondary star's mass and accretion efficiency during \case BC-RLOF}\label{sec:appendix:Mhe_Rl_q}

As initially presented in Sect.\,\ref{sec:ps:binary_grid}, there is a wide parameter space of interacting binaries which can give rise to a tight-enough HeS + MS binary where the HeS can undergo \case BC RLOF. Our models focused an initial mass ratio $q_\mathrm i=0.5$, which is only a subset of the models, and companion masses for HeS donors can have a variety of masses, ranging from a $\simle2\Msun$ to $\simgr 20\Msun$.

\begin{figure}
\centering
\includegraphics[width=1\linewidth]{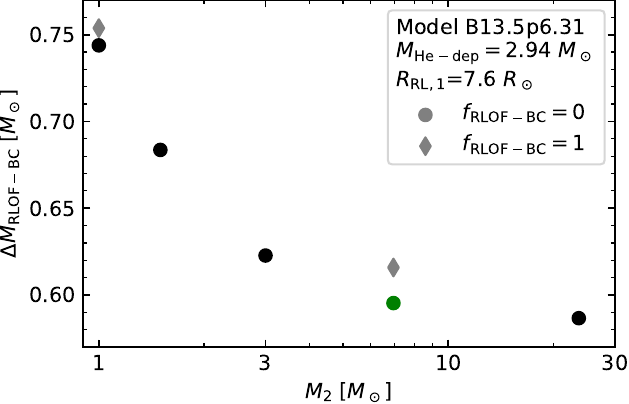}

\caption{Amount of mass shed during \case BC RLOF for model B13.5\EDIT{p6.31} (green circle) re-calculated with different companion masses (black circles) and/or accretion efficiency of unity (gray diamonds).}
    \label{fig:diff_q_eta}
\end{figure}

To measure the systematic uncertainty in our estimates on $\Delta M_\mathrm {RLOF-BC}$, we rerun model B13.5\EDIT{p6.31} (with $M_\mathrm{He-dep}=2.94\Msun$ and \EDIT{$M_2=7\Msun$}) with a secondary of a different mass $M_2$ (namely $1, 1.5, 3$ and $27\Msun$) positioned at an orbital separation such that $R_\mathrm{RL,1}=7.6\Rsun$. \EDIT{We show the results} in Fig.\,\ref{fig:diff_q_eta}. If we maintain mass transfer as inefficient, models with $M_2\geq3\Msun$ see a negligible difference in $\Delta M_\mathrm{RLOF-BC}$ compared to the original run which can be understood in the context of the orbital evolution, as RLOF is occurring from the less massive component to the more massive, causing the orbit to widen. For models where the companion is less massive, $\Delta M_\mathrm {RLOF-BC}$ increases by as much as $0.15\Msun$. These models, which would correspond to systems which initially had a mass ratio of about $0.11$ and $0.08$, would have been produced by systems that underwent unstable \case B RLOF, which would have most likely merged. As such, they are representative for the case in which the companion is a compact object, like a NS or BH. 

Our choice of $\qi=0.5$ means that in the vast majority of cases the accretor will be more massive than what we simulated, as lower $\qi$ tend to undergo unstable RLOF before producing a HeS (cf. Sect.\,\ref{sec:disc:MT_stability}). We can therefore expect that in the stable mass transferring channel, $\Delta M_\mathrm{RLOF-BC}$ will be similar to the values we estimated. We expect models with significantly less massive companions, like a NS, to have a higher $\Delta M_\mathrm{RLOF-BC}$, in the order of  $\sim\,0.15\Msun$ from our expectation. 

\begin{figure}
\centering
\includegraphics[width=1\linewidth]{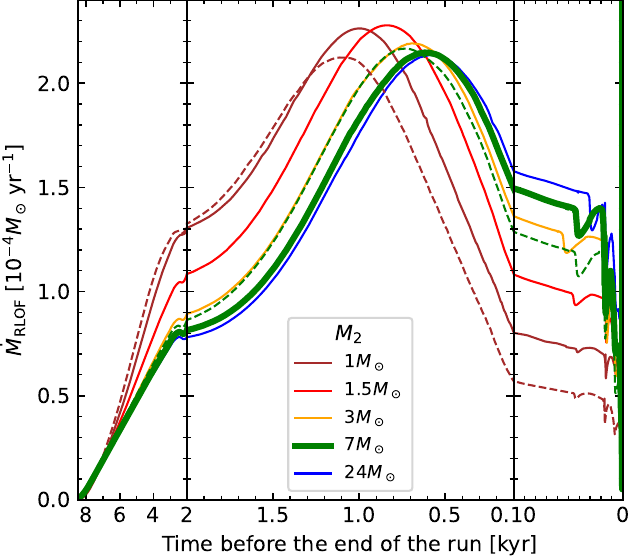}

\caption{Mass transfer rate before the end of the run for model B13.5\EDIT{p6.31} (green, thick line) with different secondary masses (see legend). For solid lines indicate the models where mass transfer is assumed inefficient, while the dashed ones assume an accretion efficiency of unity.}
    \label{fig:mdot_diff_q_eta}
\end{figure}

With regards to the mass-transfer rate before core collapse, we similarly observe small variations compared to the original run when changing the secondary's mass (cf. Fig.\,\ref{fig:mdot_diff_q_eta}). The peak of the mass transfer rate happens earlier on for more extreme mass ratios (shifting by at most $500\yr$) reaching values at most $\simle10\%$ higher. The shift of the peak at earlier times result in a more significant drop in the mass transfer rate in the last $100\yr$ before core-collapse, by as much as a factor 2, which would decrease the probability of interaction-power dominating in the light-curve of the exploding SN with a constantly-expanding CSM (cf. Sect.\,\ref{sec:SN:compare:inter:lightcurve}). The opposite is instead true for higher-mass companions.

We will focus on the effect of a higher mass transfer efficiency \case BC RLOF. We rerun model B13.5\EDIT{p6.31} two times, with a companion $M_2$ of $1\Msun$ and \EDIT{$7\Msun$} (i.e., that of the original model) and a mass transfer efficiency of 1. As the accretion of pure He on a MS star has proven difficult, we resorted to model this phase by setting the secondary star to a point mass. The amount of mass transferred \EDIT{is} increased by a negligible amount (cf. Fig.\,\ref{fig:mdot_diff_q_eta}), but this comes at the cost of no CSM being formed around the system at CC, therefore hindering the possibility of interaction features appearing in the SN. If accretion were to be efficient, the secondary star would inevitably show significant He- and N- enrichment on the surface \citep[e.g.][]{Richars2024_HeAccretion} which will be more prominent for less massive companions. 

\FloatBarrier

\section{A better parameter to calibrate the outcome of \case BC-RLOF}\label{sec:appendix:A_BETTER_PARAMETER}

\EDIT{As we have shown in Sect.\,\ref{sec:overv:caseC} (cf. Fig.\,\ref{fig:tri_deltaMC_hecore}), the amount of mass lost during \Case BC $\Delta M_\mathrm{RLOF-BC}$ qualitatively scales with the time between the onset of RLOF and CC. As such, for a given binary-stripped HeS model, we can derive $\Delta M_\mathrm{RLOF-BC}$ by estimating when a} single HeS model of the same mass \EDIT{at core He depletion $M_\mathrm{He-dep}$ reaches the donor's Roche lobe radius $R_\mathrm{RL,1}$ (cf. Fig.\,\ref{fig:tri_deltaMC_hecore}). In this section, we show that replacing $M_\mathrm{He-dep}$ with the maximum mass of the convective He-burning core $M_\mathrm{conv,He}^\mathrm{max}$ yields a more reliable estimate of $\Delta M_\mathrm{RLOF-BC}$.}

\EDIT{Since the} single- and binary-stripped HeS models \EDIT{are known to} have qualitative differences in their internal structures that affect radius evolution \citep[e.g.,][]{ Laplace2020_Stripped_Envelope_SNe}\EDIT{, it is reasonable to assume that a single- and binary-stripped HeS with the same $M_\mathrm{He-dep}$ may not actually share the same radius evolution.}

\EDIT{To identify such differences, we study the radius evolution of models B12.3\EDIT{p251}, He2.86, and He2.99 (Fig.\,\ref{fig:single_vs_binary_hestar}), which have similar $M_\mathrm{He-dep}$ (respectively $2.90$, $2.86$ and $2.99\Msun$, cf. Tables \ref{tab:data_sHeS} and \ref{tab:data}) in the range where we expect} the sharpest difference in radii, especially at later times (cf., Fig\EDIT{s.\,\ref{fig:R_vs_RL},} \ref{fig:tri_deltaMC_hecore}). \EDIT{Although we expect that the radius evolution of model B12.3p251 should be an intermediate case between that of He2.86 and He2.99,} instead \EDIT{we find that it is} almost identical to that of model He2.86 (cf. Fig.,\ref{fig:single_vs_binary_hestar})\EDIT{ with which it also} shares a very similar final CO-core mass ($M_\mathrm{CO,end}=1.62\Msun$). \EDIT{Since the mass of the core determines the luminosity and the remaining lifetime of the model, it will ultimately also determine the radius evolution.}   \EDIT{To investigate why model B12.3p251 does not exhibit an intermediate value for $M_\mathrm{CO,end}$ between that obtained in model He2.86 and He2.99, we explore the} evolution \EDIT{of each model} during core He burning. 

\begin{figure}
\centering
\includegraphics[width=1\linewidth]{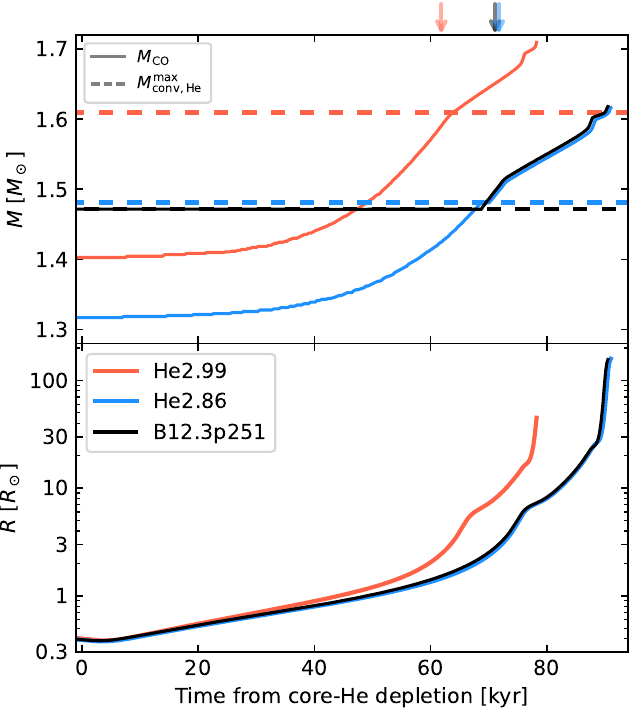}

\caption{Mass of the CO-core (\EDIT{top}) and radius (bottom) as a function of time, for two {single HeS} models (He2.97 and He2.83, in red and blue respectively) and one {binary-stripped HeS} model (B12.3\EDIT{p251} in black). In the top panel, $M_\mathrm{conv, He}^\mathrm{max}$ is shown with a dashed line for each model, and the moment of core C-ignition is highlighted with an arrow on top of the figure for each model. The models are plotted until the end of the first C-burning shell.}
    \label{fig:single_vs_binary_hestar}
\end{figure}

In \EDIT{He2.86 and He2.99}, the total mass, \EDIT{initially $3.55\Msun$ and $3.76\Msun$ respectively, decreases} steadily over time as winds erode the exposed He-rich layers. In \EDIT{B12.3p251}, which retains a thin H-rich envelope, \EDIT{the mass of the He-core is only $2.72\Msun$ at core He-ignition and grows to $3.35\Msun$ via H-shell burning before winds remove the H-rich envelope, reducing its mass (refer to Fig.\,\ref{fig:example_run} for a similar evolution). Thus, B12.3p251's He-core mass is initially smaller than He2.86 and He2.99 but later achieves an intermediate mass due to H-shell burning and winds. This impacts the He-burning core's luminosity and the extent of the convective region. In B12.3p251, the maximum mass reached by this region ($M_\mathrm{conv,He}^\mathrm{max}$) is $1.47\Msun$, smaller than He2.99's $1.61\Msun$ but again almost identical to He2.86's $1.48\Msun$.}

\EDIT{At the moment of core He-depletion, a sharp decrease in He abundance occurs below the mass coordinate $M_\mathrm{conv,He}^\mathrm{max}$, which therefore determines the region below which we expect to find the CO-core. In the models, the mass of the CO-core ($M_\mathrm{CO}$) is} defined as the outermost \EDIT{mass coordinate with} $X_\mathrm{He}<0.01$, \EDIT{and it may not initially coincide with the location of the sharp drop in He abundance at $M_\mathrm{conv,He}^\mathrm{max}$ (cf. Fig.\,\ref{fig:single_vs_binary_hestar}), due to the leftover unburnt He. %In B12.3p251, the two match since $X_\mathrm{He}=0.008$ just below the discontinuity. However, in He2.86 and He2.99, $X_\mathrm{He}$ is substantially higher, at $0.05$ and $0.15$, respectively. 
The He-shell burning that follows consumes the remaining He below $M_\mathrm{conv,He}^\mathrm{max}$, increasing $M_\mathrm{CO}$ until it coincides with $M_\mathrm{conv,He}^\mathrm{max}$, typically by the start of core C-burning. Although $M_\mathrm{CO}$ will continue growing afterwards, it is evident that its value is intimately connected with $M_\mathrm{conv,He}^\mathrm{max}$.}

\begin{figure}
\centering
\includegraphics[width=\linewidth]{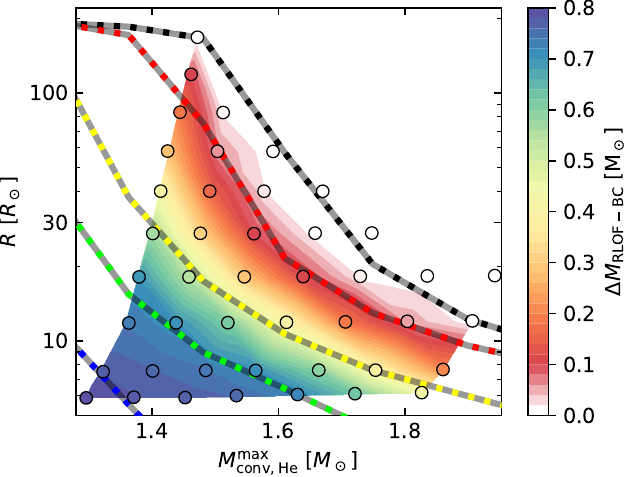}
\includegraphics[width=\linewidth]{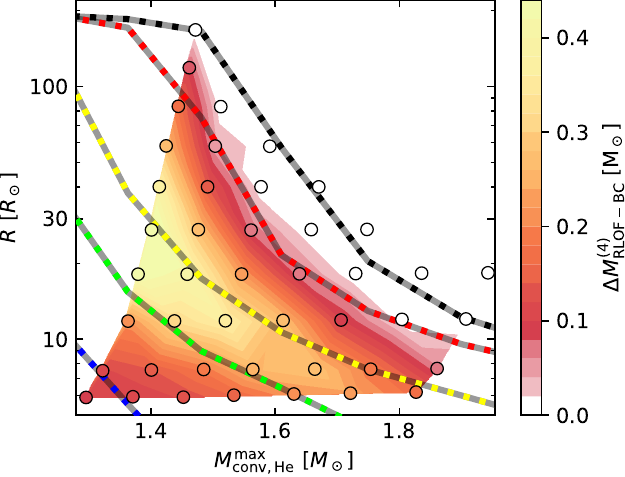}

\caption{The same maps shown in Fig.\,\ref{fig:tri_deltaMC_hecore}, where the $x$-axis is $M_\mathrm{conv,He}^\mathrm{max}$.   
}
    \label{fig:triangle_deltaMc}
\end{figure}

\EDIT{In summary, the radius evolution for a binary-stripped HeS model can be accurately predicted by single HeS models with an identical} $M_\mathrm{conv,He}^\mathrm{max}$, rather than \EDIT{those with the same} $M_\mathrm{He-dep}$. \EDIT{Re-drawing} Fig.\,\ref{fig:tri_deltaMC_hecore} \EDIT{using $M_\mathrm{conv,He}^\mathrm{max}$ in place of $M_\mathrm{He-dep}$ results} in Fig.\,\ref{fig:triangle_deltaMc}, which now shows a better agreement between the trends for $\Delta M_\mathrm{RLOF-BC}$ and the radius evolution of single HeS models. These findings, along with the \EDIT{robustness} of our estimates when \EDIT{varying} the companion mass (cf. Appendix\,\ref{sec:appendix:Mhe_Rl_q}), show the general predictive power of our method.

\section{Interaction-powered luminosity}\label{sec:appendix:L_interaction}
Let's assume that the mass un-accreted during \case BC RLOF is ejected from the system, and that it moves at a constant velocity $v_\mathrm{CSM}$. At $t=0$, the SN explosion will expel a fast moving shell (i.e., the ejecta) with initial mass $\Mej$ and initial (bulk) velocity $\vej$, which will gradually sweep the CSM as it travels through it. We will follow the evolution of the portion of the fast-moving shell that interacts with the CSM. We will refer to it as the cold dense shell \citep[CDS, ][]{Chugai_2004_SN1994W}, which has an initial mass $\Mshell(0)=\Mej\sin\theta$ (depending on the opening angle of the CSM, cf. Sect.\,\ref{sec:SN:compare:inter:cumulative}), and velocity $\Vshell(0)=\vej$. We assume that the CDS collides inelastically with the CSM, thus conserving momentum while losing kinetic energy which we assume is completely converted into radiation. We neglect any effect from photon-diffusion processes, and we only focus on the rate of energy loss from the CDS. Under this assumption, the amount of kinetic energy that is converted into radiation as the CDS sweeps through a small shell of mass $\mathrm{d}m$ is equal to
\begin{equation}\label{eq:dEkin}
    \de E = \frac{1}{2}\Mshell\Vshell^2 + \frac{1}{2}\de m \cdot v_\mathrm{CSM}^2 - \frac{1}{2}\left(\Mshell+\de m \right)\left(\frac{\Mshell\Vshell+\de m\cdot  v_\mathrm{CSM}}{\Mshell+\de m}\right)^2.
\end{equation}
Performing a series expansion for $\de m$, under the assumption that at any given time interval $\de t$ we have $\de m\ll \Mshell$, results in
\begin{equation}\label{eq:dEkin2}
    \de E = \frac12 \de m\left(\Vshell-v_\mathrm{CSM}\right)^2,
\end{equation}
which, if divided by the time interval, gives the interaction luminosity
\begin{equation}\label{eq:L_inter_general}
    L_\mathrm{inter} = \frac{\de E}{\de t} =  \frac12 \frac{\de m}{\de t}\left(\Vshell-v_\mathrm{CSM}\right)^2
\end{equation}
We can write a differential equation for the CDS's velocity $\Vshell$ per conservation of momentum, as
\begin{equation}\label{eq:diff_vej_general}
    \frac{\de\Vshell}{\de t} = -\frac{\de m}{\de t}\frac{\Vshell(t)-v_\mathrm{CSM}}{\Mshell(t)}.
\end{equation}
The CDS's mass $\Mshell$ will also increase over time, as the CDS gradually sweeps material from the CSM
\begin{equation}\label{eq:diff_Mej_general}
    \frac{\de \Mshell}{\de t} = \frac{\de m}{\de t}.
\end{equation}
Once $\frac{\de m}{\de t}$ is known, Eq.\,\ref{eq:diff_vej_general} and \ref{eq:diff_Mej_general} form a complete set of differential equations, once we include the initial conditions $\Mshell(0)$ and $\Vshell(0)$. Solving for $\Mshell(t)$ and $\Vshell(t)$, then $L_\mathrm{inter}(t)$ can be evaluated using Eq.\,\ref{eq:L_inter_general}. Note that the information on the geometry of the CSM is found inside $\de m$ and the initial condition on the CDS's mass.

We will now describe the mass-element of the CSM, assuming this was either constantly expanding following mass loss from the inner binary or a CBD, and derive $L_\mathrm{inter}$ for each case.

\subsection{Free-moving CSM}\label{sec:appendix:L_interaction_vCSM}
If we assume that the CSM is polluted following mass-loss, with the material moving away from the system, we can use the mass-loss history of the system to infer the swept up mass $\de m$ per unit time $\de t$. 

To know the time at which the material $\de m$ was ejected before CC ($t_\mathrm{RLOF}$), one needs to solve the following integral equation, where the moment of CC is set to 0,
 \begin{equation}\label{eq:integral_RLOF_CSM_TIMESCALE}
    \int_{-t_\mathrm{RLOF}}^{t} v_\mathrm{CSM}(t')\de t' = \int_0^{t}\Vshell(t')\de t'.
\end{equation}
As we have no constrain on $v_\mathrm{CSM}$, we assume it is for simplicity constant. With this, we can take the derivative of the previous equation by $t$ to obtain the following differential equation
\begin{equation}\label{eq:RLOF_CSM_TIMESCALE_general}
\frac{\de t_\mathrm{RLOF}}{\de t} = \frac{\Vshell(t)-v_\mathrm{CSM}}{v_\mathrm{CSM}},
\end{equation}
where a physical solution exists as long as  $\Vshell(t)>v_\mathrm{CSM}$. This provides a scaling factor between the mass-loss rate from the system $\dot M_\mathrm{RLOF}=\frac{\de m}{\de t_\mathrm{RLOF}}$ and the amount of mass swept by the CDS as it travels through the CSM per unit time 
\begin{equation}
\frac{\de m}{\de t} = \frac{\de m}{\de t_\mathrm{RLOF}}\frac{\Vshell(t)-v_\mathrm{CSM}}{v_\mathrm{CSM}}=\dot M_\mathrm{RLOF}(t_\mathrm{RLOF})\frac{\Vshell(t)-v_\mathrm{CSM}}{v_\mathrm{CSM}},
\end{equation}
where we retrieve $\dot M_\mathrm{RLOF}$ from the model by looking up the mass-loss rate at a time $t_\mathrm{RLOF}$ prior to CC by solving Eq.\,\eqref{eq:RLOF_CSM_TIMESCALE_general}. Substituting this back in Eq.\,\eqref{eq:L_inter_general} yields
\begin{equation}\label{eq:interaction_luminosity}
    L_\mathrm{inter}(t) = \frac12 \dot M_\mathrm{RLOF}\left(t_\mathrm{RLOF}(t)\right)\frac{ \left(\Vshell(t)-v_\mathrm{CSM}\right)^3}{v_\mathrm{CSM}}.
\end{equation}

For the generic solution,  Eq.\,\eqref{eq:interaction_luminosity} needs to be coupled with Eq.\,\eqref{eq:RLOF_CSM_TIMESCALE_general}, and the differential equations for the velocity and mass of the CDS (Eq.\,\eqref{eq:diff_vej_general} and \eqref{eq:diff_Mej_general}) which are given by
\begin{equation}\label{eq:system}
    \left\{ 
    \begin{aligned}
    \frac{\de \Vshell}{\de t} &= -\frac{\left(\Vshell(t)-v_\mathrm{CSM}\right)^2}{v_\mathrm{CSM}}\frac{\dot M_\mathrm{RLOF}\left(t_\mathrm{RLOF}(t)\right)}{\Mshell(t)}\\
    \frac{\de \Mshell}{\de t} &= \dot M_\mathrm{RLOF}\left(t_\mathrm{RLOF}(t)\right)\frac{\Vshell(t)-v_\mathrm{CSM}}{v_\mathrm{CSM}}
    \end{aligned}
    \right.,
\end{equation}
and we have to add the initial condition for the differential equation Eq.\,\eqref{eq:RLOF_CSM_TIMESCALE_general}, that is $t_\mathrm{RLOF}(0)=0$. 

\subsection{CSM bounded in a CBD}\label{sec:appendix:L_interaction:CBD}

We can assume that $\Mcsm$ is bounded within a CBD (or, more precisely, a torus) of internal radius $\Rin$, outer radius $\Rout$ and opening angle $\theta$. We assume that the material inside the CBD has no bulk movement, such that $v_\mathrm{CSM}=0$. We write the mass-element as follows:
\begin{equation}
    \de m = 4\pi \sin\theta \,\rho(r)\, r^{2}\de r.
\end{equation}
Let us work under the assumption that $\rho(r)=\rho_0\left(\frac r {\Rin}\right)^{s}$, with $\rho_0$ as the characteristic density at a distance $\Rin$, which is equal to 
\begin{equation} 
\rho_0= \begin{cases}
\Mcsm  \left[\frac {4\pi} {3+s} \Rin^{-s} \left(\Rout^{3+s}-\Rin^{3+s}\right)\sin\theta\right]^{-1} & s \neq -3 \\
\Mcsm  \left[ 4\pi \Rin^3\left( \ln\frac{\Rout}{\Rin}\right)\sin\theta\right]^{-1} & s = -3 \\

\end{cases}.
\end{equation}
We define the following quantity
\begin{equation} 
\Rp p = \begin{cases}
 \frac {1} {p} \left(\Rout^{p}- \Rin^{p}\right)  & p \neq 0\\\ln\left({\Rout}/{\Rin}\right)  & p = 0
\end{cases}
\end{equation}
which carries the information of a length to the power of $p$. We thus obtain the following form for the mass-element
\begin{equation} \label{eq:dm_CBD}
\de m =   \Mcsm \frac{r^{2+s}\de r}{\Rp {3+s}}
\end{equation}
We now have all the information necessary to rebuild the differential equations Eqs.\,\ref{eq:diff_vej_general}-\ref{eq:diff_Mej_general}, by writing $v(t)=\frac{\de r}{\de t}$, which are
\begin{equation}\label{eq:dmej_vej_CBM}
    \left\{ 
\begin{aligned}
    \frac{\de \Mshell}{\de t} &=\Mcsm \Vshell(t)\frac{r(t)^{2+s}}{\Rp {3+s}}\\
     \frac{\de \Vshell}{\de t} &= -\frac{\Mcsm}{\Mshell(t)}\Vshell^2(t) \frac{r(t)^{2+s}}{\Rp {3+s}}
\end{aligned}.\right.\end{equation}
And finally, combining Eqs.\,\ref{eq:L_inter_general} and \ref{eq:dmej_vej_CBM} yields
\begin{equation}\label{eq:Linter_peak_vr}
L_\mathrm{inter}(t) = \frac 1 2 \Mcsm \Vshell^3(t) \frac{r(t)^{2+s}}{\Rp {3+s}}.
\end{equation}
Do note that, because of the assumption that the disk is stationary, the equation for conservation of momentum simplifies to 
\begin{equation}\label{eq:mom_cons}
    \Mshell(t) \Vshell(t) = \Mshell(0)\Vshell(0) = \Mej \sin\theta \vej
\end{equation}
making one of the equations in Eq.\,\eqref{eq:dmej_vej_CBM} redundant. We use Eq.\,\eqref{eq:mom_cons} to get rid of $\Mshell(t)$ in the equations. We therefore only need to solve the equation of motion of the CDS, which is a second-order non-linear differential equation 
\begin{equation}
    \ddot r = - \frac{\Mcsm}{\Mej\sin\theta}\frac{1}{\vej}\frac{1}{\Rp{3+s}}\dot r^3 r^{2+s}
\end{equation}
where $\ddot r=\de v/\de t$ and $\dot r=v(t)$. It is possible to obtain the generic (albeit implicit) solution
\begin{equation}
        c_1r(t) + \frac{1}{4+s}\frac{1}{3+s}\frac{\Mcsm}{\Mej\sin\theta}\frac{1}{\vej}\frac{r(t)^{4+s}}{\Rp{3+s}}=t+c_2
\end{equation}
where $c_1$ and $c_2$ can be determined using the initial conditions at $t_0=\frac{\Rin}{\vej}$ which are $r(t_0)=\Rin$ and $v(t_0)=\vej$. 
We can therefore write the analytical expressions for $r(t)$ and $v(t)$ as
\begin{equation}\label{eq:solution_Linter}
    \left\{\begin{aligned}
       r(t)-\vej t &= \frac{\Mcsm}{\Mej\sin\theta}\left[ \frac{\Rin^ {s+3}(r(t)-\Rin)}{\Rout^{3+s}-\Rin^ {3+s}}-\frac{1}{4+s}\frac{r(t)^ {4+s}-\Rin^{4+s}}{\Rout^{3+s}-\Rin^ {3+s}}\right] \\
        v(t) &= \vej\left( 1+\frac{\Mcsm}{\Mej\sin\theta}\frac{r(t)^{3+s}-\Rin^{3+s}}{\Rout^{3+s}-\Rin^{3+s}}\right)^{-1}
    \end{aligned}\right. .
\end{equation}

%\section{Other SNe}\label{sec:appendix:otherSNe}
%In the main text (Sect.\,\ref{sec:SN:compare}) we only focused on \Type Ibn and Icn SNe. Here, we will also report other SNe that show interaction features, namely the low-energy USSNe and Ca-rich transients (CaRTs, Appendix\,\ref{sec:SN:obs:US}) and SLSNe (Appendix\,\ref{sec:SN:obs:SL}).

%\subsection{USSNe and CaRTs}\label{sec:SN:obs:US}

\multilinecomment{
\begin{figure*}
\centering
\includegraphics[width=1\linewidth]{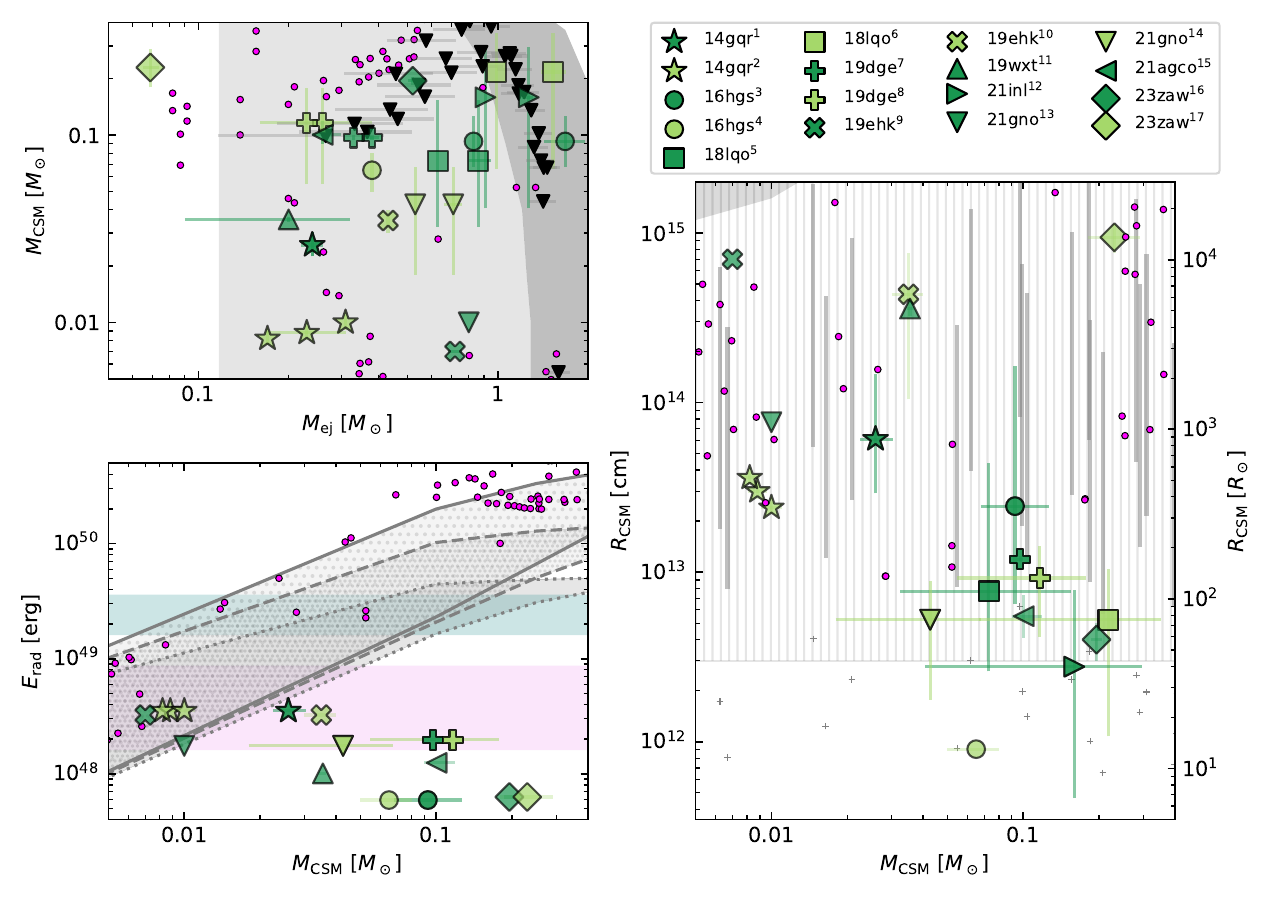}
\caption{The same as Fig.\,\ref{fig:Mej_Mcsm_Ibn} but for USSNe and CaRTs. \\
 {References:}; 
\object{iPTF14gqr}: (1) \citealt{Yao2020_2019dge, De2018_iPTF14gqr}
(2) \citealt{De2018_iPTF14gqr}; 
\object{iPTF16hgs}:
(3) \citealt{Yao2020_2019dge, De18_iPTF16hgs}
(4) \citealt{De18_iPTF16hgs}; 
\object{SN2018lqo}:
(5) \citealt{Yao2020_2019dge}
(6) \citealt{Das2023_DPSNe_with_Mext}; 
\object{SN2019dge}:
(7) \citealt{Yao2020_2019dge}
(8) \citealt{Das2023_DPSNe_with_Mext, Yao2020_2019dge}; 
\object{SN2019ehk}:
(9) \citealt{JG2020_2019ehk}
(10) \citealt{Nakoka2021_2019ehk, JG2020_2019ehk}; 
\object{SN2019wxt}:
(11) \citealt{SN2019wxt_USSNe_candidate}; 
\object{SN2021inl}:  (12) \citealt{Das2023_DPSNe_with_Mext}; 
\object{SN2021gno}:
(13) \citealt{Ertini2023_SN2021gno}
(14) \citealt{Das2023_DPSNe_with_Mext}; 
\object{SN2021agco}:
(15) \citealt{Yan2023_USSNeIb_2021agco}; 
\object{SN2023zaw}:
(16) \citealt{Das2024_SN2023zaw}
(17) \citealt{Moore2024_SN2023zaw}
}
    \label{fig:Mej_Mcsm_US}
\end{figure*}
}

\end{appendix}

\end{document}